\documentclass{article}
\usepackage[utf8]{inputenc}
\usepackage[sort&compress,numbers]{natbib}
\usepackage{amsmath}
\usepackage{mathtools}
\usepackage{amssymb}
\usepackage{hyperref}
\usepackage{cleveref}
\usepackage{caption}
\usepackage{subcaption}
\usepackage{fullpage}
\usepackage{afterpage}
\usepackage{graphicx}
\usepackage{fancyvrb}
\usepackage{booktabs}
\usepackage{placeins}
\usepackage{soul}
\usepackage{xcolor}
\usepackage[a4paper, total={6.5in, 9.5in}]{geometry}
\usepackage{setspace}
\usepackage[version=4]{mhchem}
\usepackage{mymacros_11Sep20}
\usepackage{tikz,pgfplots}
\usepgfplotslibrary{colormaps}
\usepackage{tikz-3dplot}
\usetikzlibrary{arrows.meta}
\usepackage[affil-it]{authblk}
\usepackage{comment}
\usepackage[title, toc]{appendix}

\AtBeginEnvironment{appendices}{\crefalias{section}{appendix}}

\title{On first-order thermodynamic equilibrium conditions for fluid-fluid and solid-phase interfaces}
\author[1]{Nicodemo Di Pasquale  \thanks{Corresponding author: nicodemo.dipasquale@unibo.it}}
\author[2]{Thomas Hudson}
\affil[1]{Department of Industrial Chemistry, University of Bologna, Bologna, Italy}
\affil[2]{Warwick Mathematics Institute, University of Warwick, Coventry, CV4 7AL, United Kingdom}
\date{}

\newcommand{\id}{\mbox{\textbf{I}}}

\newcommand{\grandpzero}{\Omega}

\newcommand{\totmolconc}{\rho^{\mathrm{tot}}}
\newcommand{\helm}{\psi}
\newcommand{\area}{\Sigma}
\newcommand{\bline}{\Gamma}
\newcommand{\vol}{V}
\newcommand{\ncomp}{n}

\newcommand{\press}{P}

\newcommand{\cauchy}{S}
\newcommand{\bfcauchy}{\boldsymbol{\cauchy}}
\newcommand{\eshelby}{\mathbb{E}}

\newcommand{\jacob}{J}

\newcommand{\dxi}{\boldsymbol{\delta\xi}}
\newcommand{\dxin}{\delta\xi_n}

\newcommand{\interst}{\mathcal{I}}
\newcommand{\subst}{\mathcal{S}}

\renewcommand{\bfn}{{\boldsymbol{n}}}
\renewcommand{\bft}{{\boldsymbol{t}}}
\renewcommand{\bfx}{{\boldsymbol{x}}}

\renewcommand{\bfnu}{{\boldsymbol{\nu}}}

\newcommand{\strain}{F}
\newcommand{\bfstrain}{\mathbf{F}}

\newcommand{\sstress}{\bfsigma^\area}  
\newcommand{\lstress}{\bfsigma^\bline}  

\newcommand{\diffpot}{\Lambda}

\newcommand{\CHv}{\bfX} 

\DeclareMathOperator{\tr}{tr}

\begin{document}

\maketitle

\begin{abstract}
    Starting from the constrained variational formulation of Larch\'{e} and Cahn for solids in contact with fluids, a unified thermodynamic framework for such systems is developed. Both bulk and interfacial equilibrium conditions arise as stationarity conditions of a single thermodynamic functional. While earlier theory neglects a full treatment of interfacial contributions and therefore cannot describe systems in which surface effects are significant, the present formulation incorporates interfacial thermodynamics directly into the variational principle. The framework is introduced first for fluid-fluid systems to establish the underlying mathematical structure, and is then extended to solid-fluid interfaces. Within this setting, classical equilibrium relations for heterogeneous systems emerge naturally from different classes of admissible variations, providing a common theoretical basis for bulk, interfacial, and configurational thermodynamics.
\end{abstract}

\section{Introduction}
The determination of interfacial properties in solid-fluid systems is critical to our understanding of a plethora of phenomena which have profound implications in materials science, metallurgy, geophysics, and biomaterials engineering. An important example is the nucleation mechanism which drives solid-fluid phase transitions \citep{Hoose2012,Kalikmanov2012,Sosso2016,Sleutel2014}, which is essential in understanding and predicting phenomena such as water freezing \citep{Zhang2016,espinosa2016interfacial,Montero2023,Espinosa2016,Zhang2018Ice}. Another arises in the solidification of metals \citep{Wang2020Metal}, where the final quality of casting depends on the interfacial properties among the solid in contact with its melt \citep{Asta2009}. For further discussion and other examples, we refer the interested reader to the review \citep{DiPasquale2025}, which discusses the importance of solid-fluid interfaces and the challenges of determining their properties using Molecular Dynamics simulations.

Although there were many earlier contributions to the thermodynamic theory of systems containing multiple phases in equilibrium, it was only with Gibbs and his monumental work \citep{Gibbs1957} that the theory in this area received firm foundations. In particular, the insight of Gibbs was based on the assumption that at equilibrium, the energy of a system is minimal while other macroscopic properties of the system are kept constant, for example the total entropy and total mass of each of each chemical component making up the system.  Taking this viewpoint, Gibbs derived the necessary conditions that a system at thermodynammic equilibrium must satisfy, expressed as relations between the various different physical observables of the system. 

When considering the interface between different phases of matter, Gibbs observed that the energy of a system containing an interface is not generally equal to the sum of the energies contained within its component subsystems. Instead, as the interfacial region between the phases is a region of inhomogeneity, it must be treated differently from bulk phases. Gibbs addressed this challenge by introducing the concept of dividing surface and formulated interfacial thermodynamics using excess quantities in \citep{Gibbs1957}. Within this framework, the total thermodynamic properties are decomposed into contributions from the bulk phases and an interfacial excess. Gibbs assigned the excess quantities to a two-dimensional surface which has since been referred to as the \textit{Gibbs dividing surface}. The Gibbs dividing surface is responsible for its own contribution to the total extensive thermodynamic variables (entropy, energy, and moles of components) and is characterized by an extra thermodynamic quantity representing the extra (reversible) work required to create a unit area of interface. Notably, Gibbs obtained a complete description of the fluid-fluid interface (see also \citep{Defay1966} for a more modern account of the theory), whereas for solid-fluid interfaces, he observed that additional complications arise. 

Since the time of Gibbs' work, various studies have addressed the problem of the thermodynamic equilibrium for solid-fluid and solid-solid interfaces. One of the main observations which differentiates systems involving solid phases from those containing only fluid phases was reported by Shuttleworth, who derived the equation that now bears his name \citep{Shuttleworth1950,Herring1951}. This equation was recently re-derived in a statistical mechanical context; see \citep{DiPasquale2020}.
Subsequent developments in the thermodynamic theory of heterogeneous systems containing solids were presented by Larch\'{e} and Cahn in a series of papers \citep{Larche1973,Larche1978,Larche1978B}, where they derive the thermodynamical equilibrium relations as a solution of a constrained minimization problem. In their analysis, they introduced an extra constraint to which systems including a solid phase must obey, which they term the `conservation of the network'. One of the consequences of their theory is a rigorous derivation of a relation between stress  and the material equilibrium within a solid material \citep{Larche1985,Larche1982}, which was described decades earlier \citep{Herring1950}.

Larch\'{e} and Cahn's analysis rigorously extends the work of Gibbs by giving the thermodynamic equilibrium relation in cases where solids are involved; their theory was recently tested against experiments giving a consistent description of the experimental results \citep{Shi2018}. 

However, their formulation does not include an explicit interfacial excess contribution in the constrained functional, which is the focus of the present work. Although interfaces enter their analysis through phase-equilibrium conditions and the transfer of lattice sites across phase boundaries, they are not treated as distinct thermodynamic subsystems endowed with an explicit excess free energy and the associated capillary stresses. Given that their primary aim was the analysis of macroscopic systems, for which interfacial contributions may be comparatively small, this decision may be viewed as reasonable. Since then, other works have appeared that extend Larch\'{e} and Cahn's analysis, see for example \citep{Leo1989,Mullins1985}, including some analysis of phase boundaries.

Nevertheless, there are contexts (primarily close to the nanoscale, illustrated by the examples discussed above) in which the surface terms required to correctly describe equilibrium are important. In this work, we therefore extend the work of Larch\'{e} and Cahn by including the effect of interfaces in the constrained minimization problem. We also extend and generalize their analysis to include more general interfaces, which need not be flat or isotropic. The main novelty of our work when compared with previous works is the fact that we only assume the energy to depend on some properties of the system which could be thermodynamic (entropy or moles density) or geometric (deformation state, direction of the interface normal) or a combination of both, depending on the system we are considering, (e.g., if we are discussing the fluid phase we are not making the energy depending on the deformation state of the system). By providing a general geometric approach to variation, we show that all the well-known relations used in the thermodynamic of interfaces (Young-Laplace equation, Shuttleworth equation, Cahn-Hoffman vector) naturally emerge from a unified formalism. In particular, we carefully describe the different types of variation employed, including the class of \textit{inner variations}, which pertains to the deformation of the physical domain considered, as well as \textit{configurational variation}, which are especially relevant to solids because phase-boundary motion changes the material configuration and may be coupled to the generation or relaxation of elastic strain \citep{AlexanderJohnson1985,Gurtin2000}. The inner variation formalism represents the bridge between the thermodynamic and the geometric description of the systems and we will show the generality of this geometrical thermodynamics approach by re-deriving the known equilibrium equations which are usually scattered throughout the literature where they are obtained with mostly \textit{ad hoc} assumptions and derivations.

One feature of the literature on heterogeneous systems is that the various equilibrium relations governing bulk phases, interfaces and material configurations have historically been developed within largely independent theoretical frameworks. Gibbs' excess formalism provides the thermodynamic description of fluid interfaces \citep{Gibbs1957}, Shuttleworth established the relationship between surface free energy and surface stress in deformable solids \citep{Shuttleworth1950}, Cahn and Hoffman introduced the vectorial formulation required to describe anisotropic interfaces \citep{Hoffman1972,Cahn1974}, Kirkwood and Buff derived a mechanical route to the determination of interfacial free energies in fluids \citep{Kirkwood1949}, while Eshelby and, later, Gurtin developed configurational mechanics to describe the equilibrium and evolution of material configurations \citep{Eshelby75,Gurtin2000}. Although these theories address closely related physical phenomena, they are generally presented as separate developments, each relying on its own assumptions, variational arguments and mathematical formalism.

The present work suggests a different perspective. We formulate the equilibrium of heterogeneous systems as a single constrained thermodynamic minimization problem in which different classes of admissible variations naturally give rise to different equilibrium conditions. Variations of the thermodynamic fields recover the classical bulk equilibrium relations, mechanical variations of the spatial configuration lead to force balance and interfacial equilibrium, variations of the interface geometry generate the generalized Young-Laplace, Shuttleworth and Cahn-Hoffman relations, while configurational variations associated with lattice-label conservation give rise to the Eshelby energy-momentum tensor. Within this unified framework, classical equilibrium relations that have historically been developed within separate theoretical settings appear as complementary stationarity conditions of the same thermodynamic functional.

Besides providing a unified interpretation of these classical theories, the proposed framework also clarifies the assumptions under which each result is valid and provides a systematic basis for extending equilibrium thermodynamics to more general heterogeneous systems, in the spirit of the theory of rational mechanics.
The present formulation is developed for equilibrium solid–liquid systems containing a single crystalline phase; extensions to coherent solid–solid interfaces are only briefly discussed and will be left for future works. We further note that the scope of this work is to focus only on first-order necessary conditions, rather than verifying local convexity (i.e. positivity of second variations) at equilibrium.

The remainder of the paper is organized as follows; we first present the general constraint minimization problem to be solved. We therefore proceed to solve it in specific situations, starting from the well-known case of the fluid-fluid systems and we then move to the more general case of three separated phases in contact. Once the methodological analysis is established, we move to the solid-fluid cases, deriving all the relevant first-order equilibrium conditions. We then discuss the relations we found and connect them with the known literature results before drawing some conclusions.

\section{General Variational Problem}

In this section we will describe the constrained variational problem we want to solve and fix the notation for the rest of the work.

We start by expressing the total internal energy $U$ and the total entropy $S$ of a system as the integral over the volume of the internal energy density field $u$ and the entropy density field $\eta$:
\begin{subequations}\label{eq:def} 
\begin{align}
    U & = \int_{\vol}{u \,\de v} \label{eq:defU}  \\
    S & = \int_{\vol}{\eta \,\de v.} \label{eq:defS} 
\end{align}
\end{subequations}
For a system of $\ncomp$ different chemical components, we assign to the $i$-th component a molar density $\rho_i$, so that the total moles of component $i$ are
\begin{equation}
\label{eq:def2}
        M_i  = \int_{\vol}{\rho_i \,\de v.} 
\end{equation}
Throughout this work, the amount of matter is expressed on a molar basis. For a fluid phase, the choice between molar and mass densities is largely a matter of convention, since the two formulations are directly related through the molar masses of the components. In a crystalline solid, however, molar quantities are more natural for the present formulation, as they provide a more direct description of the material structure and composition of the solid, in a sense that will be made more precise in the following sections. A mass-based formulation would nevertheless remain possible, provided that all related thermodynamic quantities were transformed consistently. Accordingly, molar densities are used throughout the following derivations.

As a general rule, we will use lower case letters for density fields and upper case letters for total quantities which are integrated over a physical domain. All fields are functions of the spatial three coordinates $\bfx=(x,y,z)$, even if in the following we will not typically acknowledge this dependence to ease notation.

Our fundamental assumption is then that the internal energy density can be expressed as a function of local state variables, including at least the local entropy and the local composition of the system:
\begin{equation}\label{eq:energy}
    u = u(\eta,\rho_1,\ldots,\rho_\ncomp).
\end{equation}
As we will show in the next section, a full constitutive description of the system may require further dependencies: the energy may also depend on the geometric features of the system such as local measures of strain and other internal variables.

In order to study the minimisation problem further, we will need to distinguish between a system in some reference state and a system in the current state. Although the latter distinction turns out to be immaterial for simple fluids, since such fluids do not resist static shearing, it becomes essential when considering solids. 
We will therefore use, when needed, the subscripts $0$ to indicate a system in the reference state. \Cref{eq:def,eq:def2} will therefore become for solids:
\begin{subequations}\label{eq:defSol} 
    \begin{align}
    U_0 & = \int_{\vol_0}{u_0 \,\de v_0} \\
    S_0 & = \int_{\vol_0}{\eta_0 \,\de v_0}\\
    M_{0,i}&=\int_{\vol_0}\rho_{0,i}\,\de v_0.
\end{align}
\end{subequations}
Here, $\vol_0$ represents the volume of the system in the reference state. We note that we will discuss the various implications of our implicit assumption that such a reference configuration can be determined in \cref{sec:solid-fluid}.

Following Gibbs, an \emph{equilibrium state} is one that minimises the internal energy subject to appropriate constraints. We can find such equilibrium states by considering a variational problem:
\begin{equation}
    \min_{\eta,\rho_1,\ldots,\rho_\ncomp} \int_{\vol}{u(\eta,\rho_1,\ldots,\rho_\ncomp) \,\de v}
\end{equation}
subject to the constraints that the material has:
\begin{itemize}
    \item[(A)] Fixed total entropy: 
        \begin{equation}
            S = \int_{\vol}{\eta\,\de v} = \mbox{constant;} 
        \end{equation}
    \item[(B)] Fixed total moles of each component: 
        \begin{equation}
            M_i = \int_{\vol}{\rho_i\,\de v} = \mbox{constant; and} 
        \end{equation}
    \item[(C)] Fixed total volume, so the domain occupied by the phases, $\vol$, has fixed boundaries.
\end{itemize}
The classical approach to enforcing these constraint is to introduce an extended functional that includes Lagrange multipliers to enforce these constraints, and we will follow this approach in the developments below.

\subsection{Notions of variation}
In order to derive the local conditions necessary for thermodynamic equilibrium to hold, we will assume that the system of interest is free to change its properties by several possible mechanisms, which we now detail.

The first possible route is for the dependent variables (i.e. the thermodynamic fields considered) to change their values at a given spatial point. For example, the density of component $i$ may change from the value $\rho_i(x)$ to a new value $\rho_i(\bfx)+\delta \rho_i(\bfx)$ at the spatial point $\bfx=(x,y,z)$, where $\delta\rho_i(\bfx)$ is the perturbation. One possible physical mechanism that could lead to such change would be a chemical reaction occurring at the point $\bfx$.

A second possible route for variation is more geometric in nature, reflecting that fact that the matter may deform and flow in space. Instead of changing the value of a dependent field, the location at which the field is evaluated is instead perturbed by a displacement $\dxi(\bfx)$; variations of this sort allow for the motion of material points within the body, and hence for the transport of the physical fields. Under such variation, scalar quantities attached to material particles at a spatial point $\bfx$ are transported to $\tilde{\bfx} = \bfx+\dxi(\bfx)$ according to $\tilde{q}(\tilde{\bfx}) = q(\bfx)$. Volumetric densities such as the species moles densities and the entropy density transform according to
conservation of their associated material measures:
\[
	\tilde\eta(\tilde{\bfx})\de \tilde{v} = \eta(\bfx) \de v,  \qquad\qquad \tilde\rho_i(\tilde{\bfx})\de \tilde{v} = \rho_i(\bfx) \de v,\quad i=1,\dots,\ncomp.
\]
where $\de \tilde{v}$ is the element of volume in the perturbed (or deformed) configuration and $\de v$ is the element of volume in the current configuration.
Consequently, their pointwise values generally change  when the local volume changes. The internal-energy density is not independently transported, but is evaluated from the constitutive relation using the transported entropy and moles densities.

A third possible route by which the system may vary is through the transformation of one phase into another. If the phases are immiscible, this can only happen through changes at the boundary of the volume, or by the generation of inclusions of one phase within another.


In the mathematical theory of the Calculus of Variations, these classes of change to the system are respectively known as outer variations, inner variations and shape derivatives \cite{GH95}. We will see that a proper treatment of the class of inner variations and shape derivatives is particularly relevant when considering the separation of phases and possible geometric changes to the phase boundaries, and this will be a particular focus here. Assuming that the system is free to change through these two possible classes of variation, it is clear that at equilibrium, any such variations should lead to local increases in the internal energy. This leads us naturally to necessary conditions for equilibrium that we outline here. A third class of variation will be introduced, the \textit{configurational variation}, which represents a variation of the identity of the phases involved.

\subsubsection{Outer variations}\label{sec:outvarLL}
Suppose that smooth local variations in each of dependent fields $\eta$, $\rho_i$ are possible at any spatial point, subject to appropriate physical constraints. As mentioned above, it is standard to introduce Lagrange multipliers to enforce the constraints (A)--(C) above, and to therefore consider the augmented functional
\begin{equation}\label{eq:augFunct}
E[\eta,\rho_1,\ldots,\rho_n;V] := \int_{\vol} u\,\de v + T\left(S-\int_{\vol} \eta\,\de v\right)+\sum_{i=1}^n\mu_i\left(M_i-\int_{\vol}\rho_i \,\de v\right),
\end{equation}
where we introduced the unknown coefficients $T$, $\mu_i$, as the Lagrange multipliers required to enforce the constraints on the entropy and moles of each species. At equilibrium, these coefficients are identified with intensive thermodynamic quantities. The total volume is not enforced by an additional multiplier, since the external domain $V$ is prescribed and the admissible variations are restricted so that normal displacements vanish on $\partial V$.

If equilibrium is attained for the fields $\eta$ and $\rho_i$, admissible variations in these fields must lead to an increase in the constrained internal energy, and hence
\[
E[\eta+\delta\eta,\rho_1+\delta \rho_1,\ldots,\rho_\ncomp+\delta\rho_\ncomp]-E[\eta,\rho_1,\ldots,\rho_\ncomp]\geq 0.
\]
Under assumptions of sufficient smoothness of the fields involved, the requirement that the augmented functional increases leads naturally to the first-order requirement that
\[
\int_{\vol} \left(\frac{\partial u}{\partial \eta}-T\right)\delta\eta\,\de v = 0
\]
for any small perturbation $\delta \eta$; since the sign of this perturbation is arbitrary, this yields the necessary condition that
\begin{equation}\label{eq:EqT}
    \frac{\partial u}{\partial \eta} = T
\end{equation}
throughout the system. Similar considerations show that for the densities of each chemical component, we must also have that
\begin{equation}\label{eq:EqMu}
    \frac{\partial u}{\partial \rho_i} = \mu_i
\end{equation}
throughout the system. At equilibrium, \cref{eq:EqT,eq:EqMu} allow us to identify $T$ with the temperature, and $\mu_i$ with the chemical potential of the species $i$.

\subsubsection{Inner variations}
\label{sec:innervariations_onephase}
Next, we consider perturbations of the system through local displacement of the material points, which enable changes in the local volume through local strain. In particular, we will consider a diffeomorphism $\bfphi:\bfx\mapsto\tilde{\bfx} = \bfx+\dxi(\bfx)$, where $\dxi$ is a small displacement field of the material points.
To preserve the overall volume, a natural requirement is that the displacement normal to the boundary vanishes on the boundary of the volume, $\partial \vol$. As we discussed at the beginning of the section, we suppose that the values of the moles and entropy fields are transported as densities. 

The minimization of the augmented functional in \cref{eq:augFunct} entails that the energy after such a perturbation must increase, i.e.:
\[
\tilde E[\tilde\eta(\tilde\bfx),\tilde\rho_1(\tilde\bfx),\ldots,\tilde\rho_\ncomp(\tilde\bfx)]-E[\eta(\bfx),\rho_1(\bfx),\ldots,\rho_\ncomp(\bfx)]\geq 0.
\]
We now determine the difference of the fields we considered between their original value and their value after the application of the diffeomorphism $\bfphi(\bfx)$.

Under the form of perturbation above, the volume element transforms as follows:
\begin{equation}\label{eq:transf_vol}
\de \tilde{v}=\det\big(\id+\nabla\dxi(\bfx)\big)\de v=\Big(1+\nabla\cdot\dxi(\bfx)+O(|\dxi|^2)\Big)\de v.
\end{equation}
Next, we note that for a field $a$ expressed as a density per unit volume, under inner variation we consider $\tilde{a}(\tilde{\bfx})\,\de \tilde{v} = a(\bfx)\,\de v$. Informally-speaking, this means that
$$
\tilde{a}(\tilde{\bfx}) = a(\bfx)\frac{\de v}{\de \tilde{v}}(\bfx),
$$
where the `derivative' on the right should be understood as a Jacobian determinant. Assuming $\dxi$ is small, we have
\[
\frac{\de v}{\de \tilde{v}} = \det(\id+\nabla\dxi)^{-1} = 1-\nabla\cdot\dxi + O(|\dxi|^2).
\]
Using this result, we find that
in particular, for the entropy and moles density fields, we have
\begin{equation}\label{eq:mass_ent_transformation}
 \begin{aligned}
 	\tilde\eta(\tilde{\bfx}) &= \eta(\bfx)-\eta(\bfx)\nabla\cdot\dxi(\bfx)+O(|\dxi|^2) \\
  	\tilde\rho_i(\tilde{\bfx}) &= \rho_i(\bfx)-\rho_i(\bfx)\nabla\cdot\dxi(\bfx)+O(|\dxi|^2).
\end{aligned}
\end{equation}
We will combine these results below to derive equilibrium conditions.



%
To proceed further, we suppose that the energy density per unit current volume of phase $\alpha$, $u^\alpha$, is expressed as some fixed thermodynamic function of the local component densities and entropy, so that $u^\alpha(\bfx) = u^\alpha(\rho_1(\bfx),\dots\rho_n(\bfx),\eta(\bfx))$ and therefore also $\tilde u^\alpha(\tilde{\bfx})\de\tilde{v} = \tilde u^\alpha\big(\tilde\rho_1(\tilde{\bfx}),\dots,\tilde \rho_n(\tilde{\bfx}),\tilde\eta(\tilde{\bfx})\big)$ after the deformation. For the sake of concision, we write $u^\alpha(\rho_i,\eta)$, since the densities $\rho_i$ play an interchangeable role in what follows.

Using \cref{eq:transf_vol,eq:mass_ent_transformation} and Taylor expanding appropriately, we now find that
\begin{equation*}
\tilde u^\alpha\big(\tilde\rho_i,\tilde\eta\big)\de\tilde{v}=u^\alpha(\rho_i,\eta)\,\de v+ \bigg(u^\alpha -\Dpartial{u^\alpha}{\eta} \eta
-\sum_{i=1}^n\Dpartial{u^\alpha}{\rho_i} \rho_i\bigg)\nabla\cdot\dxi\,\de v + O(|\dxi|^2).
\end{equation*}
Since we supposed that moles and entropy fields are transported as material densities, their integrals do not contribute directly to the variation of the augmented functional $E$, and so upon integrating, we have
\begin{align*}
\tilde E[\tilde\eta,\tilde\rho_i] -E[\eta,\rho_i]&=  \int_{\widetilde \vol}u^\alpha(\tilde{\bfx})\de\tilde{v}-\int_{\vol}u^\alpha(\bfx)\de v \\
& = \int_{\vol} \bigg[u^\alpha - \Dpartial{u^\alpha}{\eta} \eta -  \sum_{i=1}^\ncomp\Dpartial{u^\alpha}{\rho_i} \rho_i\bigg]\nabla\cdot\dxi\,\de v + O(|\dxi|^2).
\end{align*}
In view of the equilibrium relations discussed in \cref{sec:outvarLL}, we have $\Dpartial{u}{\rho_i} = \mu_i$ and $\Dpartial{u}{\eta} = T$, where $T$ and $\mu_i$ are the Lagrange multipliers spatially which are uniform throughout the system, and we replace these derivatives by these constant values below.
Our next step is to perform an integration by parts. Since admissible displacement fields must satisfy $\dxi\cdot \bfn=0$ on $\partial \vol$ with normal $\bfn$, we have that boundary terms vanish:
\begin{align}\label{eq:extFuncLL}
& \int_{\vol} \sqp{ u^\alpha - \Dpartial{u^\alpha}{\eta} \eta - \sum_{i=1}^\ncomp \Dpartial{u^\alpha}{\rho_i} 
\rho_i} \nabla\cdot\dxi\,\de v    \nonumber \\ 
& \qquad \qquad  = \int_{\partial \vol}\cip{u^\alpha - T\eta- \sum_{i=1}^n\mu_i\rho_i}\dxi \cdot\bfn \,\de a - \int_{\vol} \nabla\cip{u^\alpha - T\eta - \sum_{i=1}^n\mu_i\rho_i} \cdot\dxi\,\de v \nonumber \\
    & \qquad \qquad  = \int_{\vol} \nabla\cip{\sum_{i=1}^n\mu_i\rho_i+T\eta-u^\alpha} \cdot\dxi \,\de v.
\end{align}
For convenience, we introduce the grand potential density for the phase, which is defined to be
\begin{equation}
\omega^\alpha:=u^\alpha-T\eta-\sum_{i=1}^n\mu_i\rho_i.
\label{eq:grand_pot_def}
\end{equation}
Now, we make an equilibrium argument: at equilibrium any such variation must increase the internal energy under the constraints, and hence this linear term must vanish.
As such, we obtain that
\begin{equation*} 
    \int_{\vol}\nabla\omega^\alpha\cdot \dxi\,\de v = 0.
\end{equation*}
Since this condition must hold for any displacement field $\dxi$ with $\dxi\cdot\bfn=0$ on the boundary, we can deduce that the gradient field must vanish on the entire volume $\vol$, i.e.
\begin{equation}
    \nabla \omega^\alpha=\bfzero,
    \label{eq:LLinvar_eq}
\end{equation}
and so the grand potential $\omega^\alpha$ must be constant.
Applying the usual thermodynamic convention, we denote this constant $-P$, i.e. the negative of the pressure in the system, obtaining:
\begin{equation}\label{eq:defP}
    \omega^\alpha = -P\quad\text{or equivalently}\quad u^\alpha = T\eta+\sum_{i=1}^n\mu_i\rho_i-P
\end{equation}
throughout the volume $\vol$. In this case we define the Cauchy stress tensor
\begin{equation}\label{eq:bulk_fluid_cauchy}
	\bfcauchy := \left(u^\alpha-T\eta-\sum_{i=1}^n\mu_i\rho_i\right)\id =  -P\id,
\end{equation}
and note that \cref{eq:LLinvar_eq} can be equivalently expressed as the requirement that
\begin{equation*}
\nabla\cdot\bfcauchy = \bfzero.
\end{equation*}
We will see that this identification is consistent with the more general definition for a solid system (see next section). 


\subsection{Variations in the presence of two fluid phases}\label{sec:2fluids}
We now consider a fixed volume $\vol$ containing two phases, labelled $\alpha$ and $\beta$. We suppose that phase $\alpha$ occupies a subvolume $V^\alpha$, phase $\beta$ occupies a subvolume $V^\beta$, and the boundary between the phases forms a surface $\area$. This surface may have its own boundary curve $\bline = \partial \area$; when only two phases are present, any such boundary curve must lie within the boundary of the full volume $\vol$.

For notational convenience, we assume that intensive properties of the system can be expressed as separate fields on the different regions of the volume $\vol$, so the entropy per unit volume in phase $\alpha$ is $\eta^\alpha$, defined on $V^\alpha$, the entropy per unit volume in phase $\beta$ is $\eta^\beta$, defined on $V^\beta$. The entropy per unit area on the boundary is $\eta^\area$, defined on $\area$, and the entropy per unit length on the boundary curve $\bline$ is $\eta^\bline$. We use similar notation for the internal energy per unit volume, area and length, and the moles density of each species per unit volume, area and length.

The interfacial quantities introduced in this work are to be intended as Gibbs surface excess quantities. Their numerical value therefore depend on the location chosen for the dividing surface and on the bulk reference states used to define the corresponding bulk contributions. The Gibbs construction is understood here as a rule for identifying, in each configuration, an ideal sharp surface representing the finite interfacial region. Thus, the geometrical surface may move and deform during the considered transformation, while the convention used to locate it relative to the interfacial region is kept fixed. A different choice of dividing surface would redistribute entropy, composition, and the other extensive quantities between the bulk phases and the interface, without altering the corresponding total quantities or the complete thermodynamic balance, provided that all contributions are transformed consistently.

Assuming once more that we prescribe the total entropy and the moles of the species in the fixed volume $\vol$, we consider the extended functional

\begin{align}\label{eq:enMultiPh}
    E[\eta,\rho_1,\ldots,\rho_\ncomp,V^\alpha,V^\beta]
     &= \int_{V^\alpha}u^\alpha\,\de v+\int_{V^\beta}u^\beta\,\de v+\int_{\area} u^\area\,\de a + \int_{\Gamma} u^\Gamma \,\de\ell\nonumber \\
     &\qquad\qquad+T\left(S-\int_{V^\alpha}\eta^\alpha\,\de v-\int_{V^\beta}\eta^\beta\,\de v-\int_\area\eta^\area\,\de a- \int_{\Gamma} \eta^\Gamma \,\de\ell\right) \nonumber \\
     &\qquad\qquad+\sum_{i=1}^\ncomp\mu_i\left(M_i-\int_{V^\alpha}\rho_i^\alpha\,\de v-\int_{V^\beta}\rho_i^\beta\,\de v-\int_\area\rho_i^\area\,\de a- \int_{\Gamma} \rho_i^\Gamma \,\de\ell\right).
\end{align}

\paragraph{Outer variations.} Perturbing the volumetric fields $\eta^\alpha$, $\eta^\beta$, $\rho_i^\alpha$ and $\rho_i^\beta$, the surface fields $\eta^\area$ and $\rho_i^\area$ the line field $\eta^\Gamma$ and $\rho_i^\Gamma$ through outer variations, we can deduce identical conditions to those found in the single phase case above, i.e. that 
\begin{equation}\label{eq:equilibriumabs}
    \frac{\partial u^\alpha}{\partial\eta^\alpha}
=\frac{\partial u^\beta}{\partial\eta^\beta} 
=\frac{\partial u^\area}{\partial\eta^\area}=\frac{\partial u^\Gamma}{\partial\eta^\Gamma}
=T,\quad\text{and}\quad\frac{\partial u^\alpha}{\partial\rho_i^\alpha}  = \frac{\partial u^\beta}{\partial\rho_i^\beta}= \frac{\partial u^\area}{\partial\rho_i^\area} =\frac{\partial u^\Gamma}{\partial\rho_i^\Gamma}= \mu_i.
\end{equation}

\paragraph{Inner variations in bulk phases.} Turning to the inner variation, we can consider a displacement of points in each of the bulk volumes $V^\alpha$ and $V^\beta$ only which vanish on the boundary of the volume $\partial \vol$ and the inner phase boundary $\area$; they must also necessarily vanish on $\bline$. By identical arguments to those made to that for a single phase in \cref{sec:innervariations_onephase} we deduce that at equilibrium
\begin{equation}\label{eq:gradBulk}
   \nabla\omega^\alpha = \nabla\left(u^\alpha-T\eta^\alpha-\sum_{i=1}^n\mu_i\rho_i^\alpha\right) = \nabla\omega^\beta = \nabla\left(u^\beta-T\eta^\beta-\sum_{i=1}^n\mu_i\rho_i^\beta\right)=0, 
\end{equation}
on the interior of the phases, where we recall the definition of the grand potential in \cref{eq:grand_pot_def}.
As such, these quantities must be constant throughout their domain. Defining the relevant constants to be $-P^\alpha$ and $-P^\beta$, the thermodynamic pressure in each phase, we find that
\begin{equation}\label{eq:PaPb}
    u^\alpha=T\eta^\alpha+\sum_{i=1}^n\mu_i\rho_i^\alpha-P^\alpha\quad\text{and}\quad
u^\beta=T\eta^\beta+\sum_{i=1}^n\mu_i\rho_i^\beta-P^\beta
\end{equation}
in each of the two phases.

\paragraph{Variation through phase change at the boundary.}
A natural consequence of \cref{eq:gradBulk,eq:PaPb} is that the pressure difference between the two phases is constant, i.e.
\begin{equation}\label{eq:pressDiff}
    P^\alpha - P^\beta = \mbox{const}. 
\end{equation}
Here, we recover Larch\'e and Cahn's result, showing that $P^\alpha=P^\beta$ (see \citep{Larche1973,Larche1978B,Larche1978}) if there is no energy associated to the interface between the phases, i.e., we consider the disturbance given by the interface is negligible and the phases are homogeneous up to the interface of separation: this is equivalent to assuming that $u^\area=0$ and $u^\Gamma=0$ in \cref{eq:enMultiPh}. 

To show this, we now introduce a third class of admissible variations together with the inner and outer variations. In particular, outer variations correspond to local variations in the field values themselves, while inner variations correspond to mechanical variation of the current configuration generated by a smooth displacement field  $\dxi$ deforming the existing material, but leaving the phase identity of material points intact. Both of these variations therefore preserve the phase identity of material points under their action.
However, in a multiphase system, a further possibility arises: a phase change may also take place, typically at a boundary between phases. As particular examples, components of a fluid can precipitate, becoming part of the solid part of the system, or some components in the solid may dissolve in the fluid.

Therefore, for a complete treatment of the equilibrium conditions in a multiphase system we need to include a different, independent variation which is associated with the displacement of the phase boundary. Let \(\area\) denote the interface between the $\alpha$ and $\beta$ phases, with unit normal $\bfn^{\alpha\beta}$ oriented from phase $\alpha$ to $\beta$ such that $\bfn^{\alpha\beta}=-\bfn^{\beta\alpha}$. We write the normal displacement of the interface as
\begin{equation}\label{eq:int_deform}
	\dxi^\area = \delta\xi\, \bfn^{\alpha\beta}.
\end{equation}
This variation changes the partition of the domain phases $\alpha$ and $\beta$. In particular, a positive or negative value of $\dxi^\area$ corresponds to the conversion of an infinitesimal layer of one phase into the other. The corresponding variation is therefore configurational rather than purely mechanical. If at least one of the phases is solid, this variation implies the modification of the reference configuration of the solid, and in general, the total molar content in each phase is changing. This last observation, in turn, will imply that we need to include an extra constraint on the solid structure, namely that the variation does not involve the bulk, but we will leave this analysis for the developments in the next section. 

If the interface between the phases moves normally by $\dxi^\area$ through a phase change, the leading-order contribution due to the local volume change at the boundary of phase $\alpha$ is
$\dxi\cdot\bfn^{\alpha\beta}\,\de a$, with a similar expression for the local change in volume of phase $\beta$, $\dxi\cdot\bfn^{\beta\alpha}\,\de a$. Assuming that $\dxi^\area$ vanishes on $\Gamma$, we can consider the variation of the extended functional under this phase change variation. In particular, the first-order term which results is
\begin{equation}\label{eq:shiftInt}
\int_\area
\cip{u^\alpha-T\eta^\alpha-\sum_{i=1}^\ncomp \mu_i\rho_i^\alpha}\dxi^\area\cdot\bfn^{\alpha\beta}\,\de a
+\int_\area\cip{u^\beta-T\eta^\beta-\sum_{i=1}^\ncomp \mu_i\rho_i^\beta}\dxi^\area\cdot\bfn^{\beta\alpha}\de a.
\end{equation}
In terms of the grand potential for each phase, $\omega^\alpha$ and $\omega^\beta$ (see \cref{eq:grand_pot_def}), we find that at equilibrium, we must have
$$
\int_\area (\omega^\alpha-\omega^\beta)\dxi^\area\cdot\bfn^{\alpha\beta}\,\de a = 0
$$
for all phase perturbations at the boundary $\area$. Since this is true for all possible perturbations, we have that $\omega^\alpha=\omega^\beta$, or equivalently $P^\alpha = P^\beta$, as $\omega^\alpha=-P^\alpha$ and $\omega^\beta=-P^\beta$.
Therefore, in the absence of additional interfacial energy, the equilibrium with respect to arbitrary admissible normal shifts of the interface necessarily entails that
\[
P^\alpha=P^\beta .
\]
In this case, we see therefore there is no jump associated with the pressure in passing from system $\alpha$ to system $\beta$ and we can define a single pressure for the entire multi-phase system $P$:
\[
    P^\alpha = P^\beta = P. 
\]
However, as we will demonstrate below, the constant in \cref{eq:pressDiff} may vanish even when $u^\area$ are non-zero, provided the interface assumes specific geometrical configurations. These findings are captured by the classical Young–Laplace equation, which, as we will show, emerges naturally from our analysis.

\paragraph{Inner variations at two-phase boundary.} Next, we consider the effect of an inner variation in the vicinity of the phase boundary $\area$ where some internal energy is indeed stored in the boundary region, so that $u^\area$ is not identically zero. In particular, we note that given a displacement $\dxi$, we must consider the deformation of the surface area element under the variations we consider.
We can express the change of surface area element as follows:
\[
\de \tilde{a} = \det(\id+\nabla \dxi)\left|(\id+\nabla\dxi)^{-T}\bfn\right|\de a,
\]
where $\bfn$ is the normal to the surface, again we assume to point from inside the $\alpha$ phase into the $\beta$ phase but we drop the superscripts to ease the notation. Taylor expanding in the quantity $\dxi$, we have
\[
(\id+\nabla\dxi)^{-T} = I - \nabla\dxi^T + O(|\dxi|^2),
\]
and so
\[
\left|(\id+\nabla\dxi)^{-T}\bfn\right| = \left|\bfn-\nabla(\dxi)^T\bfn+O(|\dxi|^2)\right| = 1-\bfn\cdot\left(\nabla\dxi^T\bfn\right)+O(|\dxi|^2).
\]
Putting this together with the Taylor expansion of the determinant, we find
\begin{align}\label{eq:varA}
\de \tilde{a} & = \det(\id+\nabla \dxi)\left|(\id+\nabla\dxi)^{-T}\bfn\right|\de a = \Big(1+\nabla\cdot\dxi-\bfn\cdot\left(\nabla\dxi\,\bfn\right)+O(|\dxi|^2)\Big)\,\de a \nonumber \\
&\qquad  = \cip{1+\nabla_\area\cdot \dxi + O(|\dxi|^2}\de a
\end{align}
where in the latter expression we introduce the surface divergence operator:
\[
\nabla_\area\cdot\dxi := \bft_1\cdot(\nabla\dxi\,\bft_1)+ \bft_2\cdot(\nabla\dxi\,\bft_2) = \nabla\cdot\dxi-\bfn\cdot\left(\nabla\dxi\,\bfn\right);
\]
with $\bft_1,\bft_2,\bfn$ forming an orthonormal frame of tangent vectors and normal on the surface.
By definition, this operator does not depend on the particular local tangent vectors chosen, only the local orientation of the surface $\area$, as expressed through the normal $\bfn$.

We now wish to follow a similar line of argument to that made for the bulk; however an important subtlety is that to make the application of the divergence theorem on the surface $\area$ clearer, we first decompose the perturbation $\dxi$ into normal and tangential components along the surface. 
For convenience, we introduce the normal projection onto the surface, $\id_\area:=\id-\bfn\otimes\bfn$, and we split $\dxi$ into normal and tangential components at the surface. In particular, let $\dxi_n$ be the normal component of $\dxi$, and $\dxi_t$ the tangential component, i.e.
$$
\dxi_n:= (\dxi\cdot\bfn)\bfn  = \delta\xi_n \bfn\quad\text{and}\quad\dxi_t:=\id_\area \dxi.
$$
With this decomposition, a classic geometric result (the variation of area formula) states that
$$
\nabla_\area\cdot\dxi = \nabla_\area \cdot\dxi_t+\nabla_\area\cdot\dxi_n = \nabla_\area\cdot\dxi_t -2H\,\delta \xi_n,
$$
where $H(\bfx):=-\tfrac12\nabla_\area\cdot\bfn(\bfx)$ is the mean curvature on the surface.

We again suppose here that the values of the molar and entropy fields per unit area on the surface are transported as surface densities and so their transport laws are
\[
\tilde\eta^{\area}(\tilde{\bfx}) = \eta^\area(\bfx) \frac{\de a}{\de \tilde{a}}(\bfx),  \qquad\text{and}\qquad \tilde\rho^\area(\tilde{\bfx}) = \rho^\area(\bfx) \frac{\de a}{\de \tilde{a}}(\bfx).
\]
The appropriate area factor in this case can be derived from \cref{eq:varA} to be
\[
\frac{\de a}{\de \tilde{a}} = 1-\nabla_\area\cdot\dxi+O(|\dxi|^2),
\]
and hence we obtain that:
\begin{equation}
\begin{aligned}
    \tilde\eta^\area(\tilde{\bfx}) & = \eta^\area(\bfx)-\eta^\area(\bfx)\nabla_\area\cdot\dxi(\bfx)+O(|\dxi|^2) \\
    \tilde\rho_i^\area(\tilde{\bfx}) & = \rho_i^\area(\bfx)-\rho_i^\area(\bfx)\nabla_\area\cdot\dxi(\bfx)+O(|\dxi|^2)    
\end{aligned}
\label{eq:mass_ent_surface_transformation}
\end{equation}
in analogy with the volumetric perturbations, see \cref{eq:mass_ent_transformation}.

As in the bulk phase argument pursued in \Cref{sec:innervariations_onephase}, we suppose that the energy density per unit current area of the interface, $u^\area$,  can be expressed through some fixed thermodynamic function of the local component densities per unit area and the entropy per unit area, so that $u^\area(\bfx) = u^\area(\rho^\area_1(\bfx),\dots,\rho^\area_n(\bfx),\eta^\area(\bfx))$. 
Then using \cref{eq:varA,eq:mass_ent_surface_transformation}, we can Taylor expand in a manner analogous to the steps used in the inner variation of the bulk  to obtain
\begin{equation*}
       \int_{\tilde{\area}}\tilde{u}^\area\,\de \tilde{a}-\int_{\area} u^\area\,\de a 
       =\int_{\area}\cip{u^\area - \Dpartial{u^\area}{\eta^\area}\eta^\area - \sum_{i=1}^\ncomp \Dpartial{u^\area}{\rho_i^\area}\rho_i^\area} 
       \nabla_\area\cdot \dxi\,\de a + O(|\dxi|^2)\,.
\end{equation*}
Similar arguments applied to the remaining terms in the extended functional entail that
\begin{equation*}
\begin{aligned}
    T\int_{\tilde{\area}} \tilde{\eta}^\area\,\de \tilde{a}
    &=T\int_{\area}\eta^\area\,\de a+O(|\dxi|^2)\\
\mu_i\int_{\tilde{\area}} \tilde{\rho}_i^\area\,\de \tilde{a}
    &=\mu_i\int_{\area}\rho_i^\area\,\de a+O(|\dxi|^2).
\end{aligned}
\end{equation*}

We now consider two cases separately. Let us assume first that $\dxi_n$ vanishes everywhere on $\area$ and that $\dxi$ vanishes completely on $\Gamma$. Upon using the conditions that $\Dpartial{u^\area}{\rho_i^\eta} = \mu_i$ and $\Dpartial{u^\area}{\eta^\area} = T$, the standard first-order argument requires that
\begin{align}
   \int_\area \cip{u^\area - \Dpartial{u^\area}{\eta^\area}\eta^\area - \sum_{i=1}^\ncomp \Dpartial{u^\area}{\rho_i^\area}\rho_i^\area} 
       \nabla_\area\cdot \dxi_t \de a 
       & = \int_\area \cip{u^\area - T\eta^\area - \sum_{i=1}^\ncomp \mu_i\rho_i^\area} 
       \nabla_\area\cdot \dxi_t \de a \notag\\
       & = \int_\area \nabla_\area \bigg[T\eta^\area+ \sum_{i=1}^n\mu_i \rho^\area_i-u^\area\bigg]\cdot\dxi_t\de a = 0.\label{eq:LL_tang_var}
\end{align}
In the last equality we have integrated by parts, discarding the line contribution as we assumed no variation on the boundary of the interface $\area$, and we have replaced the derivatives of the surface energy with the equilibrium condition derived in \cref{eq:equilibriumabs}.

If we now define the surface grand potential density $\omega^\area$: 
\begin{equation}\label{eq:sigma}
   \omega^\area := u^\area-T\eta^\area-\sum_{i=1}^n\mu_i\rho^\area_i,
\end{equation}
then, since $\dxi_t$ is an arbitrary tangential perturbation, \cref{eq:LL_tang_var} entails that there exist a constant, $\gamma$ on $\area$ such that
\begin{equation}\label{eq:definitiongamma}
    \omega^\area = \gamma \,.
\end{equation}
The constant $\gamma$ has been known historically as the surface tension, and is the surface analogue of the bulk pressure constants $P^\alpha$ and $P^\beta$. However, despite its historically accepted name, we prefer to use the term \textit{surface free energy}, following the discussion in \citep{DiPasquale2020,DiPasquale2025} where it is highlighted that calling this quantity ``surface tension'' is source of ambiguity when solid systems are considered (as we will consider in the next sections). We highlight here that in previous similar works \cite{Leo1989} the surface grand potential density $\omega^\area$ was equated to the surface free energy $\gamma$ without deriving the proper equilibrium condition which we instead derived here via \cref{eq:LL_tang_var}.

By next considering variations in the normal direction to the surface, we get an additional condition on the surface $\area$. In particular, at equilibrium the linear terms in $\dxi_n$ must also vanish, which entails that
\begin{equation}\label{eq:YL}
	\left(u^\area-T\eta^\area-\sum_{i=1}^n\mu_i\rho_i^\area\right)\nabla_\area\cdot\bfn^{\alpha\beta}-P^\alpha+P^\beta= 0.
\end{equation}
with our convention on the normal, i.e. $\bfn^{\alpha\beta}$ is the normal to the surface $\area$ pointing from phase $\alpha$ to phase $\beta$.
Since the term in parentheses above is exactly $\gamma$, and the negative of the surface divergence of the normal vector when properly interpreted is twice the mean curvature of the surface, denoted $H$, we find that the surface $\area$ must satisfy
\[
2\gamma H+P^\alpha-P^\beta = 0,
\]
which we can interpret as a requirement that the surface tension forces must balance the pressure change across the interface. This is the classical expression of the Young-Laplace equation. Since the pressures are constant in each phase, this is furthermore a requirement that the surface $\area$ has constant mean curvature at equilibrium.
Examples of surfaces with constant mean curvature include (but are not limited to) spheres.

\subsubsection{Orientation-dependent generalization of the energy term}

\noindent In the previous section, the surface excesses were treated as materially transported densities. This variation accounts for the change of the surface element and leads to the corresponding equilibrium conditions involving the surface grand potential density
\[
\omega^{\area} = \gamma=u^\area-T\eta^\area-\sum_{i=1}^n\mu_i\rho_i^\area .
\]
In this section we extend the analysis by assuming that the energy, entropy and molar densities of the surface depend on the (local) normal of the surface $\bfn$. The same geometrical deformation we considered in previous sections also changes the local normal to the interface. Therefore, if the equilibrium surface excesses are orientation-dependent constitutive quantities,
\[
u^\area = u^\area(\bfn); \,\;\; 
\eta^\area = \eta^\area(\bfn); \, \,\;\; 
\rho^\area = \rho^\area(\bfn) \, .
\]
then their values in the deformed configuration are obtained by evaluating the same constitutive functions at the new normal \(\tilde \bfn\). This is not a material transport law for \(\eta^\area \de a\) or \(\rho_i^\area \de a\), but an orientation-dependent constitutive update. Here $u^\area(\bfn)$, $\eta^\area(\bfn)$, $\rho^\area(\bfn)$ denote the values of the interfacial equilibrium fields along a family of equilibrium states parametrized by the interface orientation. The underlying constitutive dependence of $u^\area=u^\area(\bfn,\eta^\area,
\rho_i^\area)$ is understood.

If $u^\area(\tilde{\bfn})$, where $\tilde{\bfn}$ is the new normal of the surface after the deformation $\dxi$, we can Taylor expand it to the first order in the deformation $\dxi$:
\[
\tilde{u}^\area(\tilde{\bfn}) = u^\area(\bfn)+\frac{\partial u^\area(\bfn)}{\partial\bfn}\cdot(\tilde{\bfn}-\bfn)+O(|\dxi|^2).
\]
The new normal $\tilde{\bfn}$ is related to the normal of the interface in the reference configuration $\bfn$ by
\[
\tilde{\bfn} = \frac{(\id+\nabla\dxi)^{-T}\bfn}{|(\id+\nabla\dxi)^{-T}\bfn|}.
\]
We can again Taylor expand the previous relation to first order to get
\[
\tilde{\bfn} = \bfn-\id_\area\nabla\dxi^T\bfn+O(|\dxi|^2),
\]
where we recall that $\id_\area = \id-\bfn\otimes\bfn$.
When included in the Taylor expansion of the energy $\tilde{u}^\area(\tilde{\bfn})$, this gives additional terms originated from variation in the normal:
\[
\tilde{u}^\area(\tilde{\bfn}) = u^\area(\bfn)-\frac{\partial u^\area(\bfn)}{\partial\bfn}\cdot(\id-\bfn\otimes\bfn)\nabla\dxi^T\bfn+O(|\dxi|^2).
\]
When we combine this new expression for the energy with the variation of the area element of the interface (see \cref{eq:varA}) we obtain, after decomposing the variation in its normal and tangential component:
\begin{align}\label{eq:normUsigma}
\int_{\tilde{\area}} \tilde{u}^\area\,\de \tilde{a}
= & \int_{\area} \cip{u^\area(\bfn)-\frac{\partial u^\area(\bfn)}{\partial\bfn}\cdot \id_\area\nabla\dxi_t^T\bfn - \frac{\partial u^\area(\bfn)}{\partial\bfn}\cdot\id_\area\nabla\dxi_n^T\bfn} \,\de a \nonumber \\
& +\int_{\area}u^\area\,\nabla_\area\cdot\dxi_n\,\de a+\int_{\area}u^\area\,\nabla_\area \cdot\dxi_t\,\de a+O(|\dxi|^2).
\end{align}
Similar arguments apply to the remaining terms in the extended functional, entailing that
\begin{subequations} \label{eq:surface-energy-var-norm}
\begin{align}
    T\int_{\tilde{\area}} \tilde{\eta}^\area\,\de \tilde{a}
    & =T\int_{\area}\cip{\eta^\area(\bfn)-\frac{\partial \eta^\area(\bfn)}{\partial\bfn}\cdot\id_\area(\nabla\dxi_t^T+\nabla\dxi_n^T)\bfn}\,\de a  \nonumber \\ 
    & +T\int_\area \eta^\area\nabla_\area\cdot\dxi_n\,\de a+T\int_\area\eta^\area\nabla_\area\cdot\dxi_t\,\de a+O(|\dxi|^2) \\
\mu_i\int_{\tilde{\area}} \tilde{\rho}_i^\area\,\de \tilde{a}
    &=\mu_i\int_{\area}\cip{\rho^\area(\bfn)-\frac{\partial \rho^\area(\bfn)}{\partial\bfn}\cdot\id_\area(\nabla\dxi_t^T+\nabla\dxi_n^T)\bfn}\,\de a  \nonumber \\
    & +\mu_i\int_\area \rho^\area_i\nabla_\area\cdot\dxi_n\,\de a+\mu_i\int_\area\rho_i^\area\nabla_\area\cdot\dxi_t\,\de a+O(|\dxi|^2).
\end{align}
\end{subequations}
Upon summation of the terms including the gradient of the deformation $\nabla\dxi_t^T$, we obtain the term (a similar result applies for the term including $\nabla\dxi_n^T$):
\[
\int_\area\pard{\gamma(\bfn)}{\bfn}\cdot\id_\area\nabla \dxi_t^T\bfn\,\, \de a
\]
where we used \cref{eq:sigma} to replace the sum of the thermodynamic quantities in \cref{eq:surface-energy-var-norm} with the surface free energy $\gamma$. We note that, in this case, $\gamma$ depends on the normal $\bfn$.

Let us now rewrite the argument of the integral as follows:
\[
\pard{\gamma(\bfn)}{\bfn}\cdot\id_\area\nabla \dxi_t^T\bfn = \id_\area\pard{\gamma(\bfn)}{\bfn}\cdot \nabla \dxi_t^T\bfn
\]
where we used the fact that the operator $\id_\area$ is symmetric.

By using \cref{eq:nbTens_a} we can rewrite the previous expression as:
\[  
 \pard{\gamma(\bfn)}{\bfn}\cdot\id_\area\nabla \dxi_t^T\bfn =\bfn \otimes \bigg(\id_\area\pard{\gamma(\bfn)}{\bfn}\bigg) : \nabla \dxi_t \,\,.
\]
%
From \cref{eq:normUsigma} we can write
\[
    \gamma(\bfn)\,\nabla_\area\cdot\dxi_t = 
    \gamma(\bfn)\id_\area \,:\,\nabla\dxi_t
\]
and we then obtain
\begin{equation}
\begin{aligned}
&\int_{\area}\sqp{\gamma(\bfn)\,\nabla_\area \cdot\dxi_t - \frac{\partial \gamma(\bfn)}{\partial\bfn}\cdot\id_\area\nabla\dxi_t^T\bfn}\,\de a  = \int_{\area}\sqp{\gamma(\bfn)\id_\area - \bfn \otimes \cip{\id_\area\pard{\gamma(\bfn)}{\bfn}}}\,:\,\nabla\dxi_t\,\de a \nonumber \\
    &\qquad\qquad = \int_{\area}\sqp{\gamma(\bfn)\id - \bfn \otimes \cip{\pard{\gamma(\bfn)}{\bfn}}}\id_\area \,:\,\nabla\dxi_t\,\de a 
     \\
    &\qquad\qquad = \int_{\area}\sqp{\gamma(\bfn)\id - \bfn \otimes\frac{\partial \gamma(\bfn)}{\partial\bfn}}\,:\,\nabla_\area\dxi_t\,\de a
\end{aligned}
\end{equation}
where in the second equality, we use the property of the tensor product reported in \cref{eq:SymmDot}, along with the definition of the surface gradient operator $\nabla_\area\dxi_t = \nabla\dxi_t\id_\area$.

We can now apply the surface-form of the divergence theorem to obtain 
\begin{equation}\label{eq:norm}
     \int_{\area}\sqp{\gamma(\bfn)\id - \bfn \otimes\frac{\partial \gamma(\bfn)}{\partial\bfn}}\,:\,\nabla_\area\dxi_t\,\de a = -\int_{\area}\cip{\nabla_\area\cdot\sqp{\gamma(\bfn)\id - \bfn \otimes\frac{\partial \gamma(\bfn)}{\partial\bfn}}}\cdot \dxi_t\,\,\de a = 0
\end{equation}
where the boundary terms vanish as there are no deformation at the boundary of the surface $\area$ since $\partial \area \subset \partial \vol$. 

\Cref{eq:norm} defines the equilibrium at the interface, and leads us to define the tensor field $\sstress(\bfn)$ on the phase boundary $\area$, depending on the position on the surface and the normal $\bfn$, which we call the \emph{surface stress}, defined to be:
\begin{equation} \label{eq:def_surf_stress_n}
 \sstress(\bfn) := \gamma(\bfn)\id_{\area} - \bfn \otimes\frac{\partial \gamma(\bfn)}{\partial\bfn},
\end{equation}
which is a condition analogous to the vanishing of the divergence of the Cauchy stress we found in the bulk.
\Cref{eq:norm} entails that the projection of the surface divergence of this tensor is zero at the interface:
\begin{equation}\label{eq:surface_stress_balance}
\id_\area\left(\nabla_\area \cdot \sstress(\bfn)\right) = 0 \, .
\end{equation}
We note that the surface stress tensor $\sstress(\bfn)$ can alternatively be rewritten as
\begin{align}\label{eq:CH_vector}
  \sstress(\bfn) & = \gamma(\bfn) (\mathcal \id - \bfn \otimes \bfn) - \bfn \otimes \frac{\partial \gamma(\bfn)}{\partial\bfn} = \gamma(\bfn) \mathcal \id  - \bfn \otimes \cip{\gamma(\bfn)\bfn +  \pard{\gamma(\bfn)}{\bfn}} = \gamma(\bfn) \mathcal \id - \bfn \otimes \CHv
\end{align}
where 
\begin{equation}\label{eq:defn_CHv}
    \CHv = \gamma(\bfn) \bfn + \pard{\gamma}{\bfn}
\end{equation}
is the Cahn--Hoffman capillarity vector, see \citep{Hoffman1972}.
This quantity emerges naturally as part of our formulation. We note that in their original work Cahn and Hoffman used the symbol $\boldsymbol \xi$ \citep{Hoffman1972}: Here, we have used the symbol $\bfX$ to avoid confusion with the notion of variation $\dxi$ we are employing in this work.

We now show that our formulation is physically equivalent to the Cahn-Hoffman one.  Let us now consider a variation in the normal direction $\dxi_{n}=(\dxi \cdot \bfn)\bfn=\delta\xi_n\bfn$ and write all the terms depending on $\dxi_n$:
\begin{align}
    \int_\area \sqp{\cip{P^\beta - P^\alpha} \dxi \cdot \bfn + \sstress : \nabla_{\area}\dxi_n} \de a = 0
\end{align}
where we collected the terms coming from \cref{eq:normUsigma,eq:surface-energy-var-norm} in the form of the surface stress tensor $\sstress(\bfn)$ as expressed by \cref{eq:def_surf_stress_n}, and we added the surface terms coming from the variation of the bulk of the two phases $\alpha$ and $\beta$; see the discussion of \cref{eq:extFuncLL}.

We now focus on the term depending on the surface stress, which we expand using its alternative definition in terms of the Cahn-Hoffman vector, \cref{eq:CH_vector}:
\[
    \sstress : \nabla_{\area}\dxi_n =  \cip{\gamma(\bfn) \mathcal \id - \bfn \otimes \CHv}: \nabla_{\area}\dxi_n 
\]
Let us rewrite the second term of the dyadic product in the previous equation as:
\[
    \nabla_{\area}\dxi_n = \nabla_{\area}(\delta \xi_n \bfn) = \bfn \otimes \nabla_{\area} \dxin + \delta \xi_n\nabla_{\area} \bfn
\]
We can now analyse the expression we obtained term by term:
\begin{gather*}
\gamma(\bfn) \id : \bfn  \otimes \nabla_{\area} \dxin  = \tr(\gamma(\bfn)  \bfn  \otimes  \nabla_{\area} \dxin   ) = \gamma(\bfn) n_i\partial_i \dxin = \gamma(\bfn) \bfn \cdot \nabla_{\area}\dxin \\[1mm]
\bfn \otimes \CHv : \bfn  \otimes  \nabla_{\area} \dxin = \tr((\bfn \otimes \CHv)( \bfn  \otimes  \nabla_{\area} \dxin)^T)=n_iX_jn_i\partial_j \dxin = \bfX \cdot \nabla_{\area} \dxin \\[1mm]
\gamma(\bfn) \id : \delta \xi_n\nabla_{\area} \bfn = \tr(\gamma(\bfn)\dxin \nabla_{\area}\bfn) = \gamma(\bfn)\dxin \partial_i n_i = \gamma(\bfn)\dxin \nabla_{\area}\cdot\bfn \\[1mm]
\bfn \otimes \CHv : \delta \xi_n\nabla_{\area} \bfn = \tr((\bfn \otimes \CHv )(\delta \xi_n\nabla_{\area} \bfn)^T) = \dxin n_i X_j \partial_j n_i = \dxin X_j \tfrac 12 \partial_j(n_i n_i) = 0
\end{gather*}
where in the last equation we used the fact that $|\bfn|=1$.
Now, we consider the first two terms and we integrate by parts, remembering that we assumed that the variation vanishes on the boundary line to the surface, $\partial\area$:
\begin{equation*}
    \int_{\area} \sqp{\cip{\gamma(\bfn) \bfn -\bfX }\cdot \nabla_{\area} \dxin }\de a = \int_{\area}\dxin \nabla_\area \cdot \cip{ \bfX - \gamma(\bfn)\bfn }  \de a 
\end{equation*}
We can now put everything together:
\begin{align}\label{eq:norm_dep_norm_xi}
    & \int_\area \dxin \sqp{\cip{P^\beta - P^\alpha}  + \nabla_\area \cdot \cip{ \bfX - \gamma(\bfn)\bfn } +\gamma(\bfn)\nabla_{\area}\cdot\bfn} \de a = \int_\area \dxin \sqp{\cip{P^\beta - P^\alpha}  + \nabla_\area \cdot \bfX} \de a = 0,
\end{align}
where we have used the product rule $\nabla_{\area}\cdot(\gamma(\bfn)\bfn) = \gamma(\bfn)\nabla_{\area}\cdot\bfn + \bfn \cdot \nabla_\area \gamma(\bfn)$ together with the fact that $\nabla_\area \gamma(\bfn)$ is tangential to $\area$, such that $\bfn \cdot \nabla_\area \gamma(\bfn) =0$.
\Cref{eq:norm_dep_norm_xi} therefore entails the equilibrium condition:
\begin{equation}\label{eq:eq_cond_norm}
    \cip{P^\beta - P^\alpha}  + \nabla_\area \cdot \bfX = 0.
\end{equation}
The equilibrium condition \cref{eq:eq_cond_norm} is identical to that obtained by Cahn and Hoffman for their vector $\CHv$ (see Eq. 13 in \citep{Cahn1974}), and as such, the physics described by our formalism is identical to that obtained by Cahn and Hoffman. 

Finally, we make one further observation about the definition of the surface stress introduced in \cref{eq:def_surf_stress_n}, which is that in the case that $\gamma$ does not depend on $\bfn$, the definition reduces to
\[
\sstress := \gamma\id_{\area},
\]
and hence the formalism reduces naturally to the surface tension case. We note the analogy in this case with the form of the Cauchy stress in the bulk fluid being $\bfcauchy=-P\id$, see \cref{eq:bulk_fluid_cauchy}.


\subsubsection{Variations in the presence of three fluid phases}

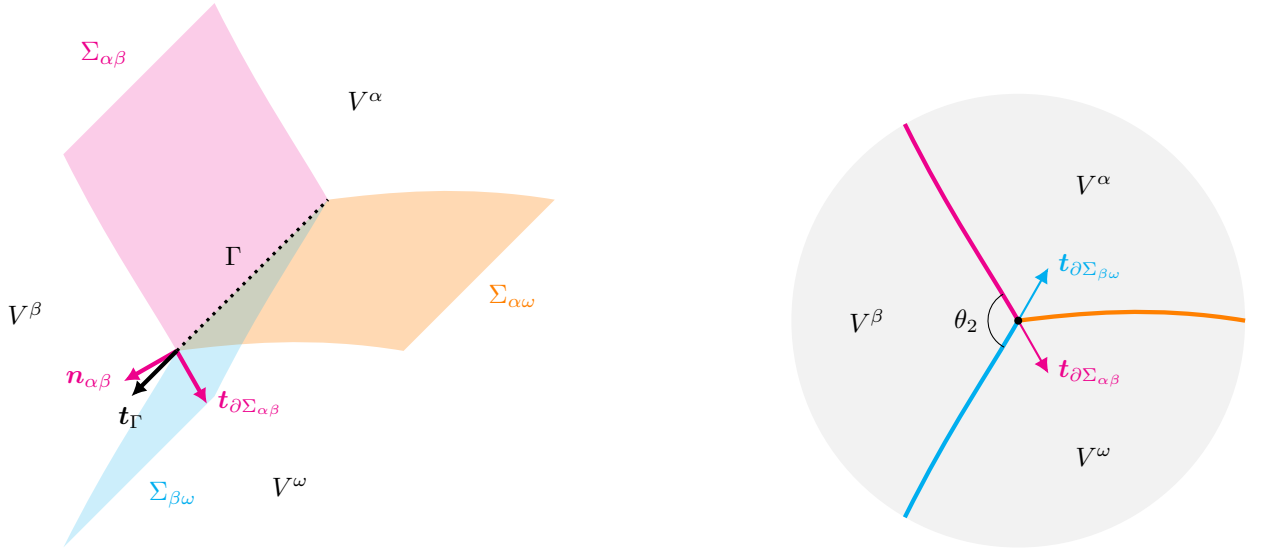
\begin{figure}[htbp]

    \begin{center}
         \begin{tikzpicture}[]

\coordinate (O) at (0,0);




\fill[orange,opacity=0.3]
  (O) .. controls (0.8,0.1) and (1.8,0.2) .. (3,0) -- node[below right,opacity=1] {$\Sigma_{\alpha\omega}$} (5,2) .. controls (3.8,2.2) and (2.8,2.1) .. (2,2) -- (O);

\fill[magenta,opacity=0.2,ultra thick]
  (O) .. controls (-0.4,0.7) and (-0.9,1.4) .. (-1.5,2.6) -- node[above left,opacity=1] {$\Sigma_{\alpha\beta}$}(0.5,4.6) .. controls (1.1,3.4) and (1.6,2.7) .. (2,2) -- (O);

\fill[cyan,opacity=0.2,thick]
  (O) .. controls (-0.4,-0.7) and (-0.9,-1.4) .. (-1.5,-2.6) -- node[below right,opacity=1] {$\Sigma_{\beta\omega}$} (0.5,-0.6) .. controls (1.1,0.6) and (1.6,1.3) .. (2,2);

\node[] at (2.5,3.3) {$\vol^\alpha$};
\node[] at (-2,0.5) {$\vol^\beta$};
\node[] at (1.5,-1.8) {$\vol^\omega$};

\draw[magenta,-{Latex[length=2mm,width=2mm]},ultra thick]
  (O) -- ($(O)-(-0.4,0.7)$) node[right] {$\bft_{\partial\area_{\alpha\beta}}$};

\draw[magenta,-{Latex[length=2mm,width=2mm]},ultra thick]
  (O) -- ($(O)-(0.7,0.4)$) node[left] {$\bfn_{\alpha\beta}$};

\draw[-{Latex[length=2mm,width=2mm]},ultra thick]
  (O) -- ($(O)-(0.6,0.6)$) node[below] {$\bft_{\bline}$};

\draw[dotted,very thick] (O) -- node[above left] {$\bline$} (2,2);

\end{tikzpicture}
\hfill
\begin{tikzpicture}[]

\coordinate (O) at (0,0);

\fill[gray!10] (O) circle (3.0);



\draw[ultra thick, orange]
  (O) .. controls (0.8,0.1) and (1.8,0.2) .. (3,0);

\draw[ultra thick, magenta]
  (O) .. controls (-0.4,0.7) and (-0.9,1.4) .. (-1.5,2.6);

\draw[ultra thick, cyan]
  (O) .. controls (-0.4,-0.7) and (-0.9,-1.4) .. (-1.5,-2.6);

\node[] at (1,1.8) {$\vol^\alpha$};
\node[] at (-2,0) {$\vol^\beta$};
\node[] at (1,-1.8) {$\vol^\omega$};

\draw[magenta,-{Latex[length=2mm,width=2mm]},thick]
  (O) -- ($(O)-(-0.4,0.7)$) node[right] {$\bft_{\partial\area_{\alpha\beta}}$};

\draw[cyan,-{Latex[length=2mm,width=2mm]},thick]
  (O) -- ($(O)-(-0.4,-0.7)$) node[right] {$\bft_{\partial\area_{\beta\omega}}$};

\fill (O) circle (1.5pt);

\draw (-0.2,0.35)
  arc[start angle=120,end angle=242,radius=0.4];
\node[left] at (-0.4,0) {$\theta_2$};

\end{tikzpicture}
    \end{center}
 \caption{Illustrations of a three-phase contact line $\bline$ between phases $\alpha$, $\beta$, and $\omega$. In the left figure, the phase boundary and contact line are illustrated. The surface normal $\bfn_{\alpha\beta}$, outward-pointing tangent $\bft_{\partial\area_{\alpha\beta}}$ and the tangent to the contact line $\bft_\bline$ are all shown for the $\alpha$--$\beta$ phase boundary. On the right, looking directly along the boundary line tangent direction, we illustrate the definition of one of the three dihedral angles between the interfaces at the contact line.}
\label{fig:AIC}
\end{figure}

We now consider a fixed volume $\vol$ containing three phases, labelled $\alpha$, $\beta$ and $\omega$, illustrated in \cref{fig:AIC}. Phase $\alpha$ occupies a subvolume $V^\alpha$, phase $\beta$ occupies a subvolume $V^\beta$, and phase $\omega$ a subvolume $V^\omega$. The boundary between the phases is composed by three surfaces $\area_{\alpha\beta}$, $\area_{\alpha\omega}$, $\area_{\beta\omega}$. These three surfaces meet on a curve belonging to the boundary of each interface $\bline = \partial \area_{\alpha\beta} \cap \partial \area_{\beta\omega} \cap \partial \area_{\alpha\omega}$. When only two phases are present, the contact line must either be empty, or else it coincides with the boundary of the entire volume considered, $\partial\vol$. However, in the case of three phases in contact, connected components of $\bline$ are either closed curves or intersect the boundary of the entire volume $\partial \vol$ at two points. 

As for the previous case, we assume that intensive properties of the system can be expressed as separate fields on the different regions of the volume $\vol$, e.g. $\eta^\alpha$ for entropy per unit volume in phase $\alpha$ and the same for phase $\beta$ and $\omega$. The entropy per unit area on the boundary between phases $\alpha$ and $\beta$ is $\eta^\area_{\alpha\beta}$, defined on $\area_{\alpha\beta}$, and we use analogous notation for the other two interfaces. The entropy per unit length on the boundary curve $\bline$ is $\eta^\bline$. We will use similar notation for the internal energy per unit volume, area and length, and the molar density of each species per unit volume, area and length.

Assuming once more that we prescribe the total entropy and the moles of the species in the fixed volume $\vol$, we consider the extended functional
\begin{equation}
\label{eq:enMulti3Ph}
\begin{aligned}
    &E[\eta,\rho_1,\ldots,\rho_\ncomp,\vol^\alpha,\vol^\beta,\vol^\omega]\\
     &\qquad\qquad= \sum_{j\in\{\alpha,\beta,\omega\}}\int_{\vol^j}u^j\,\de v+\sum_{jk\in\{\alpha\beta,\beta\omega,\alpha\omega\}}\int_{\area_{jk}} u^{\area_{jk}}\,\de a+\int_\bline u^\bline\,\de\ell \nonumber \\
     &\qquad\qquad\qquad\qquad+T\left(S-\sum_{j\in\{\alpha,\beta,\omega\}}\int_{\vol^j}\eta^j\,\de v-\sum_{jk\in\{\alpha\beta,\beta\omega,\alpha\omega\}}\int_{\area_{jk}}\eta^{\area_{jk}}\,\de a-\int_\bline \eta^\bline\,\de\ell\right) \nonumber \\
     &\qquad\qquad\qquad\qquad+\sum_{i=1}^\ncomp\mu_i\left(M_i-\sum_{j\in\{\alpha,\beta,\omega\}}\int_{\vol^j}\rho_i^j\,\de v-\sum_{jk\in\{\alpha\beta,\beta\omega,\alpha\omega\}}\int_{\area_{jk}}\rho_i^{\area_{jk}}\,\de a-\int_\bline \rho_i^\bline\,\de\ell\right).
\end{aligned}
\end{equation}
The derivation of the volume and surfaces equilibrium conditions for this case is completely analogous to argument performed above; all we must do is consider each pair of phases separately for each of the surface conditions. 
As such, the conditions for the volumes are:
\begin{align*}
  \frac{\partial u^\alpha}{\partial\eta^\alpha}
&=\frac{\partial u^\beta}{\partial\eta^\beta} = \frac{\partial u^\omega}{\partial\eta^\omega}
=\frac{\partial u^{\area_{\alpha\beta}}}{\partial\eta^{\area_{\alpha\beta}}}
=\frac{\partial u^{\area_{\beta\omega}}}{\partial\eta^{\area_{\beta\omega}}}
=\frac{\partial u^{\area_{\alpha\omega}}}{\partial\eta^{\area_{\alpha\omega}}}
=\frac{\partial u^\bline}{\partial\eta^\bline}
=T,\\
\text{and}\qquad\frac{\partial u^\alpha}{\partial\rho_i^\alpha}  
&= \frac{\partial u^\beta}{\partial\rho_i^\beta}= \frac{\partial u^\omega}{\partial\rho_i^\omega}
=\frac{\partial u^{\area_{\alpha\beta}}}{\partial\rho_i^{\area_{\alpha\beta}}}
=\frac{\partial u^{\area_{\beta\omega}}}{\partial\rho_i^{\area_{\beta\omega}}}
=\frac{\partial u^{\area_{\alpha\omega}}}{\partial\rho_i^{\area_{\alpha\omega}}}
=\frac{\partial u^\bline}{\partial\rho_i^\bline} =  \mu_i.  
\end{align*}
The equilibrium conditions for the three interfaces are given by \cref{eq:sigma,eq:YL} for each interface separately. We will therefore have three different \cref{eq:sigma,eq:YL}, each characterized by the difference in the value of the pressure of two phases in contact and the normal to the particular interface:
\begin{gather*}
    u^{\area_{jk}} = T\eta^{\area_{jk}}+\sum_{i=1}^n\mu_i\rho^{\area_{jk}}_i +\gamma_{jk} \\
    \left(u^{\area_{jk}}-T\eta^{\area_{jk}}-\sum_{i=1}^n\mu_i\rho_i^{\area_{jk}}\right)\nabla_{\area_{jk}}\cdot\bfn^{\area_{jk}}-P^j+P^k=0.
\end{gather*}
where $i,j=\alpha,\,\beta,\,\omega$ and $k\neq j$. One thing to notice is that each interface is characterized by its own surface free energy. This observation will become useful in what we discuss next. 

\paragraph{Inner variations at three-phase boundary line.}
Finally, we consider variations in the neighbourhood of the curve $\bline$ that bounds the phase boundary surfaces. In the three phase case considered, any components of the boundary curves $\bline$ must lie at the intersection of the three interfaces; again we refer the reader to \cref{fig:AIC} for an illustration. On this boundary, we can define a frame for each interface in which $\bfn_{jk}$ is the normal vector to $\area_{jk}$ close to the boundary, $\mathbf{t}_{\partial\area_{jk}} = \bft_\bline \times \bfn_{jk}$ is the tangent vector to $\area_{jk}$ which points outwards on $\bline$, and $\mathbf{t}_\bline$ is the tangent vector along the curve $\bline$ oriented in a right-handed fashion.

In this case, we note that the length element along $\bline$ transforms under the inner variation $\dxi$ as
\[
\de\tilde{\ell} = |(\id+\nabla\dxi)\bft_\bline|\,\de \ell = \Big(1+\bft_\bline\cdot(\nabla\dxi\bft_\bline)+O(|\dxi|^2)\Big)\,\de\ell.
\]
Let us also define the divergence along the curve $\bline$, setting
\[
\nabla_\bline \cdot \dxi := \frac{\de}{\de s}(\bft_\bline\cdot\dxi),
\]
where $\tfrac{\de}{\de s}$ denotes the derivative with respect to an arc-length parametrisation. Using the product rule and the Frenet-Serr\'e formula for the derivative of the tangent with respect to arc-length, we note that we have
\[
\bft_\bline\cdot(\nabla\dxi\,\bft_\bline) = \bft_\bline\cdot\frac{\de \dxi}{\de s}=\frac{\de}{\de s}\left(\bft_\bline\cdot\dxi\right)-\frac{\de \bft_\bline}{\de s}\cdot \dxi = \nabla_\bline\cdot\dxi-\bfk_\bline \cdot\dxi,
\]
where $\bfk_\bline$ is the curvature vector for $\bline$, often expressed as the product of the scalar curvature $\kappa_\bline$ and the normal vector $\bfn_\bline$ (as long as the normal is well-defined).
Next, under an inner variation, for any density field we  have that
\begin{align*}
	\tilde{\eta}^\bline(\tilde{\bfx})= \eta^\bline(\bfx) \frac{\de \ell}{\de \tilde{\ell}}(\bfx),
\end{align*}
and thus by expanding the change of length element in analogy to the previous sections (compare \cref{eq:mass_ent_transformation,eq:mass_ent_surface_transformation}), we have
\begin{subequations}
	\begin{align}
		\tilde{\eta}^\bline(\bfx) & =   \eta^\bline(\bfx) - \eta^\bline(\bfx) \nabla_\bline\cdot\dxi +\eta^\bline(\bfx)\bfk_\bline\cdot\dxi+O(|\dxi|^2),\\
	\tilde{\rho}^\bline_i(\bfx) & =   \rho^\bline_i(\bfx) - \rho^\bline_i(\bfx) \nabla_\bline\cdot\dxi+ \rho^\bline_i(\bfx) \bfk_\bline\cdot\dxi +O(|\dxi|^2).
	\end{align}
\end{subequations}
Writing the deformation $\dxi$ as the sum of the variation tangential to the curve $\bline$ and its orthogonal component,
\begin{equation*}
\dxi_t  := (\bft_\bline \otimes \bft_\bline) \dxi \qquad\text{and}\qquad
\dxi_{\perp} := \dxi - \dxi_t,
\end{equation*}
we note that
\[
\dxi_t\cdot\bfn_{jk} = 0\quad\text{and}\quad\dxi_t\cdot\bft_{\partial\area_{jk}} = 0,\quad j,k=\alpha,\beta,\omega;\, j \neq k.
\]
Note that $\bfk_\bline\cdot\dxi = \bfk_\bline\cdot\dxi_\perp$, as the curvature vector $\bfk_\bline$ is always orthogonal to the tangent $\bft_\bline$.
We can expand the perturbation to the infinitesimal line energy $\tilde{u}^\bline\,\de\tilde{\ell}-u^\bline\,\de\ell$ under the assumption that $\dxi$ is small to find that
 %
\begin{align*} 
\tilde u^\bline \de \tilde{\ell} - u^\bline \de \ell & = \left(u^\bline - \rho\frac{\partial u^\bline}{\partial\rho^\bline}-\eta\frac{\partial u^\bline}{\partial\eta^\bline}\right)\nabla_\bline\cdot\dxi\,\de\ell
-\left(u^\bline - \rho\frac{\partial u^\bline}{\partial\rho^\bline}-\eta\frac{\partial u^\bline}{\partial\eta^\bline}\right)\bfk_\bline\cdot\dxi_\perp\,\de\ell
+O(|\dxi|^2).
\end{align*}
This allows us, finally, to write the perturbation of the functional $E^\bline$ as
\begin{align*}
     \tilde{E}^\bline-E^\bline & = \int_{\bline}  \left(u^\bline - \sum_{j=1}^\ncomp\rho_j\frac{\partial u^\bline}{\partial\rho^\bline_j}-\eta\frac{\partial u^\bline}{\partial\eta^\bline}\right)\Big(\nabla_\bline\cdot\dxi -\bfk_\bline\cdot\dxi_\perp\Big)\de \ell +O(|\dxi|^2).
\end{align*}
Assuming first that $\dxi_\perp$ vanishes, and using integration by parts in a manner completely analogous to the argument made for variations tangent to a two-phase boundary surface $\area$, and assuming no variation at any boundaries of $\bline$, we find that
\[
\nabla_\bline\left(u^\bline-T\eta^\bline-\sum_{i=1}^n\mu_i\rho_i^\bline\right) = 0\,.
\]
It follows that there exists a constant $\tau$ called the \emph{line tension} along the curve such that
\[
u^\bline = T\eta^\bline+\sum_{i=1}^n\mu_i\rho_i^\bline+\tau,
\]
which is the analogue of the bulk pressures $P^\alpha$ and $P^\beta$ and the surface tension $\gamma$.

We can use the line tension to define a \emph{line stress} $\lstress$ as
\begin{equation}\label{eq:line_stress}
\lstress := \tau\,\bft_\bline\otimes\bft_\bline
\end{equation}
in analogue with the surface and bulk cases.

We now consider a non-zero component of the deformation orthogonal to the triple line, which, upon using the definition of $\tau$, now yields the first-order contribution
\[
\int_\bline
-\tau\,\bfk_\bline\cdot\dxi_\perp\,d\ell,
\]
where we recall that $\bfk_\bline$ is the curvature vector for $\bline$.
Using this form, we are now free to specify the component of the deformation $\dxi_\perp$ perpendicular to $\bline$. In particular, we consider variations tangent to the interfaces, i.e., proportional to $\bft_{\partial \area_{jk}}$ for $j,k=\alpha, \beta, \omega$ and $j \neq k$, so that
\[
\dxi_\perp = \delta\xi\,\bft_{\partial \area_{jk}}
\]
for each of the outward tangent directions. We also define the angles at the contact line between pairs of two-phase interfaces:
\begin{subequations}\label{eq:tangent_relations}
\begin{align}
    \bft_{\partial \area_{\alpha\beta}} \cdot \bft_{\partial \area_{\alpha\omega}} & = \cos{\theta}_{1}, \\ 
    \bft_{\partial \area_{\alpha\beta}} \cdot \bft_{\partial \area_{\beta\omega}} & = \cos{\theta}_{2}, \\ 
    \bft_{\partial \area_{\alpha\omega}} \cdot \bft_{\partial \area_{\beta\omega}} & = \cos{\theta}_{3};   
\end{align}
\end{subequations}
see the right of \cref{fig:AIC} for an illustration.
Each adjoining interface contributes a conormal force at the triple line, directed along $\bft_{\partial \area_{jk}}$, which is tangent to $\area_{jk}$ and orthogonal to $\Gamma$, see \cref{eq:LL_tang_var}. Let us start with $\dxi_\perp = \delta\xi\,\bft_{\partial \area_{\alpha\beta}}$, where the key part of the expansion of the internal energy takes the form
\begin{align}\label{eq:threeLine}
& \sum_{\substack{jk\in\{\alpha\beta,\beta\omega,\alpha\omega\}}} \sqp{\int_{\widetilde{\area}_{jk}}\tilde{u}^\area\,\de \tilde{a} - \int_{\area_{jk}}u^\area\,\de a} \nonumber \\
& \qquad \qquad =  \int_\bline \left(\sum_{\substack{jk\in\{\alpha\beta,\beta\omega,\alpha\omega\}}}\gamma_{jk}\,\bft_{\partial \area_{jk}}-  \tau  \bfk_\bline\right)\cdot \bft_{\partial \area_{\alpha\beta}}\delta\xi\,\de \ell  + O(|\dxi|^2).
\end{align}
The summed terms in parentheses on the right hand side come from the term dependent on the surface divergence we derived in the previous section, \cref{eq:LL_tang_var}. This gives a contribution on $\bline$ for each pair of phases in contact. As $\delta\xi$ is arbitrary, we obtain that the integral terms linear in $\delta\xi$ must vanish, so that:
\begin{equation}\label{eq:N1}
   \gamma_{\alpha\beta} + \gamma_{\alpha\omega}\cos{\theta_1} + \gamma_{\beta\omega}\cos{\theta_2} = \tau\bfk_\bline\cdot \bft_{\partial \area_{\alpha\beta}}.
\end{equation}
By applying the same reasoning for perturbations in the other outward-pointing tangent directions, we obtain:
\begin{align}
    \gamma_{\alpha\beta}\cos{\theta_1} + \gamma_{\alpha\omega} + \gamma_{\beta\omega}\cos{\theta_3} & =\tau \bfk_\bline\cdot \bft_{\partial \area_{\alpha\omega}}, \label{eq:N2} \\
    \gamma_{\alpha\beta}\cos{\theta_2}  + \gamma_{\alpha\omega}\cos{\theta_3} + \gamma_{\beta\omega} & = \tau \bfk_\bline\cdot \bft_{\partial \area_{\beta\omega}}. \label{eq:N3}
\end{align}
The three equations \cref{eq:N1,eq:N2,eq:N3} represent the generalized Neumann equations \citep{Li1989phase}. These three equations are not independent as the angles between interfaces satisfy $\theta_1 + \theta_2 + \theta_3 = 2\pi$. The \cref{eq:N1,eq:N2,eq:N3} can equivalently be rewritten as a single vectorial equation as:
\begin{equation}\label{eq:young}
    \gamma_{\alpha\beta}\bft_{\partial \area_{\alpha\beta}} + \gamma_{\alpha\omega}\bft_{\partial \area_{\alpha\omega}} + \gamma_{\beta\omega}\bft_{\partial \area_{\beta\omega}} = \tau \bfk_\bline\,.
\end{equation}
It is easy to verify that taking the scalar product of \cref{eq:young} with one of the three outward-pointing tangent vectors will yield each of the \cref{eq:N1,eq:N2,eq:N3}. We note that if the line tension $\tau$ is negligible, or $\bline$ is a straight line (having curvature $\bfk_\bline=\bfzero$), then we obtain the standard form of the Neumann equation \citep{Rowlinson2013}. This result was also obtained in \citet{Hoffman1972} as a special case for isotropic energy in their discussion of anisotropic energy; see eq. (20) therein.

Let us consider an illustrative special case to show that this result is a generalisation of another well-known case. Suppose that one of the phases, $\omega$, forms flat, rigid interfaces with the other two phases, $\area_{\alpha\omega}$ and $\area_{\beta\omega}$. This is a common assumption used to model a fluid droplet on a solid surface surrounded by an ambient gas. Suppose also that the line tension $\tau$ of the three-phase line $\bline$ is negligible relative to the surface tension. In this case, $\theta_3=\pi$ and $\theta_1+\theta_2=\pi$. Of the three deformations considered before (i.e., along $\bft_{\partial \area_{kj}}$), there is now no deformation possible in the direction $\bft_{\partial \area_{\alpha\beta}}$, as we assumed the interface of the $\omega$ phase  to be rigid. This constraint means that only \cref{eq:N2,eq:N3} remain: in turn, these can be shown to be the same equation by using the addition formula for the cosine $\cos{\theta_1}
=\cos{(\pi-\theta_{2})}=-\cos{\theta_2}$:
\[
\gamma_{lv}\cos{\theta_2} - \gamma_{sv} + \gamma_{ls} = 0.
\]
Here, for clarity, we have replaced $\alpha \rightarrow v$, $\beta \rightarrow l$, $\omega \rightarrow s$, to indicate vapour, fluid and solid respectively. This last equation is the well-known Young equation \citep{Rowlinson2013,Zhou2025} which again arises as a particular case of the framework presented here.

We note that several curvature corrections to the classical Young–Laplace equation have been proposed, beginning with the famous Tolman correction \citep{Tolman1949,Blokhuis2006}, in which the surface free energy of a spherical interface is not identified with its planar value $\gamma_\infty$, but depends on the droplet radius and may be expressed asymptotically as an expansion in inverse powers of that radius of curvature, with the leading correction governed by the Tolman length. The corresponding equilibrium condition then contains additional curvature-dependent terms and surface differential operators, and no longer reduces to the classical Young–Laplace relation with constant $\gamma$. In the present work, however, the interfacial energy is assumed to depend only on the interfacial entropy density, the surface molar densities, and the interface orientation, i.e., $u^\Sigma(\eta,\rho_i,\bfn)$, with no explicit dependence on the mean or Gaussian curvature. Tolman and higher-order curvature contributions are therefore outside the scope of the present formulation, but it is of course possible to extend our analysis to include such terms.

As a final derivation of this section, we will show that Herring's formula \citep{Herring1999} also emerges naturally from this framework. Herring's formula can be seen as extension of the Young equation to the case where $\gamma(\bfn)$ depends also on the normal to the interface. Herring's formula arises in the case where the line tension is negligible; we could also consider a case where the line energy $u^\bline$ varies with the tangent direction of the curve $\bline$, but for simplicity of presentation and comparison with Herring's formula, we will not discuss this case for now.
Therefore, let us again consider \cref{eq:threeLine}, for a variation $\dxi = \delta\xi\,\bft_{\partial \area_{ij}}$, with negligible line tension and where we now include the dependence on the orientation of the surface $\bfn$: 
\begin{align}
&\sum_{\substack{jk\in\{\alpha\beta,\beta\omega,\alpha\omega\}}}  \int_{\tilde{\area}_{jk}}\tilde{u}_{jk}^\area(\tilde{\bfn})\,\de \tilde{a} =   \sum_{\substack{jk\in\{\alpha\beta,\beta\omega,\alpha\omega\}}}  \int_{\area_{jk}} \sqp{u_{jk}^\area(\bfn) -\frac{\partial u_{jk}^\area(\bfn)}{\partial\bfn}\cdot(\id-\bfn\otimes\bfn)\nabla\dxi^T\bfn + u_{jk}^{\area}(\bfn) \nabla_{\area_{jk}} \cdot  \dxi}\,\de a  \nonumber \\
&\qquad\qquad\qquad + \int_\bline \sum_{\substack{jk\in\{\alpha\beta,\beta\omega,\alpha\omega\}}} \sqp{u_{jk}^\area(\bfn)\bft_{\partial \area_{jk}} }\cdot \dxi\,\de \ell + O(|\dxi|^2) \nonumber \\
&\qquad = \sum_{\substack{jk\in\{\alpha\beta,\beta\omega,\alpha\omega\}}}  \int_{\area_{jk}} \sqp{u_{jk}^\area - \cip{\nabla_\area\cdot\sqp{u^\area_{jk}(\bfn)\id - \bfn \otimes\frac{\partial u^\area_{jk}(\bfn)}{\partial\bfn}}}\cdot \dxi_t }\,\de a \nonumber \\
&\qquad\qquad\qquad+ \int_\bline \sum_{\substack{jk\in\{\alpha\beta,\beta\omega,\alpha\omega\}}} \sqp{\cip{u_{jk}^\area(\bfn)- \bfn \otimes\frac{\partial u^\area_{jk}(\bfn)}{\partial\bfn}}\bft_{\partial \area_{jk}}\cdot \dxi }\,\de \ell + O(|\dxi|^2)
\end{align}
where the second equality is obtained by using the same reasoning used to obtain \cref{eq:norm}, but we now include the boundary term along $\bline$. We can apply the same derivation to the other thermodynamic quantities and sum all the equations together which brings in the terms $\gamma_{kl}(\bfn)$. We focus on the boundary term which now reads as:
\begin{align}\label{eq:N_gamma_CH}
    \int_{\bline}\sum_{\substack{jk\in\{\alpha\beta,\beta\omega,\alpha\omega\}}}  \sqp{\gamma_{kl}(\bfn)\id - \bfn_{kl}\otimes \pard{\gamma_{kl}}{\bfn_{kl}}}\bft_{\partial\area_{kl}} \cdot \dxi  \,\de \ell
\end{align} 
from which we obtain the equilibrium condition:
\begin{equation}\label{eq:preherr}
    \sum_{\substack{jk\in\{\alpha\beta,\beta\omega,\alpha\omega\}}}  \sqp{\gamma_{kl}(\bfn)\id - \bfn_{kl}\otimes \pard{\gamma_{kl}}{\bfn_{kl}}}\bft_{\partial\area_{kl}} = \bfzero.
\end{equation}
For now, let us consider one of these terms in this sum, dropping the $kl$ subscripts for conciseness for the moment. We will show that following relation holds for each term:
\begin{equation}
\gamma\cip{\bfn}\bft_{\partial\area} - \cip{\pard{\gamma\cip{\bfn}}{\bfn}\cdot \bft_{\partial\area}} \bfn = \bft_{\bline} \times \CHv,
\end{equation}
where $\CHv$ is the Cahn--Hoffman capillarity vector, as defined in \cref{eq:defn_CHv}.
Considering one of the summands and multiplying by the appropriate outward-pointing tangent vector, we have
\begin{align}
\cip{ \bfn\otimes \pard{\gamma\cip{\bfn}}{\bfn}\bft_{\partial\area}} = n_i (\partial_j\gamma)  t_j = \cip{\pard{\gamma\cip{\bfn}}{\bfn}\cdot \bft_{\partial\area}} \bfn.
\end{align}
Next, using \cref{eq:CH_vector}, the vectorial relation that $\bft_{\bline} \times \bfn = \bft_{\partial \area}$ and the fact that $\pard{\gamma(\bfn)}{\bfn}$ is tangent to the interface $\area$, so can be decomposed in the two directions, $\bft_{\bline}$ and $\bft_{\partial \area}$, we have
\begin{align}
 \bft_{\bline} \times \CHv & = \bft_{\bline} \times \cip{\gamma(\bfn) \bfn + \pard{\gamma}{\bfn} } \nonumber \\
 & = \gamma(\bfn) \bft_{\partial\area} + \bft_{\bline} \times \cip{\pard{\gamma}{\bfn}\cdot \bft_{\bline}}\bft_{\bline} + \bft_{\bline} \times \cip{\pard{\gamma}{\bfn}\cdot \bft_{\partial\area}}\bft_{\partial\area}  \nonumber \\
 & = \gamma(\bfn) \bft_{\partial\area} + \bft_{\bline} \times \cip{\pard{\gamma}{\bfn}\cdot \bft_{\partial\area}}\bft_{\partial\area} = \gamma(\bfn) \bft_{\partial\area} -  \cip{\pard{\gamma\cip{\bfn}}{\bfn}\cdot \bft_{\partial\area}} \bfn.
\end{align}
As a result, we can therefore rewrite the equilibrium conditions in \cref{eq:preherr} as:
\begin{equation}\label{eq:herr}
    \bft_{\bline} \times \sum_{\substack{jk\in\{\alpha\beta,\beta\omega,\alpha\omega\}}} \bfX_{kl} = \bfzero,
\end{equation}
where $\CHv_{kl}$ is simply the appropriate Cahn--Hoffman vector for each of the interfaces.
This condition entails that the vector sum of the Cahn-Hoffman vectors for all three interfaces at the three-phase boundary is parallel to the direction of the tangent to the curve $\bline$. This can be seen as a generalized force balance at the intersection curve, where the components of the forces includes terms depending on the anisotropy of the surface free energy at the interfaces; this latter fact is clearly shown in \citep{Herring1999}, see Appendix 2 in the cited reference. The relation for the Cahn-Hoffman vector also appears explicitly in the analysis of \citet{Wheeler1999}; see eq. 9 in \citep{Wheeler1999}.

We note that the hypothesis that the line tension is negligible is not restrictive. In more generality, consider \cref{eq:N_gamma_CH} together with \cref{eq:threeLine} obtaining:
\begin{align}
    \int_{\bline}\sqp{\sum_{\substack{jk\in\{\alpha\beta,\beta\omega,\alpha\omega\}}} \cip{\gamma_{kl}(\bfn)\id - \bfn_{kl}\otimes \pard{\gamma_{kl}}{\bfn_{kl}}}\bft_{\partial\area_{kl}} - \tau\bfk_\bline} \cdot \dxi  \de \ell
\end{align} 
Our prior analysis allows us to obtain a generalized version of the equilibrium relation:
\begin{equation}
\bft_{\bline} \times \sum_{\substack{jk\in\{\alpha\beta,\beta\omega,\alpha\omega\}}}   \bfX_{jk}   =  \tau\bfk_{\bline}.
\end{equation}

We remark that the above result arises in the case where $u^\bline$ is independent of the orientation of the three-phase curve, but we could have considered an orientation dependence for the line energy itself. The most natural choice would be to assume that $u^\bline$ is a function of the tangent $\bft_\bline$. This would yield a more complex expression for the line stress \cref{eq:line_stress}, which would now involve a derivative of the line tension $\tau$ with respect to $\bft_\bline$. While we do not pursue this in detail, arguments in close analogy to our treatment of orientation dependent surface tension would allow the correct identification of such corrections. Similarly, one could pursue equilibrium relations at a four-phase contact point, but again, a full development of the necessary equilibrium conditions is beyond the scope of this work.

\subsection{Equilibrium Equations}

We now summarise the final thermodynamic equations we obtain for the most general system we have considered, i.e. three phases in contact with each surface tension depending on the local surface orientation. This yields:
\begin{align}\label{eq:EQ1}
    U = TS + \sum_{i=1}^{\ncomp} \mu_i M_i - \sum_{j=\alpha,\beta,\omega}P_j\vol^j +\sum_{\substack{jk\in\{\alpha\beta,\beta\omega,\alpha\omega\}}} \int_{\area_{jk}}\gamma(\bfn) \de a + \tau L 
\end{align}
where $L$ is the total length of the three-phase contour, $S$ is the total entropy in the system, and $M_i$ is the total moles of each species. Previous equation is coupled with the following equilibrium conditions:
\begin{subequations}
    \begin{align}
        \nabla\cdot\bfcauchy^i & = \bfzero &&\text{in }\vol^i,\label{eq:EQNOisoPress}\\
        \id_{\area_{jk}}\left(\nabla_{\area_{jk}} \cdot \sstress(\bfn^{jk})\right) & = 0&&\text{on }\area_{jk},\\
        \nabla_{\area_{jk}}\cdot\bfX_{jk}+P^k-P^j & = 0,&&\text{on }\area_{jk},\text{ and}\\
         \bft_{\bline} \times \sum_{\substack{jk\in\{\alpha\beta,\beta\omega,\alpha\omega\}}} \bfX_{jk}  & = \tau\bfk_{\bline} &&\text{on }\bline.
    \end{align}    
\end{subequations}
with $i,j,k\in\{\alpha,\beta,\omega\}$, $j\neq k$. We recall that $\bfn^{jk}$ is the normal perpendicular to the surface $\area_{jk}$ pointing from phase $j$ to phase $k$, $\bft_\bline$ is the tangent to the boundary line, and $\bfk_\bline$ is the curvature vector of the boundary line, which is parallel to the normal to the curve when it is non-zero. We also recall the definitions of the Cauchy stress in each fluid phase \cref{eq:bulk_fluid_cauchy}, the surface stress $\sstress$ \cref{eq:def_surf_stress_n} and Cahn--Hoffman vector $\CHv$ \cref{eq:defn_CHv}.

If the internal energy of each of the phase boundaries is isotropic (i.e., there is no dependence on $\bfn$) and the line tension term is negligible then we have:
\begin{align}\label{eq:EQ2}
    U = TS + \sum_{i=1}^{\ncomp} \mu_i M_i - \sum_{j=\alpha,\beta,\omega}P_jV^j + \sum_{\substack{jk\in\{\alpha\beta,\beta\omega,\alpha\omega\}}} \gamma_{jk}A^{jk} 
\end{align}
together with:
\begin{subequations}
    \begin{align}
            \nabla\cdot\bfcauchy^i & = \bfzero &&\text{in }\vol^i,\label{eq:EQisoPress}\\
            \nabla_{\area{jk}}\gamma^{jk} & = 0&&\text{on }\area_{jk}, \label{eq:EQisoGamma}\\            
        2\gamma_{jk} H_{jk}+P^j-P^k & = 0&&\text{on }\area_{jk},\text{ and} \label{eq:YLstandard}\\ 
    \gamma_{\alpha\beta}\bft_{\partial \area_{\alpha\beta}} + \gamma_{\alpha\omega}\bft_{\partial \area_{\alpha\omega}} + \gamma_{\beta\omega}\bft_{\partial \area_{\beta\omega}} & = 0&&\text{on }\bline;
    \end{align}    
\end{subequations}
where again $i,j,k\in\{\alpha,\beta,\omega\}$ with $j\neq k$.

We note that \cref{eq:EQisoPress,eq:EQNOisoPress,eq:EQisoGamma} represent trivial identities in the three-fluid case, since the Cauchy stress is determined by a constant scalar pressure, and the surface free energy terms are constant. However, we have nevertheless chosen to write the equations in this form as these relations hold in a very similar form even in cases where we consider a solid in contact with one or more fluid phases.

Finally, before discussing the changes to the derivation in solid-fluid systems, for the reader's convenience we summarize the different variations we considered in this first part and their consequences; each is reported in \cref{tab:variation-taxonomy}.

\begin{table}[ht]
\centering
\renewcommand{\arraystretch}{1.35}
\begin{tabular}{p{0.20\textwidth} p{0.40\textwidth} p{0.30\textwidth} }
\hline
\textbf{Variation} &
\textbf{What is varied} &
\textbf{Phase identity} \\
\hline

Outer variation &
Local thermodynamic fields, such as $\eta, \rho_i$, and, when admissible, independent compositional variables. In the solid phase, this includes the constrained variations of substitutional and interstitial species. &
Fixed. No material point changes phase and the phase domains are not displaced, only thermodynamic variables change at a point.  \\

\hline

Inner variation &
The current spatial configuration $\bfx$ is mapped to the new configuration $\widetilde{\bfx}$:
\[
\bfx\mapsto \widetilde{\bfx}=\bfx+\dxi(\bfx).
\]
This variation displaces and strains the existing material. In the bulk it gives the usual mechanical equilibrium conditions; on the interface it also changes the geometry of the existing dividing surface, including its area and curvature. &
Fixed. No material point changes phase, but the existing phases are stretched, and interfaces may be geometrically deformed.  \\

\hline

Configurational variation &
The phase label of the material points themselves.
In the current configuration one may write
\[
\dxi^\area=\delta \xi^\area\bfnu,
\]
whereas in the reference configuration of the solid we have
\[
\dxi^{\area_0}=\delta \xi^{\area_0} \bfN.
\]
This variation changes the domain occupied by each phase and, for a solid, changes the reference domain occupied by the lattice. &
Varies. A layer of fluid may be converted into solid, or a layer of solid may dissolve into fluid. This is an accretion/dissolution or phase-conversion variation, not a purely mechanical deformation of the existing material. \\

\hline
\end{tabular}
\caption{
A taxonomy of the admissible variations considered.
}
\label{tab:variation-taxonomy}
\end{table}


\section{Solid-fluid Systems}
\label{sec:solid-fluid}

The main difference introduced by the presence of a solid is that, in general, the internal energy density defined in \cref{eq:energy} depends on the local deformation gradient $\bfstrain$, which characterises the elastic deformation of the solid. We view the strain we measure as being obtained by integrating increments of strain at a material point along some deformation path from a fixed reference state. We suppose that this reference state can be determined in a way which is spatially uniform across the entire volume of the solid phase, as is often assumed in the case of a purely elastic solid. In the case of a crystalline solid, the reference state could be some lattice structure, determined for example through x-ray diffraction or electron microscopy. We note that the assumption of a single reference may not be valid in cases where there are large numbers of dislocations present in a crystal, since the presence of these topological defects prevent a single global reference from being found. Nevertheless, we will assume that this is indeed possible for the arguments that follow, which will still yield a model valid in systems with low dislocation density.

Throughout this section, we label the solid phase by $s$ and assume that it occupies the volume $\vol^s$. We further assume that the solid is materially homogeneous, meaning that its constitutive energy density has no explicit dependence on the material position $\bfX$. Equivalently, a uniform translation of the material coordinates does not change the energy \citep{Kienzler1997}.


In the classic work of \citet{Larche1973}, some of the species present were identified as immobile ``substitutional'' species, while others were free to diffuse, which they called ``interstitial'' components. At this stage of our approach, no distinction between interstitial and substitutional species is required, since the preceding variational formulation, using geometric variation to allow strains to develop, applies formally to a generic set of compositional fields. However, such a distinction becomes necessary when the lattice constraint of the crystalline phase is introduced, and we will see that our results are consistent with Larch\'{e} and Cahn theory once the appropriate additional constraints are imposed.

\subsection{Constitutive assumptions and outer variation}\label{sec:outerSOL}

We begin by considering just a region of solid phase for now, as did when considering multiple fluid phases. This will lead us to equilibrium conditions within the bulk of the solid phase first. Before performing such variations, we first lay out our constitutive assumptions in greater detail.

As mentioned above, the feature which distinguishes a solid from a fluid is that the internal energy depends on the local deformation away from some fixed reference state as represented by some measure of strain. This leads us to consider a map of the coordinates from the reference configuration $\vol_0$ with coordinates $\bfX$ and the current configuration with coordinates $\bfx$. When we act further to deform a solid material, the point $\bfx$ moves to a point $\tilde{\bfx}$ under this variation. In addition, due to possible spatial variation in the amount by which material points move, this motion may alter the strain, which in turn affects the internal energy.
Following the approach of \citep{Larche1978} (and more generally the approach used in nonlinear elasticity theory), we assume there is an invertible mapping of the form $\bfx=\bfchi (\bfX)$, and consider the deformation gradient $\bfstrain=\nabla_{\bfX}\bfchi(\bfX)$. The tensor $\bfstrain$ 
 describes the transformation of a small volume elements about $\bfX$ in the reference  configuration required to bring it to the configuration $\bfx$. Other choices of strain measure are possible, but we will focus on this choice here.

Assuming therefore that the internal energy density in the solid depends on the local entropy density $\eta$, the local molar densities $\rho_i$, and the deformation gradient $\bfstrain$,
the equilibrium problem now becomes to minimise the extended functional:
\begin{equation*}
    \min_{\eta,\rho_{1},\ldots,\rho_{\ncomp},\bfchi} \left[\int_{\vol^s}u^s\,\de v + T\left(S-\int_{V^s}\eta^s\,\de v\right)+\sum_{i=1}^\ncomp\mu_i\left(M_i-\int_{V^s}\rho^s_i\,\de v\right)\right],
\end{equation*}
where we assume the bulk energy depends upon the entropy and strain, i.e. $u^s = u^s(\eta^s,\rho^s_{1},\ldots,\rho^s_{\ncomp},\bfstrain\circ\bfchi^{-1})$, and each of the fields are expressed as a function of position in the current spatial configuration, $\bfx$. 

If we instead introduce entropy and molar densities in the reference configuration, defined by
\begin{equation*}     \eta^s_0(\bfX):=\eta(\chi(\bfX))\det\bfstrain(\bfX)\quad\text{and}\quad\rho^s_{i,0}(\bfX):=\rho^s_i(\chi(\bfX))\det\bfstrain(\bfX),
\end{equation*}
with the energy in the reference configuration $u_0^s$ given by 
\[
    u^s_0(\eta^s_0,\rho^s_{0,1},\ldots,\rho^s_{0,\ncomp},\bfstrain) := (\det\bfstrain) u^s\cip{\frac{\eta^s_0}{\det\bfstrain},\frac{\rho^s_{0,1}}{\det\bfstrain},\ldots,\frac{\rho^s_{0,\ncomp}}{\det\bfstrain},\bfstrain}
\]
then the equilibrium problem can also be written
\begin{equation*}
    \min_{\eta^s_0,\rho^s_{0,1},\ldots,\rho^s_{0,\ncomp}, \bfchi} \left[\int_{\vol^s_0}u^s_0\,\de v_0 + T\left(S-\int_{\vol^s_0}\eta^s_0\,\de v_0\right)+\sum_{i=1}^\ncomp\mu_i\left(M_i-\int_{\vol^s_0}\rho^s_{0,i}\,\de v_0\right)\right].
\end{equation*}
As before the additional terms in the extended functional ensure that the minimum is subject to the constraints of:
\begin{itemize}
    \item Constant total entropy: 
        \begin{equation}
            S = \int_{\vol}\eta^s\,\de v=\int_{\vol_0}{\eta^s_0\,\de v_0};
        \end{equation}
    \item Constant total moles of each component: 
        \begin{equation}
            M_i = \int_{\vol}\rho^s_i\,\de v= \int_{\vol_0}{\rho^s_{0,i}\,\de v_0},
        \end{equation} 
        for $i=1,\ldots,\ncomp$; and
    \item Constant total volume, so the domain occupied by the (solid) phase with volume $\vol = \chi(\vol_0)$ has fixed boundaries $\partial \vol = \chi(\partial\vol_0)$.
\end{itemize}
Before imposing any additional local compositional constraint, by perturbing the volumetric fields $\eta^s$, $\rho_i^s$, through outer variation as before (see \cref{sec:outvarLL}), we can deduce:
\begin{align*}
 \frac{\partial u^s}{\partial\eta^s} =T \qquad \qquad
\frac{\partial u^s}{\partial\rho_i^s} = \mu_i.   
\end{align*}
The second relation must be replaced by the corresponding constrained
stationarity condition when the compositional fields are not locally
independent, as occurs for substitutional species in a crystalline solid.

We now specialize the general compositional description to a crystalline solid. For systems containing one (or more) solid phases we add an additional constraint, the conservation of lattice site number \citep{Larche1973}. In their work, Larch\'{e} and Cahn assume that some of the species present in the solid phase are `\textit{substitutional}', while others are `\textit{interstitial}'. The distinction between these species is that the substitutional components fill a fixed set of lattice sites in the crystal structure, and hence under the additional implicit assumption that the solid is rigid, the total molar density of these components must be the same everywhere. In our framework, we can express this condition as the requirement that for some group of indices $\subst$, the total concentration of these components, defined as:
\begin{align*}
    \totmolconc(\bfx) & := \sum_{i \in \subst }\rho^s_{i}(\bfx)\,  = N 
\end{align*}   
is a constant throughout $\vol^s$. Therefore, for a crystalline solid containing substitutional species, the corresponding outer variations are not independent. Since the local number of substitutional lattice sites is fixed, an increase in the density of one substitutional species must be accompanied by a corresponding decrease in the density of one or more of the other substitutional species. Accordingly, the admissible outer variations satisfy the local constraint
\begin{equation}\label{eq:DiffPot}
    \sum_{i\in\subst} \delta \rho^s_i(\boldsymbol{x}) = 0
    \qquad
    \text{for every } \boldsymbol{x}\in V^s.
\end{equation}
Let $r\in\mathcal{S}$ be an arbitrarily chosen reference substitutional species. The contribution of the substitutional densities to the first variation of the augmented functional is therefore
\begin{align*}
\int_{\vol^s} \sum_{i\in\subst} \left( \frac{\partial u^s}{\partial \rho^s_i}-\mu_i\right)\delta \rho^s_i\,dv = \int_{\vol^s} \sum_{\substack{i\in\mathcal{S}\\ i\neq r}}
\left[ \Dpartial {u^s}{\rho^s_i} - \frac{\partial u^s}{\partial \rho^s_r} - \left( \mu_i-\mu_r\right)\right]
\delta \rho^s_i\,dv.
\end{align*}
Since the remaining variations $\delta \rho^s_i$, with $i\neq r$, are independent, stationarity yields
\begin{equation*}
    \Dpartial {u^s}{\rho^s_i} - \Dpartial {u^s}{\rho^s_r} = \mu_i-\mu_r, \qquad i\in\mathcal{S}, \quad i\neq r.
\end{equation*}
We define the diffusion potential of substitutional species $i$ relative to the reference species $r$ as:
\begin{equation}
    \diffpot_i^{\,r} := \mu_i-\mu_r = \Dpartial {u^s}{\rho^s_i} - \Dpartial {u^s}{\rho^s_r}.
    \label{eq:diffusion_potential_definition}
\end{equation}
Note that we used the symbol $\diffpot$ instead of the symbol $M$ originally used by Larch\'{e} and Cahn to avoid ambiguities with other quantities defined in this work.
The choice of the reference species $r$ is arbitrary and only determines the representation adopted for the independent diffusion potentials. The physical equilibrium conditions are unchanged if a different substitutional reference species is selected.

\subsection{Inner variation}\label{sec:innvarsolid}

We note that in our discussion of the constitutive assumption on solids, we have chosen to express all fields by default in the current configuration, as this is the state which is directly measurable. This choice has been made to be consistent with our presentation of the fluid case, and to allow us later to deduce the correct equations for the coupling between such phases in a consistent manner. To that end, we note in particular that the deformation gradient $\bfstrain$ is expressed as
\[
\bfF(\bfx) = \nabla_{\bfX}\chi\big(\chi^{-1}(\bfx)\big).
\]
Under a further geometric variation moving the point $\bfx$ to $\tilde{\bfx} = \bfphi(\bfx) = \bfx+\dxi(\bfx)$, we note that the new strain at the new position of the material point can be expressed in any of the configurations as
\[
\widetilde{\bfF}(\tilde{\bfx}) = \nabla_{\bfX}(\bfphi\circ\chi)(\bfX) = \nabla_{\bfX}(\bfphi\circ\chi)\big(\chi^{-1}(\bfx)\big)=\nabla_{\bfX}(\bfphi\circ\chi)\Big(\chi^{-1}\big(\bfphi^{-1}(\tilde{\bfx})\big)\Big).
\]
Next, we note that by applying the chain rule, we have that
\begin{equation}\label{eq:defStrain}
\widetilde{\bfstrain}(\tilde{\bfx}) = \nabla_\bfx\bfphi(\bfx)\nabla_\bfX\chi(\bfX) = \big(\id+\nabla_{\bfx}\dxi(\bfx)\big)\bfstrain(\bfx).
\end{equation}
Note that we have already identified the transformation rule for the volume element $\de v$, and for molar and entropy densities up to terms of first-order in $\dxi$ in \cref{eq:transf_vol,eq:mass_ent_transformation}.

Therefore, combining \cref{eq:transf_vol,eq:mass_ent_transformation,eq:defStrain}, we can expand under the assumption that $\dxi$ is small to find that
\begin{align*}
	&\int_{\tilde{\vol}^s} u^s\big(\tilde{\eta}^s(\tilde{\bfx}),\tilde{\rho}^s_1(\tilde{\bfx}),\ldots,\tilde{\rho}^s_\ncomp(\tilde{\bfx}),\widetilde{\bfstrain}(\tilde{\bfx})\big)\de \tilde{v}\\
    &\qquad=\int_{\vol^s}u^s\Big(\tilde{\eta}^s\circ\bfphi(\bfx),\tilde{\rho}^s_1\circ\bfphi(\bfx),\ldots,\tilde{\rho}^s_\ncomp\circ\bfphi(\bfx),\big(\id+\nabla\dxi(\bfx)\big)\bfstrain(\bfx)\Big)\Big(1+\nabla\cdot\dxi(\bfx)\Big)\de v + O(|
    \dxi|^2)\\
    &\qquad=
    \int_{\vol^s}\left[u^s + \Dpartial{u^s}{\strain_{ia}} \Dpartial{\delta\xi_i}{x_k}F_{ka} + \left(u^s-\Dpartial{u^s}{\eta^s} \eta^s- \sum_{l=1}^\ncomp\Dpartial{u^s}{\rho^s_l}\rho^s_l\right)\Dpartial{\delta\xi_i}{x_i}\right]\de v + O(|\dxi|^2),
\end{align*}
where we employ the Einstein summation convention for clarity, and we have suppressed the evaluation at $\bfx$ in the final integrand expressions for clarity.
Using the product rule and collecting terms, we can write
\begin{equation}
	\begin{aligned}
		\int_{\tilde{\vol^s}}\tilde{u}^s\,\de \tilde{v}-\int_{\vol^s}u^s\,\de v &=\int_{\vol^s}\Dpartial{}{x_k}\left[\Dpartial{u^s}{F_{ia}}F_{ka}\delta\xi_i+\left(u^s-\Dpartial{u^s}{\eta^s}\eta^s-\sum_{l=1}^\ncomp\Dpartial{u^s}{\rho^s_l}\rho^s_l\right)\delta_{ik}\delta\xi_i\right]\de v\\ 
		&\qquad-\int_{\vol^s}\Dpartial{}{x_k}\left[\Dpartial{u^s}{F_{ia}}F_{ka}+\left(u^s-\Dpartial{u^s}{\eta^s}\eta^s-\sum_{l=1}^\ncomp\Dpartial{u^s}{\rho^s_l}\rho^s_l\right)\delta_{ik}\right]\delta\xi_i\,\de v+O(|\dxi|^2),
	\end{aligned}\label{eq:solid_energy_var}
\end{equation}
where $\delta_{ik}$ is the usual Kronecker delta symbol. We note  that under this form of variation,
$$
\int_{\tilde{\vol}^s}\tilde{\eta}^s\,\de\tilde{v}=
\int_{\vol^s}\eta^s\,\de v,
$$
and a similar result holds for the molar densities $\rho^s_i$ and the total molar concentration $\totmolconc$ as it is a linear combination of molar densities of substitutional species.

Therefore, defining the Cauchy stress
\begin{equation}\label{eq:def_cauchy_solid}
	\cauchy_{ij}^s:=\Dpartial{u^s}{F_{ia}}F_{ja}+\omega^s\delta_{ij}
\end{equation}
where the grand potential density is
\begin{equation}\label{eq:grandPotDens}
    \omega^s := u^s-T\eta^s-\sum_{i\in\interst}\mu_i \rho_i^s -\sum_{i\in\subst \setminus\{r\}}\diffpot_i^r \rho^s_i - \mu_r \totmolconc
\end{equation}
where we replaced the derivatives of the substitutional components with their equilibrium expression obtained in \cref{eq:DiffPot}.
We can now rewrite \cref{eq:def_cauchy_solid}  in tensorial form as
$$
\bfcauchy^s=\Dpartial{u^s}{\bfstrain}\bfstrain^T+\omega^s\id,
$$
and by noting that the first term on the right-hand side of \cref{eq:solid_energy_var} is in divergence form, by applying the divergence theorem, the energy difference can be written as:
\begin{equation}\label{eq:deltaEsolidbulk}
     \widetilde{E}^s-E^s =  \int_{\partial V^s}
    (\bfcauchy^s\bfn)\cdot
    \dxi\,da, -\int_{\vol}(\nabla\cdot\bfcauchy^s)\cdot\dxi\,\de v +O(|\dxi|^2).
\end{equation}
The standard argument that terms linear in $\dxi$ must vanish now yields that the tensor divergence of the Cauchy stress must vanish everywhere in the bulk at equilibrium, i.e. $\nabla\cdot\bfcauchy^s = \bfzero$ in $\vol$, or in components
\begin{equation}
	\frac{\partial \cauchy^s_{ij}}{\partial x_j} = \frac{\partial}{\partial x_j}\left(\frac{\partial u^s}{\partial F_{ia}}F_{ja}+\omega^s\delta_{ij}\right) = 0. 
\end{equation}
On $\partial \vol$, the admissible transformations satisfy only the normal constraint
\[
    \dxi\cdot\bfn=0,
\]
while their tangential component remains arbitrary. Hence, the contribution over $\partial V$ reduces to
\[
    \int_{\partial \vol}
    \left(
        \id_{\partial \vol}
        \bfcauchy^s\bfn
    \right)\cdot
    \dxi_t\,da.
\]
In the absence of prescribed tangential boundary work, stationarity yields the corresponding natural boundary condition
\[
    \id_{\partial \vol^s}
    \bfcauchy^s\bfn=\boldsymbol 0.
\]
When a portion $\area\subset\partial \vol^s$ of the solid boundary is placed in contact with a liquid, the boundary is decomposed as the union the common solid-liquid interface, $\area$, and the boundary of the solid not in contact with the liquid $\partial \vol^s /\,\,\area $. The tangential traction-free condition remains valid on $\partial \vol^s /\,\,\area $, whereas on $\area$ it is replaced by the complete solid-liquid interfacial balance derived in the following sections.

\subsubsection{Discussion}
At this point, a few comments comparing our definition of the Cauchy stress tensor $\bfcauchy^s$ to existing theories are in order. First, we note that if the internal energy of the solid is assumed to be unaffected by entropy and molar density variations, so that $\frac{\partial u^s}{\partial \eta^s}=\frac{\partial u^s}{\partial \rho^s_i}=0$, then the stress $\bfcauchy^s$ reduces to just the `elastic' part $\bfcauchy^s = \frac{\partial u^s}{\partial \bfF}\bfF^T+u^s\id$, which is the classical expression (albeit in an Eulerian context) for the stored energy in nonlinear elasticity theory. To see this directly, we note that it is more common in nonlinear elasticity theory to express the energy density per unit reference volume, rather than per unit current volume. Doing so, we have
\[
\int_{\vol} u^s(\bfx)\,\de v = \int_{\vol_0}u^s(\bfX)\det\bfstrain\,\de v_0 =\int_{\vol_0}u^s_0\,\de v_0
\]
so that the usual elastic stored energy density per unit \emph{reference} volume used for a hyperelastic material is
$$
u^s_0(\bfF) = u^s(\bfF)\det\bfF.
$$
In this case, the expression for the Cauchy stress expressed in the reference configuration is
\begin{equation}\label{eq:defCauchyStress}
\bfcauchy^s = \Dpartial{u^s_0}{\bfstrain}\frac{\bfstrain^T}{\det\bfstrain} = \Dpartial{}{\bfstrain}\Big(u^s(\bfF)\det\bfF\Big)\frac{\bfstrain^T}{\det\bfstrain}=\Dpartial{u^s}{\bfstrain}\bfstrain^T+u^s\id,
\end{equation}
where we have used the product rule, and the fact that $\Dpartial{}{\bfstrain}\det\bfstrain = \det\bfstrain\,\bfstrain^{-T}$. We therefore see that our definition is completely equivalent to this well-known case.

As another alternative, the Cauchy stress is often reported as the derivative of the Helmholtz free energy, $\helm^s(\bfF,T,\rho^s_i)$ with respect the deformation (see e.g., \citep{Gurtin2010,LandauV7,Truesdell2004}). We show that again, our expression is equivalent to those obtained in this case. Let us start from the definition of the Helmholtz specific free energy (see \citep{Truesdell2004}, eqs. 82.1--82.6):
\[
\helm^s(T,\rho^s_1,\ldots,\rho^s_{\ncomp},\bfF) = u^s(\eta^s(T),\rho^s_1,\ldots,\rho^s_{\ncomp},\bfF) - T\eta^s(T).
\]
To simplify the notation, only the temperature dependence of $\eta^s$ is shown, while its dependence on $\rho^s$ and $\bfstrain$ is left implicit. The corresponding chain-rule terms cancel identically upon differentiation because $\partial u^s/\partial \eta^s = T$, as a consequence of the Legendre transformation.

Then the appropriate derivatives of $\helm^s$ are:
\[  
\Dpartial{\helm^s}{F_{ia}} =  \Dpartial{u^s}{F_{ia}},\quad \Dpartial{\psi}{T} = \left(\Dpartial{u^s}{\eta^s}-T\right)\Dpartial{\eta^s}{T}-\eta^s = -\eta^s,\quad\text{and}\quad
\Dpartial{\psi}{\rho^s_i}=\Dpartial{u^s}{\rho^s_i}=\mu_i.
\]
where we used the equilibrium relation \cref{eq:EqT} in the expression for the derivative of $\helm^s$ with respect to the temperature, and for the derivative with respect chemical potentials, the terms including the substitutional species are understood to hold on the constrained composition space (see \cref{eq:DiffPot}). This shows that 
\begin{equation}\label{eq:Helm}
\Dpartial {u^s}{\bfF}\bfF^T=\Dpartial{\helm^s}{\bfF}\bfF^T,
\end{equation}
and moreover we can express
$$
\bfcauchy^s = \Dpartial{\helm^s}{\bfstrain}\bfstrain^T+\left(\helm^s-\sum_{i=1}^\ncomp\mu_i\rho^s_i\right)\id.
$$
As such, we are free to transform the equilibrium relations we obtained in terms of the internal energy $u^s$ to relations on the Helmholtz free energy, assuming that there is a unique mapping between the local entropy density $\eta^s$ and the local temperature $T$; the latter is always true as long as the equilibrium is locally stable. We can view the previous equation as a Legendre transformation performed with respect to the entropy. These variables are independent of the surface deformation, and therefore leave the derivative with respect to $\bfstrain$ unaffected.


Another important aspect we wish to highlight here is the definition of the pressure when considering a solid. Let us rewrite the stress $\bfcauchy^s$ as the sum of a deviatoric and a spherical part
\[  
\bfcauchy^s = \bfcauchy^s_{\text{dev}} + \bfcauchy^s_{\text{iso}} = \cip{\bfcauchy^s - \tfrac 13 \trace{\bfcauchy^s}\id} + \tfrac 13 \trace{\bfcauchy^s}\id.
\]
The pressure in the solid is the negative of the coefficient in the spherical part, which, using the Einstein summation convention becomes:
\begin{equation}\label{eq:pressureSolids}
P^s:=-\tfrac 13 \trace{\bfcauchy^s} = -\omega^s-\frac 13\Dpartial{ u^s}{F_{ij}}F_{ji} =-u^s+T\eta+\sum_{i\in\interst}\mu_i \rho^s_i +\sum_{i\in\subst \setminus\{r\}}\diffpot_i^r \rho^s_i + \mu_r \totmolconc-\frac 13\Dpartial{ u^s}{F_{ij}}F_{ji}.
\end{equation}
This shows that for a solid, the grand potential and the pressure are generally distinct quantities, as a result of the solid's resistance to strain. This last observation has important consequences for the analysis  of solid--fluid systems, as we will see in the next section.  

\subsection{Variation in a solid-fluid system}\label{sec:sol-fluid-var}

We now build upon the bulk equilibrium relations obtained for a solid above and consider the case of a solid and fluid phase in contact. In particular, consider a fixed volume $\vol$ containing two phases labelled $s$ for the solid and $l$ for the fluid. Phase $s$ occupies a subvolume $V^s$, phase $l$ occupies a subvolume $V^l$, and the boundary between the phases is a surface $\area$. This dividing surface may have its own boundary curve $\bline=\partial \area$ if there are more than two phases present, but at present we will ignore this case for simplicity.

Assuming as in \cref{sec:2fluids} that we prescribe the total entropy and the moles of the species in the fixed volume $\vol$, we consider the extended functional
\begin{align}\label{eq:enMultiPhSL}
    E[\eta,\rho_1,\ldots,\rho_\ncomp,V^s,\bfstrain,V^l]
     &= \int_{V^s}u^s\,\de v+\int_{V^l}u^l\,\de v+\int_{\area} u^\area\,\de a \nonumber \\
     &\qquad\qquad+T\left(S-\int_{V^s}\eta^s\,\de v-\int_{V^l}\eta^l\,\de v-\int_\area\eta^\area\,\de a\right) \nonumber \\
     &\qquad\qquad+\sum_{i=1}^\ncomp\mu_i\left(M_i-\int_{V^s}\rho_{i}^s\,\de v-\int_{V^l}\rho_i^l\,\de v-\int_\area\rho_i^\area\,\de a\right).
\end{align}
The Larch\'{e}–Cahn lattice-site constraint is imposed only on the substitutional composition variations in the crystalline bulk phase. It is not imposed on the interfacial excess densities, since the Gibbs dividing surface is not assumed to possess an independently conserved lattice-site structure.

Perturbing the volumetric fields $\eta^s$, $\eta^l$, $\rho_i^s$, $\rho_i^l$ and the surface fields $\eta^\area$, $\rho_i^\area$ through outer variation, we can deduce as before for the temperature:
\begin{align*}
 \frac{\partial u^s}{\partial\eta^s}
& =\frac{\partial u^l}{\partial\eta^l} 
=\frac{\partial u^\area}{\partial\eta^\area}
=T 
\end{align*}
For the $\rho_i$ fields the admissible variations depend on the type of species and on the region considered. In particular, the local lattice-site constraint in\cref{eq:DiffPot} applies only to substitutional species in the crystalline bulk. It is not imposed on the interfacial excess densities, since we do not assume that the interface possesses an independently conserved lattice-site structure. 

For interstitial species $i \in \interst$, the density variations are independent in all three regions, giving
\begin{equation*}
    \Dpartial {u^s}{ \rho_i^s} = \Dpartial {u^l}{ \rho_i^l} = \Dpartial {u^\area}{ \rho_i^\area}
    = \mu_i.
\end{equation*}
For substitutional species, the density variations are independent in the liquid and interfacial regions, so that
\begin{equation*}
    \Dpartial {u^l}{\rho_i^l} = \Dpartial {u^\area}{\rho_i^\area} = \mu_i, \qquad i\in\subst.
\end{equation*}
Within the crystalline bulk, however, the substitutional variations satisfy the local lattice-site constraint. Choosing a reference species $r\in\subst$, stationarity gives
\begin{equation}
    \Dpartial {u^s}{\rho_i^s} - \Dpartial{u^s}{\rho_r^s} = \mu_i-\mu_r = \diffpot_i^{\,r},
    \qquad i\in\subst\setminus\{r\}.
\end{equation}
%



Similarly, we have already established that the appropriate consequence of inner variations within each of the bulk phases naturally requires that the divergence of the Cauchy stress must vanish, i.e.
\[
\nabla\cdot\bfcauchy^s = 0,\quad\text{and}\quad
\nabla\cdot\bfcauchy^l = 0,
\]
where in each of the two phases we have the respective definitions
\begin{itemize}
    \item $\bfcauchy^s:=\frac{\partial u^s}{\partial\bfF}\bfF^T+\omega^s\id$ in the solid phase, and
    \item $\bfcauchy^l = \omega^l\id=-P^l\id$ in the fluid phase,
\end{itemize}
and the grand potentials $\omega^s$ and $\omega^l$ are given in \cref{eq:grand_pot_def,eq:grandPotDens}, and the pressure in the fluid, $\press^l$ is defined in \cref{eq:defP}.

We now turn to the phase boundary, where the Cauchy stresses in each phase will contribute to an appropriate force balance on $\area$. To deduce a full expression, we must make some assumptions about the internal energy per unit surface area $u^\area$ in this case. In particular, we choose to assume that the internal energy of the phase boundary is of the form $u^\area = u^\area(\eta^\area,\rho^\area_1,\ldots,\rho^\area_\ncomp,\bfn,\bfstrain)$, i.e. it depends upon the entropy density, the molar density of each component, the orientation and the strain inside the solid phase close to the boundary.

Under an inner variation, we recall the transformation of the area element from \cref{eq:varA} and the transformation of the density fields in \cref{eq:mass_ent_surface_transformation}. Noting that $\nabla_\area\cdot\dxi = \nabla_\area\cdot\dxi_t-2H\delta\xi_n$, where $H$ is the mean curvature to the surface, a similar line of argument to that pursued in \cref{sec:2fluids} leads to the expansion:
\begin{equation*}
\begin{aligned}
\int_{\tilde{\area}}\tilde{u}^\area\de \tilde{a} &= \int_{\area}u^\area(\tilde{\eta}^\area,\tilde{\rho}^\area_1,\ldots,\tilde{\rho}^\area_\ncomp,\tilde{\bfn},\tilde{\bfstrain})\Big(1+\nabla_\area\cdot\dxi_t-2H(\dxi\cdot\bfn)\Big)\de a+O(|\dxi|^2)\\
&=\int_{\area}\left[u^\area+\left(u^\area-\Dpartial{u^\area}{\eta}\eta^\area-\sum_{i=1}^\ncomp\Dpartial{u^\area}{\rho_i^\area}\rho_i^\area\right)\nabla_\area\cdot\dxi_t\right]\de a -\int_{\area}\Dpartial{u^\area}{\bfn}\id_\area\nabla\dxi^T\bfn\,\de a\\
&\qquad+\int_{\area}\Dpartial{u^\area}{\strain_{ij}}\strain_{aj}\Dpartial{\delta\xi_i}{x_a}\,\de a-\int_{\area}2H\left(u^\area-\Dpartial{u^\area}{\eta^\area}\eta-\sum_{i=1}^\ncomp\Dpartial{u^\area}{\rho_i}\rho_i\right)\dxi\cdot\bfn\,\de a + O(|\dxi|^2),
\end{aligned}
\end{equation*}
where we note that the key difference is the inclusion of a strain dependent term. Recalling the discussion in the previous section around \cref{eq:Helm}, performing a Legendre transformation with respect to the entropy and the surface excess densities only, while the deformation gradient $\strain$ is retained as an independent state variable, it follows that:
\[
	\at{\Dpartial{u^\area}{\strain_{ij}}}{\eta^\area, \rho_i^\area} = \at{\Dpartial{\omega^\area}{\strain_{ij}} }{T,\mu_i}\!\!.
\]
That is to say, any time the energy term appears in the derivative with respect the strain we can replace it with the surface grand potential density $\omega^\area$. This step is not strictly essential for the derivation, but as we will see it will make the final equilibrium equations clearer allowing to rewrite all the properties in term of the surface free energy. By considering the expansion of the entropy and molar fields in terms of the variation of the normal (see \cref{eq:surface-energy-var-norm}), the equilibrium conditions obtained using the outer variations (see \cref{sec:outerSOL}),  and including the contributions from the divergence terms in the bulk (see e.g., \cref{eq:deltaEsolidbulk}), we can finally write the perturbation of the surface contribution to the energy functional $E^\area$ as:
\begin{align}\label{eq:solid_surf_var}
\widetilde{E}^\area - E^\area & =\int_{\area}\omega^\area\nabla_\area\cdot\dxi_t\,\de a \nonumber-\int_{\area}\Dpartial{\omega^\area}{\bfn}\cdot\id_\area\nabla\dxi^T\bfn\,\de a+\int_{\area}\Dpartial{\omega^\area}{\bfstrain}\bfstrain^T:\nabla\dxi\,\de a  \nonumber \\
&\qquad-\int_{\area}\cip{2H\omega^\area\bfn - (\bfcauchy^s-\bfcauchy^l)\bfn} \cdot \dxi\,\de a + O(|\dxi|^2),
\end{align}
where $\omega^\area = u^\area-T\eta^\area-\sum_{i=1}^\ncomp\mu_i\rho_i^\area$ is again the surface grand potential density as defined in \cref{eq:sigma}. We note that in this expression, we are assuming that the normal $\bfn$ points from the solid into the fluid phase.
Let us now first assume that the local motion is purely tangential in the vicinity of the surface and is unchanging along the normal direction so that $\dxi = \dxi_t$ in a neighbourhood of $\area$. Before deriving the general equilibrium condition, let us consider a special case, in which the energy $u^\area$ is isotropic and the transformation does not change the state of strain of the system $\bfF$. In this case, the variation of the extended functional becomes:
\[
	\tilde{E}-E  = -\int_\area \nabla_\area \cdot (\omega^\area\id_\area)\cdot\dxi_t\,\de a = - \int_\area (\nabla_\area\omega^\area)\cdot\dxi_t\,\de a
\]
where we note that the term including the Cauchy stress in the extended functional contains a normal part only which does not enter in the tangential variation $\dxi_t$. We can now derive, analogously to the fluid-fluid case, that $\omega^\area$ is uniform along the interface. We denote this constant interfacial grand-potential density by $\gamma$, the solid-liquid surface free energy, so that
\begin{equation}\label{eq:omegasigmagamma}
    \gamma:=\omega^\area.
\end{equation}
Let us go back to the general expression reported in \cref{eq:solid_surf_var}, focusing on purely tangential perturbations $\dxi=\dxi_t$ for now. Working through similar steps as performed in the fluid-fluid interface case for the remaining terms and including the contributions from the divergence terms in the bulk, we find that under such tangential perturbations, we obtain (see \cref{app:perturbation} for details):
\begin{align}
\widetilde{E}^\area - E^\area & =-\int_{\area}\sqp{\nabla_\area\cdot\cip{\omega^\area \id_\area - \bfn\otimes\Dpartial{\omega^\area}{\bfn} + \Dpartial{\omega^\area}{\bfstrain}\bfstrain^T \id_\area } - (\bfcauchy^s-\bfcauchy^l)\bfn} \cdot \dxi_t \de a   \nonumber \\
& \qquad \qquad + \int_{\area} \cip{\id_\area \Dpartial{\omega^\area}{\bfstrain}\bfstrain^T \bfn}\cdot \partial_n\dxi_t \de a+ O(|\dxi|^2),
\end{align}
Since this relation hold true for all tangential perturbations, we obtain the equilibrium equation
\begin{equation} \label{eq:tangentialSigma}
\id_\area\sqp{\nabla_\area\cdot\sstress-\Big((\bfcauchy^s-\bfcauchy^l)\bfn\Big)}=\bfzero,
\end{equation}
where we introduce the surface stress $\sstress$ via the definition
\begin{equation} \label{eq:defsigma}
	\sstress:=\left(\gamma\id_\area + \Dpartial{\gamma}{\bfstrain}\bfstrain^T\id_\area-\bfn\otimes\Dpartial{\gamma}{\bfn}\right).
\end{equation}
in which we replaced the surface grand potential density with the surface free energy defined in \cref{eq:omegasigmagamma}.
The surface stress tensor naturally decomposes into three physically distinct contributions: ($i$) an isotropic surface free energy term, ($ii$) a strain-induced contribution representing the variation of the surface free energy with the state of strain of the system. This latter effect was first described by Shuttleworth \citep{Shuttleworth1950} and later discussed in terms of statistical mechanics in \citep{DiPasquale2020}; and ($iii$) an orientation-induced contribution describing the variation of the grand potential with the orientation is the Cahn-Hoffman component of the surface stress which we discussed in \cref{eq:def_surf_stress_n}.

\Cref{eq:defsigma} can be regarded as a unified generalization of the Shuttleworth and Cahn-Hoffman theories, which describes the stress acting on an interface if the energy depends simultaneously on the orientation and the state of strain. 

%
%

In the strain-dependent case, we note that we obtain two further conditions by considering $\dxi$ which varies smoothly across the surface, arising from cases where $\partial_n \dxi_t \neq0$. The normal derivative of the tangential perturbation can be prescribed independently of the tangential
gradient of the normal perturbation in a neighbourhood of the interface, and we obtain:
\begin{equation}\label{eq:tangFnorm}
     \cip{\id_\area \Dpartial{\omega^\area}{\bfstrain}\bfstrain^T \bfn} = 0
\end{equation}
Finally, we need to add to the variation of the energy, the variation of the entropy and density fields with the surface normal (see \cref{eq:surface-energy-var-norm}). Working through similar steps as performed in the fluid-fluid interface case for the remaining terms and including the contributions from the divergence terms in the bulk, we find that under such tangential perturbations, using the integration by parts (see \cref{app:vectoridentity}) we obtain:
\begin{align*}
\widetilde{E}^\area - E^\area & = \int_{\area} \cip{ \id_\area\cip{\Dpartial{\omega^\area}{\bfstrain}\bfstrain^T}^T\bfn - \Dpartial{\omega^\area}{\bfn} } \cdot \nabla_\area \phi \, \de a  + \int_{\area} \cip{\id_\area \Dpartial{\omega^\area}{\bfstrain}\bfstrain^T\id_\area : \nabla_\area\bfn} \, \phi \,\de a  \\
& \qquad + \int_{\area} \cip{\bfn \cdot  \Dpartial{\omega^\area}{\bfstrain}\bfstrain^T\bfn}\partial_n\phi \, \de a -\int_{\area}\cip{2H\omega^\area + [(\bfcauchy^s-\bfcauchy^l)\bfn] \cdot \bfn} \phi\,\de a + O(|\dxi|^2) \\
& = \int_{\area}  \sqp{ \id_\area \Dpartial{\omega^\area}{\bfstrain}\bfstrain^T\id_\area : \nabla_\area\bfn-\nabla_\area \cdot\cip{\id_\area\cip{\Dpartial{\omega^\area}{\bfstrain}\bfstrain^T}^T\bfn - \Dpartial{\omega^\area}{\bfn}}-\cip{2H\omega^\area - \bfn \cdot (\bfcauchy^s-\bfcauchy^l)\bfn }} \, \phi \,\de a \\
& \qquad + \int_{\area} \cip{\bfn \cdot  \Dpartial{\omega^\area}{\bfstrain}\bfstrain^T\bfn}\partial_n\phi \, \de a + O(|\dxi|^2)
\end{align*}
from which we obtain, using the possibility of independently varying $\partial_n \phi$ and $\phi$:
\begin{align}
   0 & =\nabla_\area \cdot\cip{\id_\area\cip{\Dpartial{\omega^\area}{\bfstrain}\bfstrain^T}^T\bfn - \Dpartial{\omega^\area}{\bfn}} - \id_\area \Dpartial{\omega^\area}{\bfstrain}\bfstrain^T\id_\area : \nabla_\area\bfn +2H\omega^\area - \bfn \cdot (\bfcauchy^s-\bfcauchy^l)\bfn  \label{eq:condInterf}  \\ 
   0 & =  \bfn \cdot  \Dpartial{\omega^\area}{\bfstrain}\bfstrain^T\bfn  \label{eq:condInterf2}
\end{align}
\Cref{eq:condInterf} represents the force balance in the direction normal to the interface and expresses the local condition required for equilibrium under arbitrary normal perturbations of the interface. The first term is the surface divergence of the tangential vector, and therefore describes the spatial variation along the interface of the anisotropic contribution associated with the dependence of the interfacial energy on the deformation gradient. The second term accounts for the coupling between the tangential part of the interfacial energetic stress, and the curvature tensor $\nabla_\area \bfn$. These interfacial terms are balanced by the capillary contribution $2H\omega^\area$
 and by the jump in the normal component of the bulk configurational stress $(\bfcauchy^s-\bfcauchy^l)$. We note that in the limit of independence of the energy of the strain we recover the well known expression for the jump of the stress across the interface, i.e. the Young-Laplace equation we obtain the liquid-liquid case (see \cref{eq:YLstandard})
\[  
2H\omega^\area - \bfn \cdot (\bfcauchy^s-\bfcauchy^l)\bfn = 0
\]

 \cref{eq:tangFnorm} and \cref{eq:condInterf2} can be combined into a single vectorial
condition. The former imposes the vanishing of the tangential component of
\[
\Dpartial {\omega^\area}{\bfstrain}\bfstrain^T\bfn,
\]
whereas the latter imposes the vanishing of its normal component. Using the decomposition of the identity along its normal and tangential component (see \cref{app:vectoridentity}) one obtains
\[
\Dpartial {\omega^\area}{\bfstrain}\bfstrain^T\bfn
= \id_\area \Dpartial{\omega^\area}{\bfstrain}\bfstrain^T\bfn
+ \left(
\bfn\cdot \Dpartial{\omega^\area}{\bfstrain}
\bfstrain^T\bfn
\right)\bfn.
\]
Therefore, \cref{eq:condInterf2} and \cref{eq:tangFnorm} can be equivalently condensed into
\begin{equation} \label{eq:tractionSigma}
    \Dpartial{\omega^\area}{\bfstrain} \mathbf F^T\bfn
= \mathbf 0.
\end{equation}
This condition states that the deformation-dependent interfacial energetic stress produces no traction in the direction normal to the interface.

Since the interfacial free energy is allowed to depend on the full three-dimensional deformation gradient, the Gibbs surface is assumed to be kinematically associated with the solid side of the interface. The value of $\bfstrain$ on $\area$ is therefore understood as the value obtained by a consistent extension of that field to the selected dividing surface. The condition in \mbox{\cref{eq:tractionSigma}} then constrains the deformation dependence of the interfacial free energy for the chosen, deforming Gibbs surface; it does not determine either the location of that surface or the convention by which it is selected.

Analogously, \cref{eq:tangentialSigma} and \cref{eq:condInterf} represents the tangential and normal projection of the vectorial identity:
\[
    \nabla_\area\cdot\sstress-\Big((\bfcauchy^s-\bfcauchy^l)\bfn\Big) = 0 \qquad \qquad \text{on}\,\, \area \,.
\]
The tangential projection of the previous equation is just  \cref{eq:tangentialSigma}, so let's prove that its normal projection is \cref{eq:condInterf}.

Let us start by using the surface product rule (see \cref{app:vectoridentity}):
\begin{equation}
    \label{eq:genYLstress}
    \cip{\nabla_\area\cdot \sstress}\cdot \bfn = \nabla_\area\cdot\sqp{(\sstress)^T \bfn} - \sstress : \nabla_\area \bfn 
\end{equation}
from which, using the definition of $\sstress$
\[
    (\sstress)^T\bfn = \id_\area \Dpartial{\omega^\area}{\bfstrain} \bfstrain^T \bfn -\Dpartial{\gamma}{\bfn}\, ,
\]
\[
    \sstress : \nabla_\area \bfn = \gamma\id_\area:\nabla_\area \bfn + \Dpartial{\omega^\area}{\bfstrain} \mathbf F^T \id_\area : \nabla_\area \bfn = -2H\gamma +  \Dpartial{\omega^\area}{\bfstrain} \bfstrain^T \nabla_\area \bfn\,.
\]
By putting everything together, we may now write:
\[
 \cip{\nabla_\area \cdot \sstress }\cdot \bfn = \nabla_\area \cdot\cip{\id_\area\cip{\Dpartial{\omega^\area}{\bfstrain}\bfstrain^T}^T\bfn - \Dpartial{\omega^\area}{\bfn}}-\id_\area \Dpartial{\omega^\area}{\bfstrain}\bfstrain^T\id_\area : \nabla_\area\bfn+2H\gamma
\]
showing that the normal component of the normal balance of \cref{eq:genYLstress} is equal to \cref{eq:condInterf}.

\subsubsection{Discussion}

Unlike the classical Cahn-Hoffman formulation, where the interface is treated as a two-dimensional domain and the virtual displacement field is prescribed only on the surface, the present formulation considers a three-dimensional virtual displacement field defined in a neighbourhood of the interface. That is to say that, although the interface is represented by a mathematical dividing surface, it physically represents the limit of a thin three-dimensional interfacial region and the constitutive response of the interface is allowed to depend on the complete deformation state of this region, represented by the deformation gradient evaluated at the interface. Consequently, prescribing the virtual displacement on the interface does not uniquely determine its extension into the surrounding bulk and, in particular, does not determine the normal derivative of its tangential component,
\[
\partial_n(\id_\area\dxi).
\]
Two virtual displacement fields may therefore coincide on the interface while possessing different normal derivatives, thus producing different variations of the deformation gradient through
\[
\delta\mathbf{F}=(\nabla\delta\boldsymbol{\xi})\,\mathbf{F}.
\]
Since the interfacial free energy is assumed to depend on the deformation gradient evaluated at the interface, these different bulk extensions correspond to distinct admissible perturbations of the total energy. The stationarity condition must therefore hold for arbitrary values of $\partial_n(\id_\area\dxi)$ which immediately yields the natural equilibrium condition \cref{eq:tractionSigma}.

On the other hand, the tangential gradient of the normal perturbation,
\[
\nabla_\area(\bfn\cdot\dxi),
\]
is not an arbitrary tangential vector field, but is constrained to be the surface gradient of the scalar field $\bfn\cdot\delta\boldsymbol{\xi}$. Consequently, the corresponding contribution must first be integrated by parts over the interface, leading to the differential equilibrium condition shown in \cref{eq:condInterf,eq:condInterf2}.

It should therefore be emphasised that the additional equilibrium conditions we obtained in this section are not a consequence of the constitutive assumption alone, but follows from the richer class of admissible virtual perturbations adopted in the present three-dimensional variational framework.

\subsection{Interfacial position variation}

Let us now consider the same variation described in \cref{eq:int_deform}, where now we allow for phase transformation between the solid and fluid phases at the surface between the two. While the core idea is the same as in the fluid-fluid case, an important aspect of creating new solid phase is that we now have to consider the strain induced in the regions of solid where the phase growth occurs with respect the initial lattice domain. 

In this section we first recover the classical Larch\'e-Cahn description of solid-fluid phase transformation. The key assumption is that the crystalline lattice created (or removed) during the advancement of the interface is simply a continuation of the lattice already present in the bulk solid. In other words, although the solid may already be elastically deformed, the newly created lattice is assumed to inherit exactly the same local deformation state, without introducing any additional energetic contribution associated with the creation or destruction of the lattice itself. Under this assumption, the advancement of the interface is entirely described by the thermodynamic balance between the bulk phases and the interfacial excess quantities. 

If the interface is displaced normally by $\dxi^\area$, then:
\begin{itemize}
    \item the infinitesimal change in the volume of the solid phase is $\dxi^\area\cdot \bfn\,\de a$;
    \item the infinitesimal change in the volume of the fluid phase is $-\dxi^\area\cdot \bfn\,\de a$; and
    \item the infinitesimal change in the surface area is $-2H\dxi^\area\cdot\bfn\,\de a$.
\end{itemize}
Taking into account the change of all terms in the extended functional yields the perturbation
\begin{equation*}
\begin{aligned}
\widetilde{E}-E &= \int_\area
\cip{u^s-T\eta^s-\sum_{i\in\interst}\mu_i \rho_i^s -\sum_{i\in\subst \setminus\{r\}}\diffpot_i^r \rho^s_i - \mu_r \totmolconc}\dxi^\area\cdot\bfn\,da
-\int_\area\cip{u^l-T\eta^l-\sum_{i=1}^\ncomp\mu_i\rho_i^l }\dxi^\area\cdot\bfn\,\de a \\
&\qquad -\int_\area 2H\left(u^\area-T\eta^\area-\sum_{i=1}^n \mu_i\rho_i^\area\right)\,\dxi^\area\cdot\bfn\,\de a +O(|\dxi|^2).
\end{aligned}
\end{equation*}
Writing this relation more succinctly using the grand potential densities we have already introduced for each phase and the interface, we find:
\begin{equation}\label{eq:shiftIntSol}
\widetilde{E}-E = \int_\area
\cip{\omega^s-\omega^l-2H\omega^\area}\dxi^\area\cdot\bfn\,da+O(|\dxi|^2).
\end{equation}
In other words, this entails that we must have
\begin{equation*}
    \omega^s-\omega^l = 2H\omega^\area = 2H\gamma
\end{equation*}
everywhere on the interface. In the fluid-fluid case, the grand potential density in the bulk can be identified with the pressure, and this equation is exactly the Young-Laplace equation. However, with a solid present, $\omega^s$ is not the same as the pressure, which instead also includes an elastic contribution (see \cref{eq:pressureSolids}). This reveals another fundamental change to the theory when solids are considered: equilibrium under phase transformation yields the balance equation above, while equilibrium under local distortion yields further distinct conditions.


In the following section we remove the Larch\'e-Cahn assumption for the variation of the solid phase and consider the more general situation in which the advancement of the interface explicitly accounts for the creation or removal of crystalline lattice labels. That is to say, the newly created lattice is no longer regarded as a trivial continuation of the existing one; instead, its configurational evolution is described through the transport of lattice labels. Consequently, the energetic cost of creating or removing a crystalline lattice depends on its deformation state, even though the lattice remains crystallographically coherent with the surrounding solid. This additional configurational contribution is introduced through the lattice-label conservation condition, which we call the ''strong lattice conservation" condition from which the Eshelby configurational stress naturally emerges as the thermodynamic quantity conjugate to the advancement of the solid-fluid interface. The resulting driving force then extends the Larch\'e-Cahn formulation, which we therefore call the ''weak-label conservation" condition, by including the energetic cost associated with the creation or removal of a deformed crystalline lattice.

\subsubsection{Lattice-label conservation and configurational driving force}

We now derive the configurational contribution associated with an interfacial displacement when the full lattice-label field is conserved. Let \(\vol_0^s\) be the reference domain occupied by the solid lattice, and let
\[
\bfx=\chi(\bfX)
\]
be the deformation map from the reference lattice coordinates $\bfX$ to the current spatial coordinates $\bfx$. The deformation gradient is
\[
\bfstrain=\nabla_\bfX\chi,
\qquad
J=\det \bfstrain.
\]
Let us introduce the configurational variation of the lattice, which can be described by a perturbation of the material coordinates through a smooth virtual material displacement $\dxi(\bfX)$:
\[
\widetilde{\bfX}=\Psi(\bfX)=\bfX+ \dxi(\bfX)\,.
\]
The strong conservation of the lattice labels means that the same lattice point is represented before and after the configurational relabelling. This material-coordinate variation is purely virtual and is introduced only to test configurational equilibrium. It may perturb both the crystalline reference structure and the boundary of the reference domain, so that the virtually perturbed domain need not coincide with the original domain $\vol_0$. This construction is not intended as a kinetic description of lattice growth or site creation. 

In the portion of the solid that remains solid after the variation, we impose
\[
\widetilde \chi (\widetilde \bfX) = \chi(\bfX)
\]
with inverse given by $\bfX=\Psi^{-1}(\widetilde{\bfX})$ which satisfies, to first order,
$$\bfX= \widetilde{\bfX} - \dxi(\widetilde{\bfX}) + O(|\dxi|^2)$$
The strong lattice-label conservation introduced here represents a constitutive assumption on the class of crystalline materials considered in the present work. Physically, it expresses the requirement that the crystal lattice remains a continuous material structure during admissible configurational variations, so that material points remain identifiable throughout the bulk even though the solid may deform or the solid-liquid interface may move. Under this assumption, configurational variations describe changes of the material configuration while preserving the connectivity of the underlying crystal lattice.

This assumption is naturally satisfied in coherent crystalline solids undergoing reversible elastic deformation, equilibrium crystal growth or dissolution, and other processes in which the lattice remains well defined throughout the bulk and it is not intended to describe other processes which modify topology of the crystal lattice (e.g., creation, annihilation or motion of lattice defects, dislocation nucleation or recrystallization). The present formulation should therefore be regarded as an equilibrium theory for coherent crystalline solids, while leaving more general defective crystals to future developments.

The variation of the energy $u_0^s$ of the solid in the reference state is given by:
\begin{align*}
    \tilde{u}_0^s(\widetilde{\bfX}) &= \tilde{u}_0^s(\eta_0^s(\widetilde{\bfX}), \rho_0^s(\widetilde{\bfX}); \widetilde \bfstrain (\widetilde{\bfX}))\\
    &= u_0^s + \Dpartial{u_0^s}{\strain_{ia}}\cip{\widetilde{\strain}_{ia} -\strain_{ia} } + \Dpartial{u_0^s}{\eta_0^s}(\tilde{\eta}^s_0 - \eta_0^s)  + \sum_{i=1}^\ncomp \Dpartial{u_0^s}{\rho_{0,i}^s}(\tilde{\rho}^s_{0,i} - \rho_{0,i}^s)+O(|\dxi|^2);
\end{align*}
see \cref{eq:defSol}  for the definition of the different quantities in the reference state. 

Let us start by deriving the induced variation of the deformation gradient. Since
\[
    \widetilde{\bfstrain}(\widetilde{\mathbf X})
    =
    \nabla_{\widetilde{\mathbf X}}\widetilde\chi(\widetilde{\mathbf X})
\]
and
\[
    \widetilde\chi(\widetilde{\mathbf X})
    =
    \chi(\Psi^{-1}(\widetilde{\mathbf X}))\,.
\]
The chain rule gives
\[
    \widetilde{\bfstrain}(\widetilde{\mathbf X})
    =
    \nabla_{\mathbf X}\chi(\mathbf X)\,
    \nabla_{\widetilde{\mathbf X}}\Psi^{-1}(\widetilde{\mathbf X})
\]
and using $\nabla_{\bfX}\chi(\bfX)=\bfstrain(\bfX)$, together with
\[
\nabla_{\widetilde{\bfX}}\Psi^{-1}(\widetilde{\bfX})
= \id - \nabla_{\widetilde{\bfX}}\dxi(\widetilde{\bfX})
+O(|\dxi|^2)= \id-\nabla_{\mathbf X}\dxi(\bfX) + O(|\dxi|^2)\,,
\]
(note that the gradient was expanded again to first order in $\dxi$ around $\bfX$), we finally obtain
\begin{equation}\label{eq:variationF}
    \widetilde{\bfstrain}(\widetilde{\mathbf X})
    =
    \bfstrain(\mathbf X)
    \left(
    \mathbf I-\nabla_{\mathbf X}\dxi
    \right)
        +
    O(|\dxi|^2).
\end{equation}
We note that no convective term of the form
\((\delta\boldsymbol{\xi}\cdot\nabla_{\mathbf X})\bfstrain\) appears, because
\(\widetilde{\bfstrain}(\widetilde{\mathbf X})\) is compared with
\(\bfstrain(\mathbf X)\). In other words, the variation is configurational,
rather than a fixed-\(\mathbf X\) variation of the field \(\bfstrain\).

The scalar fields are conserved during the variation (i.e., they change as material volumes), and therefore:
\[ \tilde{\eta}_0(\widetilde{\bfX})\de \tilde{v}_0 = \eta_0(\bfX)\de v_0\qquad\text{and}\qquad\tilde{\rho}_0(\widetilde{\bfX})\de \tilde{v}_0 = \rho_0(\bfX)\de v_0.\]
As in the case of \cref{eq:transf_vol}, we have that
\[
\de \tilde{v}_0 = \det(\id+\nabla_\bfX\dxi)\de v_0 = \cip{1+\nabla_\bfX\cdot\dxi + O(|\dxi|^2)}\de v_0.
\]
Again, we can write
\[
    \tilde{\eta}_0(\widetilde{\bfX}) = \eta_0(\bfX)\frac{\de v_0}{\de \tilde{v}_0} = \eta_0(\bfX)\cip{1-\nabla_\bfX\cdot\dxi(\bfX)} + O(|\dxi|^2),
\]
and an analogous expression for the density fields $\tilde{\rho}_i$.

By putting everything together we can therefore write:
\begin{align}\label{eq:energyref}
    \tilde{u}_0^s\de \tilde{v}_0 -  u_0^s\de v_0 & = \cip{ -\Dpartial{u_0^s}{F_{iA}}\cip{F_{iB}\cip{\nabla_\bfX \dxi}_{BA}} +\cip{u_0^s-\eta^s_0\Dpartial{u_0^s}{\eta^s_0}-\sum_{i=1}^\ncomp\rho^s_{0,i}\Dpartial{u_0^s}{\rho^s_{0,i}}}\nabla_\bfX \cdot \dxi +O(|\dxi|^2)}\de v_0\nonumber \\
    & = \cip{\sqp{- \bfstrain^T\Dpartial{u_0^s}{\bfstrain} + \cip{u^s_0-\eta^s_0\Dpartial{u_0^s}{\eta^s_0}-\sum_{i=1}^\ncomp\rho^s_{i,0}\Dpartial{u_0^s}{\rho^s_{0,i}} } \id}: \nabla_\bfX \dxi +O(|\dxi|^2) }\de v_0
\end{align}
in which we have defined $\cip{\nabla_\bfX \dxi}_{BA}=\Dpartial{\delta \xi_B}{X_A}$.

Since the scalar fields are transported as material densities, the corresponding constraint terms in the extended functional $E$ (see \cref{eq:enMultiPhSL}) do not contribute directly to the configurational variation. 
However, as shown before, the local variations of $\eta^s_0$ and $\rho^s_0$ still enter the variation of the internal-energy density $u^s_0(\bfstrain(\bfX),\eta^s_0(\bfX),\rho^s_0(\bfX))$. 
The perturbation to the extended functional $E$ therefore becomes
\begin{equation*}
    \widetilde{E}^s - E^s = \int_{\widetilde{V}_0} \tilde{u}_0^s\,\de \tilde{v}_0 - \int_{\vol_0}u_0^s\,\de v_0 = \int_{\vol_0} \left(\grandpzero^s\id - \bfstrain^T\Dpartial{u^s_0}{\bfF}\right):\nabla_\bfX\dxi \de v_\bfX = \int_{\vol} \eshelby : \nabla_\bfX\dxi \de v_0 + O(|\delta\boldsymbol\xi|^2) 
\end{equation*}
where we have introduced the grand potential density $\grandpzero^s$ in the reference configuration 
\[
\grandpzero^s:=u^s-\eta_0\Dpartial{u^s_0}{\eta_0}-\sum_{i=1}^\ncomp\rho_{0,i}\Dpartial{u^s_0}{\rho_{0,i}} = u^s_0-T\eta-\sum_{i\in\interst}\mu_i \rho_{0,i}^s -\sum_{i\in\subst \setminus\{r\}}\diffpot_i^r \rho^s_{0,i} - \mu_r \totmolconc_0\,,
\]
and defined the Eshelby tensor:
\[
    \eshelby = \grandpzero^s\id - \bfstrain^T\Dpartial{u^s_0}{\bfF} \,.
\]
Now, integrating by parts, we have
\[
\widetilde{E}^s-E^s = \int_{\partial \vol_0} (\eshelby\bfN) \cdot \dxi \de a_\bfX - \int_{\vol_0}(\nabla_\bfX\cdot \eshelby)\cdot \dxi \de v_\bfX + O(|\delta\boldsymbol\xi|^2) 
\]
where $\bfN$ is the normal to the boundary $\partial \vol_0$ in the reference configuration.
If, again, $\dxi$ vanishes on the boundary of the volume $\vol_0$ then we find that the tensor divergence of the Eshelby stress must vanish everywhere in the bulk at equilibrium, i.e., $\nabla_\bfX \cdot \eshelby=\boldsymbol{0}$, or in components:
\begin{equation}
    \Dpartial{\eshelby_{AB}}{X_B} = \Dpartial{}{X_B}\cip{\grandpzero^s\delta_{AB}-F_{iA}\Dpartial{u^s}{F_{iB}}} = 0.
\end{equation}
We note here that the boundary term generated by the configurational variation of the solid bulk is not associated with the purely geometrical deformation of the existing spatial interface (which should therefore be considered an inner variation, in the sense described in previous sections). Instead, this boundary term represents the contribution conjugate to a virtual normal displacement of the crystalline boundary and therefore describes the configurational driving force for a virtual advance or recession of the solid lattice. The present construction does not specify the physical creation, removal, or assignment of lattice labels during an actual growth process, since this would require a separate quantitative theory of lattice growth and dissolution kinetics.

We now focus on the contribution at the solid--fluid interface. We use the boundary value of the configurational vector field to generate a shape variation of the reference solid domain. This operation is distinct from the pure material relabelling introduced above: during the present boundary variation, the deformation map is held fixed, so that the spatial image of the solid domain changes. Let $\area_0\subset\vol_0^s$ be the reference interfacial surface, which for the moment we suppose flat, and let $\bfN$ be the unit
normal to $\area_0$ pointing outward from the solid.  For a normal configurational displacement of the interface, we write
\[
\dxi^{\area_0}=\delta\xi_0 \bfN
\qquad
\text{on } \area_0 
\]
in the reference (identified by the $0$ subscript) configuration. The corresponding displacement of the phase boundary in the current configuration is $\dxi^{\area} = \bfstrain\dxi^{\area_0}$.

Since the fluid is described in the current configuration, we write all the quantities in the current configuration and then we pull-back to the reference configuration. 
If the solid interface in the reference configuration is displaced normally by $\dxi^{\area_0}$, the corresponding infinitesimal current volume swept during interfacial motion is
\[
\delta v=(\bfstrain\dxi^{\area_0})\cdot \bfn \,\de a,
\]
where $\bfn$ is the current unit normal (i.e. the normal of the interface in the current configuration, $\area$). By Nanson's formula,
\[ \bfn \,\de a = J \bfstrain^{-T}\bfN\,\de a_0. \]
Therefore,
\begin{equation}\label{eq:curr2ref}
    \dxi^\Sigma\cdot \bfn \de a = (\bfstrain\dxi^{\area_0})\cdot \bfn \,\de a
=
(\bfstrain\dxi^{\area_0})\cdot\left(J\bfstrain^{-T}\bfN\right)\de a_0
= \jacob\,\dxi^{\area_0}\cdot \bfN\,\de a_0
= \jacob\delta \xi_0\,\de a_0\,,
\end{equation}
where we use the notation for the Jacobian determinant $J=\det{\bfstrain}$.

Then, if the interface is displaced normally by $\dxi^{\area_0}$, the leading-order configurational contribution of the extra-terms originating from this deformation is:
\begin{equation}\label{eq:EshshiftInt}
\int_{\area_0}\sqp{(\eshelby\bfN)\cdot \bfN\, \delta \xi_0
-\jacob\cip{u^l-T\eta^l-\sum_{i=1}^n \mu_i\rho_i^l}\delta\xi_0} \,\de a_0.
\end{equation}
The minus sign corresponds to the convention that $\dxi^\area>0$ advances the solid into the fluid and therefore removes a layer of fluid.

This is the key difference with respect to the scalar Larché-Cahn constraint: the contribution of the solid is not merely its scalar grand-potential density, but the normal configurational traction of the Eshelby tensor. Since the normal displacement $\delta \xi$ is arbitrary, equilibrium gives
\begin{equation}
    \bfN \cdot \eshelby \bfN = -J P^l \,.
\end{equation}
The scalar Larch\'{e}-Cahn condition may be interpreted as a simplified limit of the strong-label configurational balance. It corresponds to a growth process in which the local deformation state is held fixed as the phase boundary advances. Consequently, the interface displacement produces no incremental variation $\bfstrain$, and hence no elastic contribution to the configurational work, even though the solid may remain finitely stressed (i.e., no variation with respect to $\bfstrain$ in \cref{eq:energyref}).
In this limit the solid Eshelby tensor reduces to its isotropic grand-potential part, so that we have
\[
    \eshelby   = \grandpzero_0^s \id ,
\]
where $\grandpzero^s$ is the solid grand-potential density per unit reference volume. The weak-label equation can therefore be regarded as the reduced form obtained when the deformation state is not varied during interfacial growth

If $\omega^s$ denotes the corresponding grand-potential density per unit current volume, then
\[
    \grandpzero^s = \jacob\omega^s .
\]
In this case, we consequently obtain
\[
    \bfN \cdot \eshelby \bfN = \bfN \cdot \left(\grandpzero_0^s \id \right)\bfN = \grandpzero^s  = \jacob\omega^s .
\]
In this scalar, or weak-label, limit, the configurational equilibrium
condition becomes
\[
    \jacob\omega^s - \jacob\omega^l = 0,
\]
or equivalently,
\[
    \omega^s=\omega^l .
\]
This is precisely the scalar Larché-Cahn condition: the phase boundary is in equilibrium when the appropriate grand-potential densities of the two phases are equal.

Thus, the scalar Larché-Cahn result is obtained as the special case in which the normal configurational traction of the solid,
\[
    \bfN\cdot\eshelby\bfN,
\]
reduces to the scalar contribution \(\jacob\omega^s\). In the more general strong-label formulation, this scalar term is replaced by the full normal Eshelby traction, which also contains the energetic cost associated with advancing or removing a deformed crystalline lattice, which may change the work required to move the phase boundary.

We now remove the assumption that the interface is planar. The bulk contributions remain those reported in \cref{eq:EshshiftInt}, while the variation of the interfacial area produces an additional term, arising from the variation of the area due to the normal deformation $\dxi^\Sigma$ (see \cref{eq:varA}). This extra term is equal to:
\begin{equation*}
  -  \int_{\area}2H\cip{u^\Sigma-T\eta^\Sigma-\sum_{i=1}^\ncomp \mu_i\rho_i^\ncomp} \dxi^\area \cdot \bfn \,\de a = -  \int_{\area}2H\gamma \dxi^\area \cdot \bfn \,\de a =  -  \int_{\area_0}2H\jacob\gamma \delta \xi_0  \,\de a_0
\end{equation*}
where we used the transformation law in \cref{eq:curr2ref} in the last equality. 

In the isotropic case considered here, for which the interfacial grand
potential has no additional dependence on the interface orientation or on the deformation gradient, by combining the previous equation with \cref{eq:EshshiftInt} we finally obtain the configurational equilibrium of the advancing interface for a curved interface
\[
    (\eshelby\bfN)\cdot \bfN\, - \jacob\cip{u^l-T\eta^l-\sum_{i=1}^n \mu_i\rho_i^l}-2H\jacob\gamma = 0\,.
\]

\subsection{Frame indifference and other measures of strain}
We note that requiring the internal energy to be frame indifferent yields some additional conditions on the notions of stress we have introduced. In particular, let us assume that the internal energy of the body is locally invariant under a global rotation $\bfx\mapsto\tilde{\bfx} =  \bfR(\bfx-\bfx_0)+\bfx_0$ where $\bfR$ is a rotation tensor. Assuming that $\bfR$ is close to the identity, so can be written as $\bfR = \id+\delta \bfW+O(\delta^2)$, where $\bfW$ is a skew symmetric tensor, frame indifference entails that
$$
u^s(\eta,\rho_1,\ldots,\rho_\ncomp;\bfR\bfstrain) = u^s(\eta,\rho_1,\ldots,\rho_\ncomp;\bfstrain),
$$
and therefore through a perturbation, we obtain
$$
\Dpartial{u^s}{\bfstrain}\bfstrain^T: \bfW=0
$$
for any skew symmetric tensor $\bfW$, so that  $\Dpartial{u^s}{\bfstrain}\bfstrain^T$ must necessarily be symmetric. Since the remaining part of $\bfcauchy^s$ is automatically symmetric as it is diagonal, it follows that the entire Cauchy stress tensor must be symmetric.

Similar frame indifference considerations entail that the surface stress should satisfy a symmetry condition. Again, assuming invariance under global rotations, we must have that
$$
u^\area(\eta,\rho_1,\ldots,\rho_\ncomp,\bfR\bfn,\bfR\bfstrain) = u^\area(\eta,\rho_1,\ldots,\rho_\ncomp,\bfn,\bfstrain)
$$
%
for every rotation tensor $\bfR$. Differentiating the frame-indifference condition gives
\begin{equation*}
    0
    =
    \Dpartial{\omega^\area}{\bfstrain} : \bfW\bfstrain  +
    \Dpartial{\omega^\area}{\bfn} \cdot \bfW\bfn,
\end{equation*}
which can be rewritten as 
\begin{align*}
    0 = \left(
        \Dpartial{\omega^\area}{\bfstrain}
        \bfstrain^T
    \right)
    :
    \bfW  + \left( \Dpartial{\omega^\area}{\bfn}
        \otimes \bfn
    \right) : \bfW = \left(
        \Dpartial{\omega^\area}{\bfstrain}
        \bfstrain^T + \Dpartial{\omega^\area}{\bfn}
        \otimes \bfn
    \right) : \bfW.
\end{align*}
Using \cref{eq:skewMat}, we obtain the frame-indifference condition as
\begin{equation}\label{eq:frameInd}
    \left[
        \Dpartial{\omega^\area}{\bfstrain}
        \bfstrain^T - \bfn
        \otimes
        \Dpartial{\omega^\area}{\bfn}
    \right]
    : \bfW  =  0
\end{equation}
for every skew-symmetric tensor $\boldsymbol W$, which entails that
\begin{equation*}
    \frac{\partial\omega^\area}{\partial\bfstrain}
    \bfstrain^T
    -
    \bfn
    \otimes
    \Dpartial{\omega^\area}{\bfn}
\end{equation*}
must be symmetric. Previous equation does not imply that the surface stress $\sstress$ is symmetric. However, using the condition \cref{eq:tractionSigma} we can write
\[
    \frac{\partial\omega^\area}{\partial\bfstrain}
    \bfstrain^T\id_\Sigma
    =
    \frac{\partial\omega^\area}{\partial\bfstrain}
    \bfstrain^T.
\]
which implies that the surface stress tensor $\sstress$ is symmetric. 
Consequently, the condition in \cref{eq:tractionSigma}  derived from the independent normal derivatives of the virtual displacement ensures that the surface stress is symmetric at equilibrium. This result should be regarded as a consistency consequence of the variational formulation presented here.

At this point, we note that we could have used other measures of strain to encode the shape change in the solid phase, but we show that many other choices of strain measure can already be handled within our theory. For example, if we assume a constitutive dependence of the energy on right Cauchy-Green strain, $\bfC:=\bfF^T\bfF$, i.e. $u^s = u^s(\eta,\rho_1,\ldots,\rho_\ncomp,\bfC)$ in place of the deformation gradient $\bfF$, then using the chain rule, we would find that
$$
\Dpartial{u^s}{\bfstrain}\bfF^T = \bfF\left(\Dpartial{u^s}{\bfC}+\left(\Dpartial{u^s}{\bfC}\right)^T\right)\bfF^T,\qquad\text{so}\qquad\bfcauchy^s = \bfF\left(\Dpartial{u^s}{\bfC}+\left(\Dpartial{u^s}{\bfC}\right)^T\right)\bfF^T+\omega^s\id.
$$
It is well-known that such energies are automatically frame indifferent, and in this case, we indeed see that symmetry of the Cauchy stress is ensured automatically; on the other hand, we also see that in general, the Cauchy stress cannot be expressed in closed form as a function of $\bfC$ alone.

If instead we had parametrised in terms of the left Cauchy-Green strain $\bfB:=\bfF\bfF^T$, then again an exercise in index notation yields that
$$
\Dpartial{u^s}{\bfstrain}\bfstrain^T = \left(\Dpartial{u^s}{\bfB}^T+\Dpartial{u^s}{\bfB}\right)\bfB.
$$
In general, such energies are not frame indifferent without further assumptions on $u^s$, but simple examples are functions which are polynomial in the principal invariants of powers of $\bfB$ and its inverse. In this case, we see that we can express the Cauchy stress consistently as a function of $\bfB$ alone.

We note that in a completely analogous manner, any dependence of $u^\area$ on these two alternative measures of strain would also transform in exactly the same way.

If we linearise the Cauchy stress assuming that $\bfstrain$ is close to a reference strain, then $\bfstrain = \id+\boldsymbol{\varepsilon}+O(|\boldsymbol{\varepsilon}|^2)$, where we assume that $\boldsymbol{\varepsilon}$ is symmetric: note that skew symmetric perturbations (at least under linearisation we perform here) do not induce stress as long as the energy is frame indifferent. Including terms up to first order in $\boldsymbol{\varepsilon}$, we have
$$
\Dpartial{u^s}{\bfstrain}\bfstrain^T = \frac{\partial^2u^s}{\partial\bfstrain^2}\bigg|_{\bfstrain=\mathbf{I}}\boldsymbol{\varepsilon}+\Dpartial{u^s}{\bfF}\bigg|_{{\bfstrain=\mathbf{I}}}(\id+\boldsymbol{\varepsilon})+O(|\boldsymbol{\varepsilon}|^2).
$$
Defining the tensors
\[
\mathbb{C}_{ijkl}:=\frac{\partial^2 u^s}{\partial \strain_{ij}\partial \strain_{kl}}\bigg|_{\bfstrain=\mathbf{I}}\quad\text{and}\quad S^{s,0}_{ij}:=\frac{\partial u^s}{\partial \strain_{ij}},
\]
this can be written
$$
\Dpartial{u^s}{\bfstrain}\bfstrain^T = \bfcauchy^{s,0}+\mathbb{C}\boldsymbol{\varepsilon}+\bfcauchy^{s,0}\boldsymbol{\varepsilon}+O(|\boldsymbol{\varepsilon}|^2).
$$
Typically in elasticity theory, one assumes that the reference state about which we linearise is stress free, so that $\bfcauchy^{s,0}$ vanishes. We can view this as an assumption about the reference energy density value we use: note that enforcing that the leading order terms of the expansion in the linearised setting satisfy frame indifference for all perturbations $\boldsymbol{\varepsilon}$ leads to the condition that $\mathbb{C}$ satisfies the symmetry $\mathbb{C}_{ijkl} = \mathbb{C}_{jikl}$, along with the much stronger condition that $\bfcauchy^{s,0}$ is necessarily a purely spherical state, i.e. that $\bfcauchy^{s,0}=\omega^{s,0}\id$ for some $\omega^{s,0}$. We have given the coefficient this name, as the value of $\omega^{s,0}$ effectively shifts the reference value for the grand potential density $\omega^s$, and so the total Cauchy stress becomes
$$
\bfcauchy^s = \mathbb{C}\boldsymbol{\varepsilon}+\omega^{s,0}\boldsymbol{\varepsilon}+(\omega^{s,0}+\omega^s)\id.
$$
Equally, one could think of $\omega^{s,0}$ as a reference pressure values, $\omega^{s,0} = -P^{s,0}$.

As in the case of nonlinear strain measures, linearisation yields a similar expression for the surface stress in terms of $\boldsymbol{\varepsilon}$, where now we have
$$
\Dpartial{u^\area}{\bfstrain}\bfstrain^T = \bfcauchy^{\area,0}+\mathbb{C}^\area\boldsymbol{\varepsilon}+\bfcauchy^{\area,0}\boldsymbol{\varepsilon}+O(|\boldsymbol{\varepsilon}|^2),
$$
with $\mathbb{C}^\area$ and $\bfcauchy^{\area,0}$ defined in terms of the appropriate derivatives of $u^\area$ rather than $u^s$.

\subsection{Summary of thermodynamic equilibrium equations}

We first summarize the equilibrium equations obtained from the minimization of the total thermodynamic potential with respect to the outer and inner
variations introduced throughout this section. These equations describe the thermodynamic and mechanical equilibrium of the solid--fluid system and do not require the introduction of configurational variations of the crystalline lattice. Our analysis yields:
\begin{equation}\label{eq:solidEq}
U = TS + \sum_{i=1}^{\ncomp}\mu_i M_i + \sum_{i\in\interst}\mu_i M_i^s  + \sum_{i\in\subst \setminus\{r\}}\diffpot_i^r M^s_i + \mu_r M^s_{tot} +\int_{V^s}\omega^s(\bfstrain)\,dv - P^lV^l + \int_{\area}\gamma(\bfstrain,\bfn)\,da.
\end{equation}
where
\begin{align*}
	U & = U^s + U^l + U^\area \\
	S & = S^s + S^l + S^\area \\
	M_i & = M_i^l + M_i^\area \,, \qquad i=1,\ldots,\ncomp \\
	M^s_{tot} & = \sum_{i\in\subst} M^s_i
\end{align*}
with the convention that the symbol $M_i$ represents the total moles of the component $i$ in the liquid and at the interface or if the component is interstitial in the solid, $N^s_i$ is the mole number of the substitutional component in the solid, and $N^s_{tot}$ are total moles of the substitutional components in the solid. 

The grand potential densities are defined as:
\begin{align*}
    \omega^s&:= u^s-T\eta^s-\sum_{i\in\interst}\mu_i \rho_i^s -\sum_{i\in\subst \setminus\{r\}}\diffpot_i^r \rho^s_i - \mu_r \totmolconc\\
    \omega^l&:= u^l-T\eta^l-\sum_{i=1}^\ncomp \mu_i \rho_i^l = -P^l \\ 
    \omega^\area&:= u^\area-T\eta^\area-\sum_{i=1}^\ncomp \mu_i \rho_i^\area = \gamma,
\end{align*}
where
\[
    \Dpartial {u^s}{\rho_i^s} - \Dpartial{u^s}{\rho_r^s} = \mu_i-\mu_r = \diffpot_i^{\,r},
    \qquad i\in\subst\setminus\{r\}.
\]
Now, defining the following stress tensors:
\begin{align*}
    \bfcauchy^s & := \Dpartial{u^s}{\bfF}\bfF^T+\omega^s\id,&&\text{in }\vol^s,\\
    \bfcauchy^l & := \omega^l\id,&&\text{in }\vol^l,\\
    \sigma^\area &:= \Dpartial{\gamma}{\bfstrain}\bfstrain^T\id_\area-\bfn\otimes\Dpartial{\gamma}{\bfn}+\gamma\id_\area,&&\text{on }\area,
\end{align*}
the equilibrium equation \cref{eq:solidEq} is coupled with the following equilibrium conditions:
\begin{subequations}
    \begin{align}
\nabla\cdot\bfcauchy^i  & = \bfzero  && \vol^i,\,\,\mbox{for}\,\, i=s,l \label{eq:bulkEq}\\
\nabla_\area\cdot\sstress-\Big(\bfcauchy^s-\bfcauchy^l\Big)\bfn & = \bfzero && \text{on}\,\, \area \label{eq:finalGenYL} \\
\id_{\partial \vol^s / \,\,\area} \bfcauchy^s \bfn & = 0 && \text{on}\,\,   \partial \vol^s / \,\,\area \\
 \Dpartial{\gamma^\area}{\bfstrain} \mathbf F^T\bfn & = \bfzero && \text{on }\area  \label{eq:normEqArea}
     \end{align}    
\end{subequations}
\Cref{eq:bulkEq} expresses the mechanical equilibrium of the solid and fluid bulk phases. The tangential projection of \cref{eq:finalGenYL}  is the generalized tangential traction balance at the interface, in which the tangential component of the jump in bulk traction is balanced by the surface divergence of the surface stress, whereas its normal projection (see \cref{eq:condInterf}) represents  the generalized normal equilibrium condition. It extends the classical Young-Laplace equation by including the coupling between the interfacial energy and the deformation gradient, the orientation dependence of the interfacial energy, and the normal jump in bulk traction.  Finally, \cref{eq:normEqArea} is an additional natural equilibrium condition arising from the dependence of the interfacial energy on the full three-dimensional deformation gradient. It combines the conditions associated with the independent normal derivatives of the tangential and normal components of the virtual displacement, and requires that the interfacial energy perform no first-order virtual work through variations of the deformation gradient in the direction normal to the interface.

Accordingly, these equations express the requirement that, at equilibrium, the first variation of the energy vanishes under the following independent classes of admissible perturbations:
\begin{itemize}
    \item localized deformations within each bulk phase;
    \item tangential displacements of the interface;
    \item normal displacements of the interface;
    \item independent variations of the normal derivative of the interfacial displacement field.
\end{itemize}
\subsubsection{Configurational Equilibrium}

We need to add to the previous set of equilibrium conditions, the additional one which governs the equilibrium with respect to the advancement of the solid-fluid interface.

Previous formulation implicitly assumes that the crystalline lattice created (or removed) during the advancement of the interface is simply a continuation of the lattice already present in the bulk solid. Consequently, although the solid may be elastically deformed, no additional configurational
work is associated with the creation or destruction of the lattice itself. We now remove this assumption by introducing the  lattice-label conservation condition. In this description, lattice labels are treated as material entities and their transport is explicitly tracked through a configurational variation of the reference configuration. The advancement of the interface therefore becomes a configurational process, whose energetic contribution is measured by the Eshelby stress tensor
\begin{equation}\label{eq:eshelby}
        \eshelby = \grandpzero^s\id - \bfstrain^T\Dpartial{u^s}{\bfF} \,.
\end{equation}
with
\[
\grandpzero^s := u_0^s-T\eta^s-\sum_{i\in\interst}\mu_i \rho_{0,i}^s -\sum_{i\in\subst \setminus\{r\}}\diffpot_i^r \rho^s_{0,i} - \mu_r \totmolconc_0\,.
\]
The configurational equilibrium of the advancing interface is therefore governed by
\begin{equation}
     \bfN \cdot  \eshelby \,\bfN - J\omega^l - 2H\jacob\gamma = 0 \qquad \text{on }\area \label{eq:norm-tan-disp} \,.   
\end{equation}
If we weaken the constraint, i.e., we consider only the creation of the new lattice sites, without asking for their complete structure with respect to the pre-existing domain, from \cref{eq:norm-tan-disp} we recover the classical Larch\'{e}–Cahn description of phase transformation.
\begin{align}
        \omega^s-\omega^l & = 2H\gamma,\qquad \text{on }\area \label{eq:grandPotEq}
\end{align}
Unlike the thermodynamic equilibrium conditions summarised in the previous section, this equation does not arise from variations of the physical configuration, but from variations of the material configuration of the crystalline lattice. It therefore represents a configurational equilibrium equation describing the configurational work required to create or remove a coherent crystalline lattice during phase transformation.  In other words, while the thermodynamic equilibrium conditions summarized in previous section determine the equilibrium state of the physical system for a prescribed crystalline lattice, the configurational equilibrium derived here constitutes an additional requirement governing the energetic cost of creating or removing that lattice during phase transformation.

\section{Summary and discussion}

In this section, we summarise our findings for the reader's convenience, and provide some general discussion of the results.

\subsection{Summary}
For a reversible transformation of a system composed of a solid phase, a fluid phase, and a single solid-fluid interface, the differential form of the first law may be written as 
%
\begin{subequations}
    \begin{align}
        \de u^s
        &=
        T\,\de\eta^s
        +
        \sum_{i\in\interst}\mu_i\,\de\rho_i^s
        +
        \sum_{i\in\subst \setminus\{r\}}\diffpot_i^r \de\rho^s_{i} 
        +
        \left[
        (\bfcauchy^s-\omega^s\mathbf I)\bfstrain^{-T}
        \right]:\de\bfstrain,
        && \text{in }V^s, \label{eq:dus}
        \\
        \de u^l
        &=
        T\,\de\eta^l
        +
        \sum_{i=1}^{n}\mu_i\,\de\rho_i^l,
        && \text{in }V^l,
        \\
        \de u^\Sigma
        &=
        T\,\de\eta^\Sigma
        +
        \sum_{i=1}^{n}\mu_i\,\de\rho_i^\Sigma
        +
        \frac{\partial \gamma}{\partial\bfstrain}:\de\bfstrain
        +
        \frac{\partial \gamma}{\partial\bfn}\cdot \de\bfn,
        && \text{on }\Sigma.
    \end{align}
\end{subequations}
where we noted that the conjugate quantity to the strain variation $\de \bfstrain$ is (compare \cref{eq:defCauchyStress}):
\[
	\Dpartial{u^s}{\bfstrain} = (\bfcauchy^s-\omega^s\mathbf I)\bfstrain^{-T}\,
\]
and, since we imposed $\de \totmolconc=0$, the Gibbs relation for the solid determines only the diffusion potentials and contains no term proportional to $\de \totmolconc$. Consequently, \cref{eq:dus} represents a constrained Gibbs relation along the admissible weak-label compositional variations and its integration determines $\omega^s$ only up to a contribution proportional to the fixed total substitutional density. The contribution $\mu_r \totmolconc$ is not recovered by integrating \cref{eq:dus} which should be specified by thermodynamic information external to the restricted Gibbs relation. This indeterminacy does not affect admissible local compositional variations performed on a fixed solid domain, because $\de\totmolconc=0$. It instead becomes relevant in a phase-boundary shape variation. Indeed, even if the local total substitutional density remains fixed, displacement of the phase boundary changes the volume occupied by the solid and may therefore change the total amount of substitutional material contained in the solid phase. The resulting contribution to the configurational work is included in the configurational variation of the solid–liquid system shown in \cref{eq:defDomain}, as this equation includes the Eshelby tensor which is constructed from the complete grand-potential density $\grandpzero^s$,
If we now want to define the total differential for the whole system, by using previous equations we can write:
\begin{align}\label{eq:totdU}
	\de U & = T\,\de S + \sum_{i=1}^{\ncomp}\mu_i \,\de M_i + \sum_{i\in\interst}\mu_i \,\de M_i^s  + \sum_{i\in\subst \setminus\{r\}}\diffpot_i^r \,\de M^s_i + \mu_r \,\de M^s_{tot} \nonumber \\
    & \qquad + \int_{\vol^s} \left[ (\bfcauchy^s-\omega^s\id)\bfstrain^{-T} \right]:\de\bfstrain \,\de v - P^l\,\de V^l + \int_\area\left[ \Dpartial{\gamma}{\bfstrain}:\de\bfstrain + \Dpartial{\gamma}{\bfn}\cdot \de\bfn + \gamma \nabla_\area \cdot \dxi \right]\,\de a .
\end{align}
where $\dxi$ here denotes the infinitesimal displacement field
associated with the considered transformation of the system. It maps the initial configuration into a neighbouring configuration and determines, to first order, the corresponding variations of the deformation gradient, the interface normal, and the local surface measure. Accordingly, its restriction to $\area$ describes the displacement of the interface, while $\nabla_\area\cdot \dxi$ characterizes the local deformation induced by the transformation. No assumption is made here on the time dependence or kinetics of the transformation.

We now rewrite \cref{eq:totdU} in a different form, closer to the standard well-known form for liquid systems,  which should better highlight the different contribution which enters in the thermodynamics of the solid-liquid ones. For simplicity, in the following derivations we will consider a homogeneous deformation within the solid phase, so that the Jacobian is uniform over the reference configuration.
The generalization of next results is straightforward and includes the rewriting of the single-point function quantities as integrals over the solid domain. 

We start by decomposing the Cauchy stress in the solid into its deviatoric and spherical parts,

\[
\bfcauchy^s = \bfcauchy^s_{\mathrm{dev}}-P^s\id, \qquad P^s := -\tfrac13 \tr \bfcauchy^s,
\]
where the latter is the definition of the pressure in the solid.
Substituting the decomposition of the Cauchy stress into the bulk mechanical contribution yields
\begin{align*}
[(\bfcauchy^s-\omega^s\id)\bfstrain^{-T}]:\de\bfstrain & = 
[(\bfcauchy^s-\omega^s\id)]:\de\bfstrain(\bfstrain^{-1}) \\
& =\bfcauchy^s_{\mathrm{dev}}:\de\bfstrain(\bfstrain^{-1}) - (P^s+\omega^s)\id : \de\bfstrain(\bfstrain^{-1}) \\
& = \bfcauchy^s_{\mathrm{dev}}\bfstrain^{-T}:\de\bfstrain 
- (P^s+\omega^s)\,\de(\ln J)\\
& = \bfcauchy^s_{\mathrm{dev}}\bfstrain^{-T}:\de\bfstrain 
- \frac{P^s+\omega^s}{J}\,\de J.
\end{align*}
where we have used the identity for a generic tensor $\mathbf A$:
\[
(\mathbf A\bfstrain^{-T}):\de\bfstrain = \mathbf A:(\de\bfstrain\,\bfstrain^{-1}),
\]
together with the relation
\[
\bfstrain^{-T}:\de \bfstrain = \de(\ln J)=\frac{\de J}{J},
\]
where $J=\det\bfstrain$.
Let us now assume a deformation within the solid phase such that a small reference volume $V_0^s$ is brought to current volume $V^s=\jacob V_0^s$. Then assuming that no increment in the reference volume is made so that $\de V^s_0=0$, we have
\[
\frac{V^s}{J}\,\de J = V^s \frac{V^s_0}{V^s}\,\de\left(\frac{V^s}{V^s_0}\right)=\de V^s.
\]
In other words, if we consider variations as specified \cref{eq:totdU} do not include addition or removal of solid material, then when the volume variations are summed over the current volume $V^s$, we have that $V^s\de(\ln J) = \de V^s$ encodes the increment in total volume. The case where the reference configuration can change leads to the configurational variation of the energy and to \cref{eq:defDomain} (see discussion in \cref{app:physVSconf}).

In the remaining part of the section, in order to simplify the discussion, we assume homogeneous bulk phases and a homogeneous interfacial transformation, for which the local fields and their variations may be taken as uniform within the corresponding regions. The general case can be recovered by considering the total differential of the energy reported in \cref{eq:totdU}. 
With homogeneity assumption, the integral form of the total energy differential reduces to:
\begin{align*}
	\de U & = T\,\de S + \sum_{i=1}^{\ncomp}\mu_i \,\de M_i + \sum_{i\in\interst}\mu_i \,\de M_i^s  + \sum_{i\in\subst \setminus\{r\}}\diffpot_i^r \,\de M^s_i - \mu_r \,\de M^s_{tot} + V^s\bfcauchy_{\mathrm{dev}}^s\bfstrain^{-T}:\de\bfstrain \, \nonumber  \\
	& \qquad \qquad  -(P^s+\omega^s)\,\de V^s  - P^l\,\de V^l + \left[\Dpartial{\gamma}{\bfstrain} :\de \bfstrain + \Dpartial{\gamma}{\bfn} \cdot \de \bfn \right]A + \gamma\,\de A.
\end{align*}
Here, the reduction of the bulk and surface integrals to products with the corresponding total volume and area assumes homogeneous fields and homogeneous variations within each region. In particular, the solid contribution
$\bfcauchy_{\mathrm{dev}}^s\bfstrain^{-T}:\de \bfstrain$ is spatially uniform. Likewise, the pressure-like coefficients multiplying $\de \vol^s$ is assumed to be uniform in the corresponding bulk phase.
Likewise, the reduction of the surface integrals to products with the total area $\area$ assumes that $\partial\gamma/\partial\bfstrain$, $\partial\gamma/\partial\bfn$,
$\de \bfstrain$, and $\de \bfn$ are uniform over $\area$. Moreover, the term $\gamma\,\de A$ assumes a spatially uniform surface free-energy density $\gamma$.

Unlike the fluid phase, we see that the volumetric contribution of the solid cannot be expressed through the thermodynamic pressure $P^s$ only. The quantity conjugate to the volumetric deformation is instead the combination 
$P^s+\omega^s$, which reflects the fact that the bulk grand-potential density and the pressure are distinct quantities in an elastically deformable solid.


A similar isotropic-deviatoric decomposition can be performed for the
mechanical contribution associated with the interface. The surface case is
slightly more involved than the bulk one, since the normal and tangential
components must first be separated. For a homogeneous interface, the
interfacial contribution to the energy differential is 
\begin{equation}\label{eq:duarea}
\de U^\area = A \left[\Dpartial{\gamma}{\bfstrain}:\de \bfstrain + \Dpartial{\gamma}{\bfn}\cdot \de \bfn \right]+ \gamma\,\de A\,.
\end{equation}
It can be shown (see \cref{app:decomposition}), using the equilibrium condition in \cref{eq:normEqArea}, the frame indifference (\cref{eq:frameInd}), and the kinematic relation for the
variation of the interface normal (\cref{eq:kindn}), that \cref{eq:duarea} can be rewritten as :
\begin{equation*}
    \de U^\area = A \cip{\id_\area \cip{\Dpartial{\gamma}{\bfstrain}\bfstrain^T}\id_\area}:(\id_\area\cip{\de \bfstrain (\bfstrain^{-1})}\id_\area) + \gamma\,\de A\,
\end{equation*}
%
%
%
If we separate the deviatoric and isotropic components of the tensor appearing in the previous equation, the interfacial contribution to the differential of the energy may finally be written as
\begin{equation}\label{eq:finSigmaU}
\de U^\area =A\cip{\id_\area\frac{\partial\gamma}{\partial\bfstrain}\bfstrain^T \id_\area}_{\mathrm{dev}}:\bfE_{\area,\mathrm{dev}}+\gamma_{\mathrm{mech}}\,\de A.
\end{equation}
where we have defined the symmetric part of the tangent-tangent deformation increment
\[
    \bfE_{\area} := \operatorname{sym}\cip{\id_\area \de \bfstrain \bfstrain^{-1}\id_\area} 
\]
its deviatoric part:
\[
    \bfE_{\area,\mathrm{dev}} := \bfE_{\area} - \frac 12 \operatorname{tr}_\area \cip{\bfE_{\area}}\id_\area 
\]
and the \textit{isotropic surface-stress coefficient} $\gamma_{\mathrm{mech}}$:
\begin{equation}\label{eq:gamma-mech}
\gamma_{\mathrm{mech}} = \gamma + \frac12 \tr_\area \left(\id_\area\frac{\partial\gamma}{\partial\bfstrain}\bfstrain^T \id_\area\right),
\end{equation}
i.e., the surface free-energy density $\gamma$ plus the isotropic elastic
contribution to the surface stress.

It is important to distinguish the contribution to the total differential of the energy of the variation of the normal in \cref{eq:duarea}, which is a decomposition of the interfacial energy differential, from the surface-stress representation obtained from the minimization procedure (see the definition of the surface stress $\sstress$, \cref{eq:defsigma}). In \cref{eq:duarea}, the orientation dependence appears in its constitutive form,
\[
\frac{\partial\gamma}{\partial\mathbf n}\cdot \de\mathbf n.
\]
whereas in the variational derivation, using the normal deformation $\dxi_n$ we obtained what we called the Cahn-Hoffman contribution (i.e. the last term in the definition of the surface stress tensor $\sstress$, see \cref{eq:defsigma}). Thus, these two terms should not be regarded as two independent energetic contributions. The former is the orientation-dependent term in the fundamental differential of the energy, whereas the latter is its representation within the surface virtual-work balance.

The distinct role of the Cahn-Hoffman contribution is further clarified by its vanishing surface trace. If we apply the surface trace operator $\tr_\area(\cdot)$ to the Cahn-Hoffman term, we immediately obtain
\[
\tr_\area\left(-\bfn\otimes\Dpartial{\gamma}{\bfn} \right)=-\id_\area:
\left( \bfn\otimes \Dpartial{\gamma}{\bfn} \right) = - (\id_\area\bfn)
\cdot \Dpartial{\gamma}{\bfn} =0,
\]
since $\mathbf I_\Sigma\bfn=\mathbf0$.
The Cahn-Hoffman term therefore modifies the distribution of interfacial tractions without altering the mean isotropic surface stress. Consequently, recalling the definition of the surface stress $\sstress$ we have:
\[
    \tr_\area \sstress = 2\gamma + \tr_\area \left(\id_\area \Dpartial{\gamma}{\bfstrain} \bfstrain^T \id_\area \right),
\]
and therefore
\[
    \gamma_{\mathrm{mech}} = \tfrac12 \tr_{\area}\sstress\, 
\] 
The difference between the thermodynamic surface free energy $\gamma$ and the isotropic surface-stress coefficient $\gamma_{\mathrm{mech}}$ is therefore entirely due to the isotropic part of the strain-dependent Shuttleworth contribution, which justify the name given to $\gamma_{\mathrm{mech}}$. The orientation-dependent term instead enters the anisotropic and mixed response of the interface through the surface-stress balance, and the generalized Young-Laplace equation.

We can now write the differential of the energy for a solid-liquid system with an interface as:
\begin{align}\label{eq:finaldU}
	\de U & = T\,\de S + \sum_{i=1}^{\ncomp}\mu_i\,\de M_i + \sum_{i\in\interst}\mu_i \,\de M_i^s  + \sum_{i\in\subst \setminus\{r\}}\diffpot_i^r\,\de M^s_i + \mu_r \,\de M^s_{tot}  + V^s\bfcauchy_{\mathrm{dev}}^s\bfstrain^{-T}:\de\bfstrain \nonumber \\
	& \qquad \qquad  -(P^s+\omega^s)\de V^s - P^l\,\de V^l + A\cip{ \Dpartial{\gamma}{\bfstrain}\bfstrain^T}_{\mathrm{dev}}:\left(\id_\area\,\de\bfstrain\, \bfstrain^{-1}\id_\area\right)_{\mathrm{dev}}+\gamma_{\mathrm{mech}}\,\de A,
\end{align}
where we have replaced the tensor $\bfC$ introduced for convenience above with its full definition.

One important point to highlight here is that the equation derived for $\de U$ (i.e. \cref{eq:totdU} or \cref{eq:finaldU}) accounts only for changes of the physical configuration. When the material configuration itself is allowed to vary, an additional configurational work contribution appears, which is conjugate to the Eshelby stress tensor rather than the Cauchy stress:
\begin{equation}\label{eq:defDomain}
	\de_c U_c = \int_{V_0} \eshelby : \nabla_\bfX \dxi\, \de v_0
\end{equation}
where $\dxi$ is now the transformation of the reference domain $V_0$. The total variation of the energy $U$, therefore includes the two contributions $\de U$ for any variation which do not change the amount of solid and liquid phases, and $\de_c U$ if a solid transformation from the reference configuration is involved. We prefer to keep the two variations of the total energy $U$ (i.e., $\de U$ and $\de_c U$) separated, as the configurational contribution in \cref{eq:defDomain} should not be interpreted as an additional correction to the physical first law reported in \cref{eq:totdU}, but rather as an independent variation associated with changes in the material configuration through phase change.

\subsection{Discussion}
In this last part we now revisit some results reported in the literature for different applications, in the light of the present variational framework.

\paragraph{Nucleation theory:} In molecular dynamics studies of crystal nucleation, several authors have reported findings which apparently contradict with the classical Young--Laplace equation. In particular, simulations of Lennard--Jones systems \citep{Gunawardana2018} and hard-sphere systems \citep{Montero2020,deJager2024} found that the actual mechanical pressure inside the crystal nucleus is lower than that in the surrounding liquid. Direct application of the classical Young--Laplace equation would therefore imply a negative surface free energy (see \cref{eq:YLstandard}, where phase $k$ denotes the solid and phase $j$ the liquid), in contradiction with the thermodynamic definition of interfacial free energy.

To rationalize this apparent inconsistency, Montero \emph{et al.}~\citep{Montero2026} distinguished between two different Young--Laplace equations. The first is formulated in terms of the thermodynamic pressure of a bulk solid at the same chemical potential as the surrounding liquid and involves the surface free energy. The second is written in terms of the actual mechanical pressure inside the nucleus and replaces the surface free energy with the interfacial stress \citep{Cacciuto2005,tenWolde1998,deJager2024}. Within the interpretation provided, the two equations describe different physical quantities and therefore need not predict the same pressure difference across the interface. Studying ice nucleation with the TIP4P/Ice model, the authors found that the mechanical and thermodynamic pressures are remarkably close for the critical nucleus under the investigated conditions, leading to comparable values of the interfacial stress and the surface free energy. However, by analysing the planar basal interface, they showed that the interfacial stress can be nearly twice the interfacial free energy, concluding that the apparent agreement observed for the nucleus is likely system-dependent and possibly coincidental rather than a general property of solid--liquid interfaces.

The distinction introduced in \citet{Montero2026} emerges naturally within the present variational framework. In our formulation, the two Young-Laplace-type relations discussed by Montero \textit{et al.} arise from independent equilibrium conditions associated with different classes of admissible variations. The condition governing the advancement of the solid--fluid phase boundary is obtained by varying the position of the interface and involves the surface grand-potential density, thereby recovering the thermodynamic Young--Laplace equation given by \cref{eq:grandPotEq}. By contrast, the generalized Young--Laplace equation in \cref{eq:finalGenYL}, obtained from the inner variation, expresses the local mechanical balance of tractions at the interface and naturally involves the surface-stress tensor. The two equations therefore represent distinct equilibrium requirements: the former expresses thermodynamic equilibrium with respect to phase-boundary motion within the weak-label setting, whereas the latter enforces local mechanical equilibrium at the interface. This distinction disappears in the fluid limit, where the surface stress reduces to the isotropic surface free energy and the mechanical and configurational conditions collapse to the conventional Young--Laplace equation. For solid--fluid interfaces, however, they generally remain distinct.

\Cref{eq:grandPotEq} has the same formal structure as the capillary condition conventionally employed in classical nucleation theory and may therefore be regarded as its counterpart within the weak-label formulation. This correspondence does not, however, establish that crystalline nucleation must obey weak-label kinematics. The distinction between the different configurational conditions is hidden in the usual fluid formulation of nucleation because, in that limit, they reduce to the same Young--Laplace relation. For a solid nucleus, the appropriate configurational condition depends on how lattice or material labels are created, transferred, or conserved during nucleation and growth. It is therefore possible that a complete theory of crystalline nucleation should also incorporate the strong-label condition, but establishing this would require an explicit model of the corresponding label kinematics and lies beyond the scope of the present work. Accordingly, \cref{eq:grandPotEq} is presented here as the weak-label analogue of the conventional capillary nucleation condition, rather than as a definitive criterion for crystalline nucleation. Independently of this configurational issue, \cref{eq:finalGenYL} provides the local mechanical-equilibrium condition that the interfacial stress state must satisfy.

\paragraph{Configurational mechanics:} The configurational variation derived here has the same variational structure as the configurational work introduced by Eshelby \citep{Eshelby75} and subsequently developed within configurational mechanics by Gurtin \citep{Gurtin1995,Gurtin2000}. In particular, \Cref{eq:defDomain} recovers the standard bulk configurational work associated with the Eshelby energy-momentum tensor. In the present formulation, this result emerges naturally from the thermodynamic variational principle once the strong lattice-label conservation condition is imposed. The role of this constraint is to distinguish variations of the material lattice from ordinary variations of the spatial configuration, thereby providing the kinematic relation required for the appearance of the Eshelby stress.

The present formulation is restricted to a single crystalline solid in contact with a liquid. Consequently, the configurational variation is defined only within the solid, where the material lattice is preserved, and leads to the configurational traction associated with the bulk Eshelby tensor. This differs from the configurational driving force employed in theories of coherent solid-solid interfaces, where two crystalline phases possessing independent material configurations meet across the interface (see e.g, \citep{Frolov2012,Frolov2012b}) and where the driving force can be expressed as the jump of the Eshelby tensor across the interface owing to the presence of two material configurations (see e.g., \citep{Markenscoff2010,Gross2002}). 
An extension of the present framework to coherent solid-solid systems would therefore require the introduction of independent configurational fields for both solids, allowing the configurational traction to be expressed in terms of the jump of the Eshelby tensor across the interface.

Nevertheless, the present result establishes a direct connection between interfacial thermodynamics and configurational mechanics. It shows that the bulk configurational work follows naturally from the equilibrium thermodynamic formulation, without introducing configurational forces as an independent postulate. In this sense, the present framework provides a thermodynamic basis for configurational mechanics at equilibrium.

It should also be emphasized that the present theory and the configurational mechanics framework developed by Gurtin are not equivalent. Gurtin's formulation provides a general description of the kinematics and evolution of material configurations, including moving phase boundaries and other configurational changes, together with the associated configurational balance laws. By contrast, the present work is intentionally restricted to equilibrium thermodynamics and does not address interface kinetics or evolution equations. The purpose of the present comparison is therefore not to establish an equivalence between the two theories, but rather to demonstrate that the equilibrium configurational structure recovered here is consistent with the bulk equilibrium limit of configurational mechanics. This correspondence suggests a possible route toward extending the present equilibrium formulation to evolving interfaces. Such an extension would require supplementing the present thermodynamic framework with the kinematic description and configurational balance laws developed within configurational mechanics, together with appropriate kinetic relations governing interface motion.

Finally, the configurational variations considered here are assumed to be smooth and compatible with the preservation of the crystalline lattice. Consequently, the present theory does not address configurational singularities such as cracks, dislocations or other lattice defects, whose treatment would require extending the admissible configurational variations beyond the assumptions adopted in this work, following the broader framework of configurational mechanics.

\paragraph{Crystal-shape theories:} Another potential implication of the present variational framework concerns the thermodynamics of equilibrium crystal shapes. In the classical Wulff construction, the equilibrium morphology of a crystal is obtained by minimizing the total interfacial free energy under a constant-volume constraint, assuming that the surface free energy depends only on the interface orientation, $\gamma=\gamma(\bfn)$ \citep{Herring1951}. Modern formulations of anisotropic surface thermodynamics express this variational problem through the Cahn-Hoffman vector, which provides a unified description of orientation-dependent surface energetics and forms the basis for generalized Wulff constructions applicable to faceted crystals and anisotropic interfaces \citep{Wheeler1999,Cahn1974,Hoffman1972,Cahn1996}.

Although equilibrium crystal shapes are beyond the scope of the present work, the present formulation naturally recovers the Cahn-Hoffman contribution as part of the interfacial stress tensor while simultaneously accounting for the dependence of the interfacial free energy on the local deformation through $\gamma=\gamma(\bfstrain,\bfn)$. Consequently, the proposed variational framework contains the essential thermodynamic ingredients required to describe anisotropic crystalline interfaces beyond the assumptions of the classical Wulff construction. In particular, allowing the interfacial free energy to depend on both orientation and deformation suggests a possible extension of equilibrium shape theories to deformable crystalline interfaces, where elastic effects, surface stress and anisotropic surface energetics are treated within a single thermodynamic framework.

From this perspective, the local equilibrium conditions derived in the present work, including the generalized Young-Laplace equation, the Shuttleworth relation and the Cahn-Hoffman contribution, may be regarded as the local stationarity conditions that any equilibrium crystal morphology must satisfy. While a complete variational derivation of equilibrium crystal shapes is left for future work, the present formulation establishes a unified thermodynamic framework from which both local interfacial equilibrium and, potentially, global equilibrium morphologies can be consistently derived.


\paragraph{Connection with the Kirkwood-Buff mechanical route:} A further implication of the present formulation concerns the mechanical determination of interfacial free energies. As shown in \cref{sec:KB}, the classical Kirkwood-Buff expression is naturally recovered as a limiting case of the present variational framework, corresponding to a hydrostatic bulk stress together with a deformation-independent interfacial free energy. In this limit, the mechanical surface stress coincides with the surface free energy and the classical pressure-tensor route is recovered. Instead of introducing the Kirkwood-Buff expression as an independent mechanical definition of the surface free energy, the present derivation shows that it follows directly from the general thermodynamic formulation under the appropriate constitutive assumptions.

This result also clarifies the relationship between the mechanical and thermodynamic routes to the determination of interfacial free energies.  The mechanical route emerges whenever the mechanical surface stress and the interfacial free energy are identical, whereas the thermodynamic route remains valid in the general case. The origin of the discrepancy between the two approaches is therefore not a limitation of the mechanical formalism itself, but the breakdown of the assumptions under which the Kirkwood-Buff relation is derived.

This observation is particularly relevant for crystalline solid-liquid interfaces. In this case, the interfacial free energy generally depends on both the interface orientation and the local deformation, giving rise to the Shuttleworth contribution and to the Cahn-Hoffman term. As a consequence, the mechanical surface stress is no longer equal to the surface free energy, and the classical pressure-tensor expression cannot be directly interpreted as an interfacial free energy. The present theory therefore explains, within a unified thermodynamic framework, both the success of the Kirkwood-Buff route for fluid interfaces and its limited applicability to crystalline solid-liquid interfaces.

\section{Conclusion}

In this work we have presented a unified variational framework for the thermodynamics of fluid-fluid and solid-fluid interfaces. Starting from a single constrained minimization principle, we derived the equilibrium conditions associated with the different classes of admissible variations, showing that many of the classical relations of interfacial thermodynamics arise naturally within the same mathematical setting.

Starting from the same variational framework, Gibbs' thermodynamics and the Young-Laplace equation are recovered for fluid interfaces. Allowing the interfacial free energy to depend on the local interface orientation recovers the Cahn-Hoffman theory, while introducing a dependence on the deformation state naturally yields the Shuttleworth relation for solid interfaces.

Extending the analysis to systems containing a crystalline solid required the introduction of the Larch\'e-Cahn lattice-site constraint, from which the corresponding bulk  thermodynamic equilibrium relations were recovered. Unlike fluids, the presence of elasticity implies that the bulk grand potential density and the pressure are no longer equivalent quantities, leading to a generalized description of equilibrium for solid-fluid systems.

A central aspect of the present work is the distinction between thermodynamic equilibrium, obtained through variations of the physical configuration, and configurational equilibrium, obtained through variations of the material configuration of the crystalline lattice. The former governs the mechanical and thermodynamic balance of the phases and the interface, whereas the latter introduces an additional configurational driving force associated with the creation or removal of a coherent crystalline lattice. Within this framework, the Eshelby stress emerges naturally as the quantity thermodynamically conjugate to configurational variations, providing a direct connection between configurational mechanics and interfacial thermodynamics.

More generally, the present work suggests that several classical theories of heterogeneous equilibrium, including Gibbs thermodynamics, Larché-Cahn theory, Shuttleworth elasticity, Cahn-Hoffman anisotropy, the Kirkwood-Buff mechanical route and configurational thermodynamics, can be interpreted as complementary stationarity conditions associated with different classes of admissible variations of a single constrained thermodynamic functional.

The variational structure developed here provides also a natural starting point for future extensions to more complex situations, including coherent and incoherent solid-solid interfaces and other heterogeneous systems. We hope that the present framework will provide a useful theoretical basis for interpreting atomistic simulations and for developing more rigorous computational methods for the determination of interfacial thermodynamic properties.

\begin{appendices}

\makeatletter
\renewcommand{\@seccntformat}[1]{%
  \ifcsname the#1\endcsname
    \appendixname~\csname the#1\endcsname:\quad
  \fi
}
\makeatother

\section{Compilation of vectorial and tensorial properties}\label{app:vectoridentity}

We collect here the main tensorial and vectorial identities and equalities used throughout the derivation in the main text.

\begin{itemize}
    \item We write the components of the gradient of a generic vector $\bfa$, $\nabla \bfa$ as
\[
    (\nabla \bfa)_{ij} = \partial_j a_i\,. \qquad  (\nabla a)^T_{ij} = \partial_i a_j
\]
where $\partial_j$ is the derivative with respect the $j$-th coordinate. Therefore, given another vector $\bfn$ we have
\[
    [(\nabla \bfa)^T\bfn]_i = \partial_i a_j n_j
\]
The scalar product of previous expression with a vector $\bfb$ now becomes:
\begin{equation}\label{eq:nbTens_a}
\bfb \cdot (\nabla \bfa^T \bfn) = b_i \partial_i a_j n_j = b_i n_j  \partial_i a_j = n_i b_j  \partial_j a_i =\bfn \otimes \bfb : \nabla\bfa\,.
\end{equation}
In the last passage we have just exchanged $i$ with $j$ as they are both dummy indexes.

\item Let us now considering the tensor products such that given two vectors $\bfa$ and $\bfb$ and a symmetric matrix $\bfP$. We can write
\begin{equation}\label{eq:SymmDot}
     (\bfa \otimes \bfb)\bfP = (\bfa \otimes \bfP\bfb)   
\end{equation}
and in the last passage we used the fact that given three matrices $\bfA,\bfB,\bfP$ with $\bfP$ symmetric:
\[
\bfA\bfP:\bfB = A_{ij}P_{jk}B_{ik}=A_{ij}P_{jk}B_{ik}= A_{jk}B_{ik}P_{kj} = \bfA:\bfB\bfP,
\]
Other identities involving the tensor product includes:
\begin{equation*}
    \left(
        \boldsymbol a\otimes\boldsymbol b
    \right)
    :
    \boldsymbol A
    =
    \boldsymbol a\cdot
    \left(
        \boldsymbol A\boldsymbol b
    \right),
\end{equation*}
which entails, for a skew-symmetric matrix $\bfW$:
\begin{equation}\label{eq:skewMat}
    \left( \bfa\otimes\bfb \right) : \bfW
    = - \left( \bfb\otimes\bfa\right) : \bfW,
\end{equation}
and:
\begin{subequations}
    \begin{align}
        \bfC(\bfn \otimes \bfn) & = (\bfC\bfn) \otimes \bfn \label{eq:tens1}  \\
        (\bfn \otimes \bfn)\bfC & = \bfn \otimes (\bfC^T\bfn)    \label{eq:tens2}   \\ 
        (\bfa\otimes \bfb)(\bfc\otimes\bfd) & = (\bfb \cdot \bfc )(\bfa  \otimes \bfd)\,. \label{eq:tens3} 
    \end{align}
\end{subequations}
also, given two vectors $\bfa$ and $\bfb$:
\[
    (\bfa \otimes \bfn):(\bfb \otimes \bfn) = (\bfa\cdot \bfb)(\bfn \cdot \bfn) = \bfa\cdot \bfb =  (\bfn \otimes \bfa):(\bfn \otimes \bfb) 
\]

\item Given a surface $\area$ with normal field $\bfn$, we can define the projector operator on the tangential directions to the surface $\id_\area$ as:
\[
    \id_\area = \id - \bfn \otimes \bfn 
\]
with the properties
\[  
    \id_\area^2 = \id_\area \qquad \id_\area \bfn = 0 \qquad \id_\area^T = \id_\area\,.
\]
From the definition of surface gradient of the normal follows that
\begin{equation*}
    (\nabla_\area \bfn)\bfn = \nabla \bfn\,\id_\area\bfn
    = \boldsymbol 0.
\end{equation*}
Hence, $\nabla_\area\bfn$ acts only on tangential directions and may be projected on the right without being modified. Moreover, since $\bfn$ is a unit vector, the surface gradient gives
\begin{equation*}
    \nabla_\area(\bfn\cdot\bfn) = 2(\nabla_\area\bfn)^T\bfn =   \boldsymbol 0,
\end{equation*}
from which
\begin{equation*}
    (\nabla_\Sigma\bfn)^T\bfn =
    \boldsymbol 0.
\end{equation*}
Then, the image of $\nabla_\area\bfn$ is also tangential, so that a left projection does not change it. Consequently,
\begin{equation} \label{eq:idSurfident}
    \nabla_\area\bfn =\id_\area \nabla_\area\bfn \id_\area.
\end{equation}
Thus, in any double contraction involving $\nabla_\Sigma\bfn$, only the tangential-tangential part of the other tensor contributes.
Using the projector onto the interface $\id_\area$, the gradient of an arbitrary vector field $\boldsymbol{v}$ can be written as
\begin{equation*}
    \nabla\bfv = \nabla_\area\bfv +    \partial_n\bfv\otimes\bfn,
\end{equation*}
where
\begin{equation*}
    \nabla_\area\bfv := \nabla\bfv\,\id_\area,
    \qquad \partial_n\bfv := \nabla\bfv\,\bfn \,.
\end{equation*}
The operator $\nabla_\area$ is the surface gradient.
The gradient of an arbitrary scalar field $\phi$ can therefore be written as
\begin{equation*}
    \nabla \phi = \nabla_\Sigma \phi + (\partial_n \phi)\bfn,
\end{equation*}
where
\begin{equation*}
    \nabla_\Sigma \phi := \id_\area \nabla \phi,
    \qquad \partial_n \phi := \nabla \phi\cdot\bfn\,.
\end{equation*}
we can also show that
\[
    \bfv \cdot \nabla_\area \phi = \bfv \cdot \id_\area \nabla_\area \phi = (\id_\area^T \bfv) \cdot \nabla_\area \phi = (\id_\area \bfv) \cdot \nabla_\area \phi\,.
\]
\item For a tangential vector field $\bfv$ and a scalar field $\phi$ defined
on the interface $\area$, the surface product rule gives
\begin{equation*}
    \nabla_\Sigma\cdot(\phi\bfv)
    =
    \bfv\cdot\nabla_\area\phi
    +
    \phi\,\nabla_\Sigma\cdot\boldsymbol v.
\end{equation*}
Therefore,
\begin{equation*}
    \boldsymbol v\cdot\nabla_\Sigma\phi
    =
    \nabla_\Sigma\cdot(\phi\boldsymbol v)
    -
    \phi\,\nabla_\Sigma\cdot\boldsymbol v.
\end{equation*}
Integrating over $\Sigma$ and applying the surface divergence theorem yields
\begin{equation*}
    \int_\area
    \bfv\cdot\nabla_\area\phi\,da
    = -\int_\area 
    \phi\,\nabla_\area\cdot\bfv\,da
    + \int_{\partial\area}
    \phi\,
    \bfv\cdot\bfnu_\area\,d\ell,
\end{equation*}
where $\boldsymbol\nu_\area$ is the outward conormal to $\partial\area$, tangent to $\area$.

\item For a second-order tensor field $\bfA$ and a tangential vector field $\bfv$, the surface product rule reads
\begin{equation*}
    \nabla_\area\cdot \left( \bfA^T\bfv\right)=    \left( \nabla_\area\cdot\bfA 
    \right)\cdot\bfv +
    \bfA:\nabla_\area\bfv.
\end{equation*}
Integration over $\Sigma$ therefore gives
\begin{align}
    \int_\area \bfA:   \nabla_\area\bfv\,\de a = -\int_\area
    \left(
        \nabla_\area\cdot\bfA
    \right)\cdot\bfv\,\de a +
    \int_{\partial\area}
    \left(
        \bfA\bfnu_\area
    \right)\cdot\bfv\,\de\ell .
\end{align}
Since $\bfv$ is tangential, only the tangential component of $\nabla_\area\cdot\bfA$ contributes:
\begin{equation}
    \left(
        \nabla_\Sigma\cdot\boldsymbol A
    \right)\cdot\delta\boldsymbol\xi_t
    =
    \left[
        \id_\Sigma
        \left(
            \nabla_\Sigma\cdot\boldsymbol A
        \right)
    \right]\cdot\delta\boldsymbol\xi_t.
\end{equation}

\end{itemize}

\section{Perturbation field at the solid-liquid interface}\label{app:perturbation}

We now show the full decomposition of the perturbation field into its tangential and normal components with respect to the interface $\area$:
\begin{equation*}
    \dxi = \dxi_t + \phi\bfn,
    \qquad \dxi_t := \id_\area \dxi,   
    \qquad \delta\xi^t := |\dxi_t| ,       
    \qquad \phi := \bfn\cdot\dxi,
\end{equation*}
Applying the previous identities to $\dxi$ we can write
\begin{equation*}
    \nabla\dxi = \nabla_\area\dxi + \partial_n\dxi
    \otimes\bfn.
\end{equation*}
The tangential-gradient contribution is
\begin{align*}
    \nabla_\area\dxi&=\nabla_\area \left(\dxi_t +  \phi\bfn \right) = \nabla_\area\dxi_t +  \bfn\otimes\nabla_\area\phi
    + \phi\nabla_\area\bfn,
\end{align*}
where the product rule has been used. Similarly,
\begin{align*}
    \partial_n\dxi & = \partial_n \left(
        \dxi_t  + \phi\bfn \right) = \partial_n\dxi_t  +    (\partial_n\phi)\bfn +
    \phi\,\partial_n\bfn.
\end{align*}
We extend the normal field away from the interface by keeping it constant along the normal lines, so that
\begin{equation}\label{eq:extnormlines}
    \partial_n\boldsymbol{n} = \boldsymbol{0}.
\end{equation}
The complete decomposition of the perturbation gradient is therefore
\begin{equation}
    \nabla\dxi =\nabla_\area\dxi_t + \bfn\otimes\nabla_\area\phi  +  \phi\nabla_\area\bfn  +  \partial_n\dxi_t
    \otimes\bfn + (\partial_n\phi) \bfn\otimes\bfn.
    \label{eq:perturbation_gradient_decomposition}
\end{equation}
The five terms in  \cref{eq:perturbation_gradient_decomposition} respectively describe the tangential variation of the tangential displacement, the tangential variation of the normal-displacement amplitude, the geometric contribution associated with the curvature of the interface, the normal variation of the tangential displacement, and the normal variation of the normal-displacement amplitude. We now decompose the only two terms in \cref{eq:solid_surf_var} which are factors of the full deformation gradient, according to \cref{eq:perturbation_gradient_decomposition}:
\begin{itemize}
\item $\displaystyle
    \Dpartial{\omega^\area}{\bfn}\id_\area(\nabla_\area\dxi_t)^T\bfn = \Dpartial{\omega^\area}{n_i}(\id_\area)_{ij} \partial^\area_j \delta\xi^{t}_{k}  n_k = n_k \Dpartial{\omega^\area}{n_i} \partial^\area_i \delta\xi^{t}_{k} = \cip{\bfn\otimes \Dpartial{\omega^\area}{\bfn}} : \nabla_\area\dxi_t 
$ 
\item  $ \displaystyle
    \Dpartial{\omega^\area}{\bfn}\id_\area(\bfn\otimes\nabla_\area\phi)^T\bfn = \Dpartial{\omega^\area}{n_i}(\id_\area)_{ij} (\partial^\area_j \phi ) n_k n_k = \Dpartial{\omega^\area}{n_i} (\partial^\area_i \phi ) = \Dpartial{\omega^\area}{\bfn} \cdot \nabla_\area \phi  
$
\item  $ \displaystyle
    \Dpartial{\omega^\area}{\bfn}\id_\area(\phi\nabla_\area\bfn )^T\bfn = \Dpartial{\omega^\area}{n_i}(\id_\area)_{ij} \phi\, (\partial^\area_j n_k) n_k = 0
$
%
\item  $ \displaystyle
    \Dpartial{\omega^\area}{\bfn}\id_\area(\partial_n\dxi_t\otimes\bfn )^T\bfn = \Dpartial{\omega^\area}{n_i}(\id_\area)_{ij}(\partial_n\delta\xi^t)_kn_jn_k =0
$
because
$ \displaystyle
    \left[ \id_\area (\bfn\otimes\bfn) \right]_{ik} = (\id_\area)_{ij}n_jn_k
    = 0,
$
%
%
\item $ \displaystyle
    \Dpartial{\omega^\area}{\bfn}\id_\area[(\partial_n\phi) \bfn\otimes\bfn]^T\bfn = \Dpartial{\omega^\area}{n_i}(\id_\area)_{ij}(\partial_n\phi) n_j n_k n_k  = \Dpartial{\omega^\area}{n_i} (\partial_n\phi)  (\id_\area)_{ij} n_j = 0
$ 
\item 
$ \displaystyle
    \Dpartial{\omega^\area}{\bfstrain}\bfstrain^T:\nabla_\area \dxi_t = \Dpartial{\omega^\area}{\bfstrain}\bfstrain^T \id_\area :\nabla_\area \dxi_t 
$
\item 
 $ \displaystyle
    \Dpartial{\omega^\area}{\bfstrain}\bfstrain^T:\bfn\otimes\nabla_\area\phi= \Dpartial{\omega^\area}{\strain_{ia}}\strain_{ja} n_i \partial^\area_j\phi = \id_\area\cip{\Dpartial{\omega^\area}{\bfstrain}\bfstrain^T}^T\bfn \cdot \nabla_\area \phi
$ \\
where we premultiplied the expression by $\id_\area$ to highlight the fact that we are considering only the tangential components (as can be seen by the scalar product with a purely tangential quantity).
\item 
$ \displaystyle
    \Dpartial{\omega^\area}{\bfstrain}\bfstrain^T:\phi\nabla_\area\bfn=  \phi \Dpartial{\omega^\area}{\bfstrain}\bfstrain^T: \id_\area \nabla_\area\bfn\id_\area =  \cip{\id_\area \Dpartial{\omega^\area}{\bfstrain}\bfstrain^T\id_\area : \nabla_\area\bfn} \phi
$ \\
see \cref{eq:idSurfident}.
\item 
 $ \displaystyle
    \Dpartial{\omega^\area}{\bfstrain}\bfstrain^T:\partial_n\dxi_t\otimes\bfn= \Dpartial{\omega^\area}{\strain_{ia}}\strain_{ja}(\partial_n\delta\xi_t)_i n_j = \cip{\Dpartial{\omega^\area}{\bfstrain}\bfstrain^T \bfn} \cdot \partial_n\dxi_t = \cip{\id_\area\Dpartial{\omega^\area}{\bfstrain}\bfstrain^T \bfn} \cdot \partial_n\dxi_t
$ \\
where the last equality comes from the fact that, under the definition we use for $\dxi_t=\id_\area\dxi$ and the assumption on the extension of the normal field along the normal lines (see \cref{eq:extnormlines}),
then 
\[
\partial_n\id_\area = -(\partial_n\bfn)\otimes\bfn
-\bfn\otimes(\partial_n\bfn)
=
\boldsymbol 0.
\]
and
\[
    \partial_n\dxi_t = \id_\area\partial_n\dxi,
\]
so that $\partial_n\delta\boldsymbol\xi_t$ is tangential and $\partial_n\delta\boldsymbol\xi_t=\id_\area\partial_n\delta\boldsymbol\xi_t$.
\item 
$ \displaystyle
    \Dpartial{\omega^\area}{\bfstrain}\bfstrain^T:(\partial_n\phi)\bfn\otimes\bfn= \Dpartial{\omega^\area}{\strain_{ia}}\strain_{ja}(\partial_n\phi) n_in_j =  \cip{\bfn \cdot  \Dpartial{\omega^\area}{\bfstrain}\bfstrain^T\bfn}\partial_n\phi
$

\end{itemize}

\section{Physical versus configurational variations} \label{app:physVSconf}

The differential form of the first law derived in the previous section describes variations of the physical configuration while keeping the material configuration fixed. In particular, the deformation of the solid is entirely described by variations of the deformation gradient, whereas the reference configuration remains unchanged. Under these assumptions, the reference volume of the solid is constant and $\de V_0^s=0$, so that
\[
    V^s = \int_{V_0^s}\jacob\,\de V_0,
\]
gives
\[
    \de V^s = \int_{V_0^s}\de\jacob\,\de V_0.
\]
For a homogeneous deformation, this reduces to
\[
    \de V^s = V^s\,\de(\ln \jacob),
\]
which is the relation employed in the previous section to separate the volumetric and deviatoric contributions to the mechanical work.

The situation changes when the advancement of the solid-fluid interface is accompanied by the creation or removal of crystalline lattice. In this case, the material domain itself is no longer fixed, and the reference volume becomes an independent variable,
\[
dV_0^s\neq0.
\]
The variation of the current volume therefore contains two distinct contributions,
\[
\delta V^s = \int_{V_0^s}\delta J\,dV_0 + \int_{\partial V_0^s}
J\,\delta\mathbf X\cdot\mathbf N\,dA_0,
\]
where the first term originates from the deformation of the existing lattice, whereas the second accounts for the creation or removal of material points.

It is important to emphasize that the second contribution should not be introduced as an additional volumetric term in the first law. Indeed, a variation of the material domain also modifies the deformation gradient according to the kinematic relation imposed by the strong lattice-label conservation condition as shown in \cref{eq:variationF}. Consequently, the configurational variation of the bulk energy receives two
contributions: one associated with the transport of the material domain and another associated with the corresponding variation of the deformation gradient. These two terms combine naturally into the Eshelby stress tensor (see \cref{eq:eshelby}),
\begin{equation*}
        \eshelby = \grandpzero^s\id - \bfstrain^T\Dpartial{u^s}{\bfF} \,.
\end{equation*}
leading to the configurational work
\[
\de_c U_c = \int_{V_0^s} \eshelby:\nabla_X(\delta\mathbf X)
\,dV_0.
\]
The differential relation in \cref{eq:finaldU} corresponds to variations of the physical configuration only, whereas the configurational variation gives rise to a second equilibrium condition governed by the Eshelby stress tensor reported here and in \cref{eq:defDomain}. The two variational problems are therefore complementary: the former describes the mechanical and thermodynamic response of a prescribed crystalline lattice, while the latter governs the  energetic cost associated with the creation or removal of the lattice itself during phase transformation.

\section{Derivation of the decomposition of the surface term} \label{app:decomposition}

Let us start from \cref{eq:duarea}:
\begin{equation*}
    \de U^\area = A \left[\Dpartial{\gamma}{\bfstrain}:\de \bfstrain + \Dpartial{\gamma}{\bfn}\cdot \de \bfn \right]+ \gamma\,\de A 
\end{equation*}
For convenience, let us introduce
\[
    \bfC := \Dpartial{\gamma}{\bfstrain}\bfstrain^T, \qquad \bfD := \de \bfstrain (\bfstrain^{-1})\,, \qquad \bfg := \Dpartial{\gamma}{\bfn}
\]
so that
\begin{equation}\label{eq:surfTerms}
        \Dpartial{\gamma}{\bfstrain}:\de \bfstrain =  \bfC : \bfD\,.
\end{equation} 
A generic tensor can be decomposed into four components, 
\[  
    \bfC = \id_\area \bfC\id_\area + \id_\area \bfC  (\bfn \otimes \bfn) +  (\bfn \otimes \bfn) \bfC  \id_\area +  (\bfn \otimes \bfn) \bfC  (\bfn \otimes \bfn)
\]
each of these components represent the tangential-tangential, the  the tangential-normal, normal-tangential, and the normal-normal components respectively.
The contraction then becomes
\begin{align*}
\bfC :  \bfD & = \left(\id_\area\bfC \id_\area \right) : \left( \id_\area \bfD \id_\area \right) \nonumber\\
&+
\id_\area\bfC  (\bfn \otimes \bfn)
:
\id_\area \bfD (\bfn \otimes \bfn)
\nonumber\\
&+
(\bfn \otimes \bfn) \bfC  \id_\area
:
(\bfn \otimes \bfn) \bfD \id_\area
\nonumber\\
&+
 (\bfn \otimes \bfn) \bfC  (\bfn \otimes \bfn)  : (\bfn \otimes \bfn) \bfD (\bfn \otimes \bfn).
\end{align*}
where we noted that out of the 16 possible combinations of the  four terms for $\bfC$ and four for $\bfD$, all cross-contractions between distinct normal-tangential blocks vanish
because they involve the scalar product of a tangential vector with a normal vector.

We now use the properties \cref{eq:tens1,eq:tens2} to rewrite previous equation as:
\begin{align*}
    \id_\area\bfC  (\bfn \otimes \bfn)
:
    \id_\area \bfD (\bfn \otimes \bfn) &  =
    (\id_\area\bfC\bfn) \otimes \bfn
:
    (\id_\area \bfD \bfn) \otimes \bfn =
    (\id_\area\bfC\bfn) \cdot (\id_\area \bfD \bfn)\,, \\
    (\bfn \otimes \bfn) \bfC  \id_\area
    :
    (\bfn \otimes \bfn) \bfD \id_\area & =
    \bfn \otimes (\bfC  \id_\area)^T\bfn 
    :
    \bfn \otimes (\bfD  \id_\area)^T\bfn  \\
    & =
    \bfn \otimes \id_\area\bfC^T\bfn 
    :
    \bfn \otimes \id_\area \bfD^T\bfn =
    (\id_\area \bfC^T\bfn) \cdot (\id_\area \bfD^T\bfn) = \bfn\cdot \bfC\id_\area \bfD^T \bfn 
    \,, \\
  (\bfn \otimes \bfn) \bfC  (\bfn \otimes \bfn)  : (\bfn \otimes \bfn) \bfD (\bfn \otimes \bfn) & =
 (\bfn \otimes \bfn) (\bfC\bfn  \otimes \bfn) :  (\bfn \otimes \bfn) (\bfD \bfn \otimes \bfn)  \\
  & = (\bfn\cdot \bfC\bfn)(\bfn  \otimes \bfn) : (\bfn\cdot \bfD\bfn) (\bfn  \otimes \bfn) = (\bfn\cdot \bfC\bfn)(\bfn\cdot \bfD\bfn)
\end{align*}
where the last equality follows from \cref{eq:tens3}. 

We finally obtain:
\begin{align}\label{eq:surface-block-decomposition}
\bfC:  \bfD & = \left(
\id_\area\bfC\id_\area \right) : \left( \id_\area \bfD \id_\area \right) +
\left( \id_\area\bfC\bfn \right)
\cdot
\left( \id_\area \bfD \bfn \right)  +
\left( \id_\area{\bfC}^{T}\bfn \right)
\cdot
\left( \id_\area \bfD^T \bfn \right) +
\left( \bfn\cdot\bfC\bfn \right) \left( \bfn\cdot \bfD \bfn
\right).
\end{align}
Imposing the equilibrium condition derived in \cref{eq:normEqArea} (i.e. $\bfC\bfn=0$) we can see that the second and fourth terms of the previous equations are zero. 
Let us analyse the third term. We notice first that the variation of the normal $\de \bfn$ appearing in \cref{eq:duarea} can be written as:
\begin{equation}\label{eq:kindn}
        \de \bfn = - \id_\area \bfD^T \bfn \,
\end{equation}
which can be obtained by the variation relation for the normal
\[
    \bfn^\prime  = \frac{(\id + \bfD)^{-T}\bfn}{|(\id + \bfD)^{-T}\bfn|}\,,
\]
and that since $\gamma$ depends on the orientation through the unit normal, $\partial\gamma/\partial\bfn := \bfg$ is understood here as a tangent vector to $\area$, which, along the condition \cref{eq:normEqArea}, allows us to write (by defining $\bfG:=\bfC - \bfn \otimes \bfg$):
\[
    \bfG\bfn=\bfC\bfn - (\bfn \otimes \bfg)\bfn = 0
\]
Use the frame indifference property (see \cref{eq:frameInd}), then $\bfG^T=\bfG$, and therefore, using the previous equation:
\[
    \bfC^T\bfn = (\bfg \otimes \bfn)\bfn = \bfg = \Dpartial{\gamma}{\bfn}
\]
Since $\partial \gamma/\partial \bfn$ is a tangent vector, then 
\[
    \bfC^T\bfn =  \id_\area  \bfC^T\bfn
\]
We can therefore write:
\[  
    (\id_\area \bfC^T\bfn)\cdot(\id_\area\bfD^T\bfn) + \Dpartial{\gamma}{\bfn}  \cdot \de \bfn= (\id_\area \bfC^T\bfn)\cdot(\id_\area\bfD^T\bfn)  + \bfC^T\bfn \cdot (- \id_\area \bfD^T \bfn) = 0
\]
such that
\begin{equation*}
    \de U^\area = A \cip{\id_\area \bfC\id_\area}:(\id_\area\bfD\id_\area) + \gamma\,\de A\,
\end{equation*}
The remaining term in \cref{eq:surface-block-decomposition} is composed by two tensors which contain only purely tangential components (they are multiplied on the left and on the right by the surface projector $\id_\area$) and therefore acts entirely within the tangent plane. They therefore admit the conventional two-dimensional isotropic-deviatoric decomposition.

Let
\begin{equation*}
\bfC_{\parallel} := \id_\area \bfC \id_\area = \id_\area \Dpartial{\gamma}{\bfstrain}\bfstrain^T\id_\area,
\end{equation*}
and decompose it as
\begin{equation*}
    \bfC_{\parallel} = \bfC_{\mathrm{dev}}  + \frac12 \tr_\area \left(    \bfC_{\parallel}\right)\id_\area,
\end{equation*}
with
\begin{equation*}
    \bfC_{\mathrm{dev}} = \bfC_{\parallel} - \frac12 \tr_\area \left(\bfC_{\parallel}\right)\id_\area,\qquad\tr_\area\bfC_{\mathrm{dev}}=0.
\end{equation*}
and where we indicated the deviatoric part with the subscript $(\cdot)_{\mathrm{dev}}$. We note that $\bfC_{\parallel}$ is symmetric, as $\bfC_{\parallel} = \id_\area \bfG \id_\area$ 
because $\id_\area(\bfn\otimes\bfg)
\id_\area=\boldsymbol 0$, and since $\bfG$ is symmetric, $\bfC_\parallel$ is symmetric. Therefore, only the symmetric part of the tangent-tangent deformation increment contributes to the interfacial mechanical work. Introducing:
\[
    \boldsymbol E_\Sigma
    :=
    \operatorname{sym}
    \left(
        \id_\area
        \bfD
       \id_\area
    \right),
\]
we have
\[
    \bfC_{\parallel}:
    \left(
        \id_\area
        \bfD
        \id_\area
    \right)
    =
    \bfC_{\parallel}:\bfE_\Sigma,
\]
because the contraction of a symmetric tensor with an antisymmetric tensor vanishes, we have
\[
\bfC_\parallel:
\operatorname{skw}
\left(
\id_\area\bfD\id_\area
\right)
=0\,.
\]

The tensor $\boldsymbol E_\Sigma$ may then be decomposed into its tangential
deviatoric and isotropic parts as
\[
    \boldsymbol E_\Sigma
    =
    \boldsymbol E_{\Sigma,\mathrm{dev}}
    +
    \frac{1}{2}
    \operatorname{tr}_\Sigma(\boldsymbol E_\Sigma)
    \boldsymbol I_\Sigma,
\]
with
\[
    \boldsymbol E_{\Sigma,\mathrm{dev}}
    =
    \boldsymbol E_\Sigma
    -
    \frac{1}{2}
    \operatorname{tr}_\Sigma(\boldsymbol E_\Sigma)
    \boldsymbol I_\Sigma.
\]
Accordingly,
\[
    \boldsymbol C_{\parallel}:\boldsymbol E_\Sigma
    =
    \boldsymbol C_{\mathrm{dev}}:
    \bfE_{\area,\mathrm{dev}}
    +
    \frac{1}{2}
    \operatorname{tr}_\Sigma(\boldsymbol C_{\parallel})
    \operatorname{tr}_\Sigma(\boldsymbol E_\Sigma).
\]
For a homogeneous variation of the interface,
\[
    \operatorname{tr}_\Sigma(\boldsymbol E_\Sigma)
    =
    \frac{\mathrm dA}{A},
\]
so that the isotropic part of $\boldsymbol E_\Sigma$ accounts for the area
change, whereas its deviatoric part describes an in-plane distortion at fixed
area.
The tangent-tangent contribution together with the term accounting for the deformation of the area may be written as
\begin{align*}
    A\bfC_{\parallel}:\operatorname{sym}\left(\id_\area \,\bfD\id_\area\right)+\gamma\,\de A=A\bfC_{\mathrm{dev}}:\bfE_{\area,\mathrm{dev}}+\gamma_{\mathrm{mech}}\,\de A,
\end{align*}
where $\gamma_{\mathrm{mech}}$ is defined as:
\begin{equation*}
\gamma_{\mathrm{mech}} = \gamma + \frac12 \tr_\area \left(\id_\area\frac{\partial\gamma}{\partial\bfstrain}\bfstrain^T \id_\area\right),
\end{equation*}
At equilibrium, the interfacial contribution to the differential of the energy may finally be written as
\begin{equation*}
\de U^\area =A\cip{\id_\area\frac{\partial\gamma}{\partial\bfstrain}\bfstrain^T \id_\area}_{\mathrm{dev}}:\bfE_{\area,\mathrm{dev}}+\gamma_{\mathrm{mech}}\,\de A.
\end{equation*}
which is \cref{eq:finSigmaU}.

\section{The mechanical route to the surface free energy}\label{sec:KB}

The analysis we reported in the main paper also allows us to establish a connection between the present sharp-interface formulation and the classical mechanical route for the determination of interfacial properties using molecular dynamics simulations. The latter is commonly derived by considering the distribution of the three-dimensional Cauchy stress across a finite interfacial region. In contrast, the present theory represents the interface as a Gibbs dividing surface carrying its own surface stress tensor.
The two descriptions can be related by requiring that the resultant interfacial traction be the same.

For simplicity, let us consider a planar interface whose unit normal is
\(
\bfn=\mathbf e_z 
\),
and let $\id_\area$ the projector onto the interface $\area$.
Let $\bfcauchy(z)$ be the local Cauchy stress through the diffuse interfacial region. Introducing the sharp-interface reference state,
\[
\bfcauchy^{\mathrm{ref}}(z) = \bfcauchy^sH(-z) + \bfcauchy^lH(z),
\]
where \(H\) is the Heaviside function, and the location $z=0$ specifies the chosen dividing surface.  We can define an excess stress as
\[
\bfcauchy^{\mathrm{ex}}(z)=\bfcauchy(z)-\bfcauchy^{\mathrm{ref}}(z).
\]
Its tangential resultant defines the tangential--tangential part of the mechanical surface stress:
\[
    \id_\area \sstress \id_\area = \int_{-\infty}^{+\infty} \id_\area
    \bfcauchy^{\mathrm{ex}}(z) \id_\area \,dz.
\]
For a solid-fluid interface, this expression must be understood as a tensorial excess construction relative to the actual bulk stress states. In particular, the solid reference stress need not be hydrostatic.

Using the definition of the surface stress in \cref{eq:defsigma}, and its tangential-tangential projection on a plane is
\[
    \id_\area \sstress \id_\area = \gamma \id_\area + \id_\area \frac{\partial\gamma}{\partial\bfstrain}
    \bfstrain^T \id_\area,
\]
since 
\[
    \id_\area \left( \bfn\otimes \frac{\partial\gamma}{\partial\bfn} \right)\id_\area=\mathbf0\,.
\]
Thus, the orientation-dependent Cahn-Hoffman term does not contribute directly to the tangential-tangential projection of the mechanical surface stress. The strain-dependent Shuttleworth contribution, instead, generally remains.

The classical Kirkwood-Buff relation is recovered only in the special case of a planar fluid--fluid interface. In this limit, both bulk phases are hydrostatic and mechanical equilibrium requires the normal pressure to be
constant across the interface,
\begin{equation*}
    P_N(z)=P_N=P.
\end{equation*}
The local stress may be written as
\begin{equation*}
    \bfcauchy(z)  =  -P_T(z)\id_\area - P_N\bfn\otimes\bfn.
\end{equation*}
The corresponding hydrostatic bulk reference stress is
\begin{equation*}
    \bfcauchy^{\mathrm{ref}} = -P\id,
\end{equation*}
so that
\begin{equation*}
    \bfcauchy^{\mathrm{ex}}(z) = \left[P_N-P_T(z)\right]\id_\area.
\end{equation*}
Consequently,
\begin{equation*}
    \id_\area \sstress \id_\area  = \left[ \int_{-\infty}^{+\infty} \left(P_N-P_T(z)\right)\,dz
    \right] \id_\area.
\end{equation*}
If, in addition, the interfacial free energy is independent of deformation, the surface stress is isotropic,
\begin{equation*}
    \sstress      =     \gamma\id_\area,
\end{equation*}
and the preceding equation reduces to the classical Kirkwood-Buff formula \citep{Kirkwood1949}:
\begin{equation*}
    \gamma  = \int_{-\infty}^{+\infty}
    \left[P_N-P_T(z)\right]\,dz.
\end{equation*}
This identification is specific to the planar fluid-fluid limit. For a solid-fluid interface, the solid bulk may sustain a non-hydrostatic stress, and the mechanical surface stress generally differs from the surface free energy because of the Shuttleworth contribution. The mechanical route must therefore be formulated as a tensorial excess relative to the appropriate bulk reference stresses and cannot, in general, be identified directly with
$\gamma$.

This observation also clarifies the conceptual distinction between thermodynamic and mechanical routes for determining interfacial properties. For a homogeneous interface, the interfacial contribution to the grand potential is
\[
\Omega^\Sigma = \int_\Sigma \omega^\Sigma\,da = \omega^\Sigma A.
\]
Consequently, the reversible work required to create interfacial area at
fixed temperature, chemical potentials, deformation state, and orientation
satisfies
\[
\left(
\frac{\partial\Omega^\Sigma}{\partial A} \right)_{T,\mu_i,\bfstrain,\bfn} = \omega^\Sigma = \gamma.
\]
Thermodynamic-integration methods therefore yield the surface grand-potential density when the integration path creates the interface reversibly while keeping its orientation and deformation state fixed. If the creation of interfacial area is accompanied by changes in strain or orientation, additional Shuttleworth and Cahn-Hoffman contributions enter in the definition of the reversible work of the transformation (for an overview of such methods and some applications see \citep{DiPasquale2025,DiPasquale2020,DiPasquale2022}).

Mechanical methods, on the other hand, probe the surface stress tensor, or suitable projections thereof which only for fluid interfaces coincide with $\gamma$. In contrast, for solid-fluid  interfaces the strain-dependent contribution introduces a difference between the thermodynamic and mechanical descriptions of the interface, while the orientation-dependent Cahn-Hoffman contribution modifies only the anisotropic distribution of the surface stress without changing its mean isotropic value. 
\end{appendices}
 
\bibliographystyle{unsrtnat}
\bibliography{bibliography_20Sep24}

\end{document}